\documentclass[12pt]{article}
\usepackage{jheppub}
\pdfoutput=1
\usepackage{epstopdf}

\usepackage{jheppub}
\usepackage{mathrsfs}
\usepackage{psfrag}
\usepackage{color}
\usepackage{hyperref}
\usepackage{subcaption}
\usepackage{placeins}

\usepackage{slashed}
\usepackage{simplewick}
\usepackage{cancel}

\allowdisplaybreaks

\usepackage[table]{xcolor}

\usepackage{amsmath,bbm,array,amsfonts,graphicx,wrapfig,lscape,float,multirow,longtable,rotating,makecell}
\usepackage{url}

\newcommand{\smallplus}{\hspace{0.5pt}\text{{\small+}}\hspace{-0.5pt}}

\newcommand{\pl}{\smallplus}
\newcommand{\la}{\langle}
\newcommand{\ra}{\rangle}
\newcommand{\del}{\partial}
\def\lr{\leftrightarrow}

\def\to{\rightarrow}
\def\e{\epsilon}
\def\ve{\varepsilon}

\newcommand{\be}{\begin{equation}}
\newcommand{\ee}{\end{equation}}
\newcommand{\bea}{\begin{eqnarray}}
\newcommand{\eea}{\end{eqnarray}}
\newcommand{\eqn}[1]{eq.~(\ref{#1})}

\definecolor{airforceblue}{rgb}{0.36, 0.54, 0.66}
\definecolor{bananayellow}{rgb}{1.0, 0.88, 0.21}
\definecolor{bittersweet}{rgb}{1.0, 0.44, 0.37}
\definecolor{blue(ncs)}{rgb}{0.0, 0.53, 0.74}
\definecolor{bole}{rgb}{0.325, 0.27, 0.23}
\definecolor{brass}{rgb}{0.71, 0.65, 0.26}
\definecolor{bronze}{rgb}{0.8, 0.5, 0.2}
\definecolor{brgreen}{rgb}{0.0, 0.26, 0.15}
\definecolor{burgundy}{rgb}{0.5, 0.0, 0.13}
\definecolor{cherry}{rgb}{1.0, 0.72, 0.77}
\definecolor{cocao}{rgb}{0.82, 0.41, 0.12}
\definecolor{citrine}{rgb}{0.99, 0.82, 0.07}

\title{Multi-Loop Positivity of the\\
Planar \texorpdfstring{${\cal N}=4$}{N=4} SYM Six-Point Amplitude}

\author{Lance~J.~Dixon,$^1$}
\author{Matt von Hippel,$^2$}
\author{Andrew~J.~McLeod,$^1$}
\author{Jaroslav Trnka$^3$}

\affiliation{$^1$ SLAC National Accelerator Laboratory,
Stanford University, Stanford, CA 94309, USA}

\affiliation{$^2$ Perimeter Institute for Theoretical Physics, 
Waterloo, Ontario N2L 2Y5, Canada}

\affiliation{$^3$ Center for Quantum Mathematics and Physics (QMAP),\\ 
Department of Physics, University of California, Davis, CA 95616, USA}

\emailAdd{lance@slac.stanford.edu, mvonhippel@perimeterinstitute.ca}
\emailAdd{ajmcleod@stanford.edu, trnka@ucdavis.edu}

\abstract{We study the six-point NMHV ratio function in planar ${\cal N}=4$ SYM theory in the context of positive geometry. The Amplituhedron construction of the integrand for the amplitudes provides a kinematical region in which the integrand was observed to be positive. It is natural to conjecture that this property survives integration, i.e.~that the final result for the ratio function is also positive in this region. Establishing such a result would imply that preserving positivity is a surprising property of the Minkowski contour of integration and it might indicate some deeper underlying structure. We find that the ratio function is positive everywhere we have tested it, including analytic results for special kinematical regions at one and two loops, as well as robust numerical evidence through five loops. There is also evidence for not just positivity, but monotonicity in a ``radial'' direction.  We also investigate positivity of the MHV six-gluon amplitude.  While the remainder function ceases to be positive at four loops, the BDS-like normalized MHV amplitude appears to be positive through five loops.}

\dedicated{Dedicated to John Schwarz on his 75$^{th}$ birthday}

\preprint{
\begin{flushright} SLAC--PUB--16873 \end{flushright}
}

\begin{document}
\maketitle

\section{Introduction}

There has been substantial progress from many different perspectives in understanding and calculating perturbative scattering amplitudes in ${\cal N}=4$ super-Yang-Mills theory~\cite{Neq4SYML}, particularly in the planar limit of a large number of colors. The standard Feynman diagram expansion, as well as more modern methods such as generalized unitarity, are based on the expansion of the (multi)loop amplitude in terms of different sets of building blocks. These pieces are then individually integrated over the loop momenta, and the final amplitude corresponds to the sum over all terms. In recent years, it was shown that both the total integrand and the final amplitudes enjoy some extraordinary properties. As it turns out, there is a completely different way to think about each quantity, holistically and without reference to any expansion in building blocks.

For the integrand there exists a complete geometric reformulation in terms of the Amplituhedron, which is a generalization of projective polygons into Grassmannians~\cite{ArkaniHamed2013jha,ArkaniHamed2013kca} (see also refs.~\cite{Bai2014cna,Franco2014csa,Bai2015qoa,Ferro2015grk,Galloni2016iuj,Karp2016uax} for recent progress). The idea is to rewrite the kinematical and helicity variables in terms of bosonized momentum twistors $Z$ serving as vertices of a geometric object -- the Amplituhedron -- whose volume is equal to the integrand of scattering amplitudes in planar ${\cal N}=4$ SYM. The definition of this space involves a generalization of the positive Grassmannian that appears in the context of on-shell diagrams~\cite{ArkaniHamed2012nw}.

On the other hand, there has also been great progress in understanding the space of transcendental functions that contains the final amplitudes.  In many cases these functions are iterated integrals~\cite{Chen}, also known as multiple polylogarithms~\cite{FBThesis,Gonch}.  The weight, or number of integrations, is $2\ell$ for perturbative amplitudes at loop order $\ell$. While the origin of these functions comes from the ``dlog'' structure of the integrand, the precise connection is still not understood in general.  For example, there may be obstructions to carrying out the dlog integrations in terms of iterated integrals.  The two-loop equal-mass sunrise integral is in this elliptic class~\cite{Laporta2004rb}, as is an integral entering the N$^3$MHV 10-point scattering amplitude in planar ${\cal N}=4$ SYM~\cite{CaronHuot2012ab}. However, it has been argued that MHV and NMHV amplitudes in this theory should be expressible solely in terms of multiple polylogarithms~\cite{ArkaniHamed2012nw,LipsteinMason}.

A function composed of multiple polylogarithms has a {\it symbol}~\cite{Goncharov2010jf}, which is constructed essentially by repeated differentiation of the function.  The alphabet, or set of letters appearing in the symbol, characterizes the function space.  These letters seem to be closely related to cluster algebras~\cite{ClusterCoordinates,SpradlinTwoLoopMHV}.  Once one knows the alphabet, as well as where the branch cuts are located, one can construct the function space iteratively.  The number of such functions turns out to be much smaller than the number of independent physical constraints on them, allowing for a unique determination of the amplitude as a whole without ever inspecting the precise integrand or its decomposition into building blocks.  This program has been carried out for the six-point amplitude through five loops~\cite{Dixon2011pw,Dixon2011nj,Dixon2014iba,Dixon2014voa,Dixon2015iva,CaronHuot2016owq}, and for the symbol of the seven-point amplitude through three loops~\cite{Drummond2014ffa}.

Given this excellent progress in understanding both the integrand and amplitude holistically, it would be great to bring them together.  It is not clear yet how the properties of the Amplituhedron extend from the integrand to the final amplitudes.  However, there is an extension of the Amplituhedron conjecture, namely the existence of the {\it dual} Amplituhedron, which we will test indirectly in this paper. In ref.~\cite{ArkaniHamed2014dca} it was argued that if the original Amplituhedron can be reformulated into a dual picture where the integrand is directly a volume of this space, then this function should be positive when evaluated inside the Amplituhedron. This positivity property has been verified explicitly for various integrands up to high loop order.  It also turns out to be true for the integrand of the ratio function -- a ratio of amplitudes with different helicities which is free of infrared (IR) divergences. 

It was then conjectured that this positivity property might also hold for the final transcendental function, rather than just the integrand.  In general, the transcendental functions that determine scattering amplitudes are complex-valued. However, there exists a Euclidean region in which the amplitude is real-valued, and thus it is possible to define positivity consistently. For the six-point amplitude, the cross-ratios $u,v,w$ are all real and positive in this Euclidean region. The conjecture is that the quantities under consideration are positive in a subregion of this Euclidean region that is selected by the properties of the Amplituhedron.

This conjecture was explicitly verified at one loop.  In this paper we will check the statement through five loops for the NMHV case, providing strong evidence that the conjecture is indeed true. In addition, we show that the same is true for the IR-finite BDS-like normalized MHV amplitude. There are many ways to subtract IR divergences but the positivity conjecture more or less singles out this function. The positivity property is very non-trivial and we do not know how to prove it in full generality even at one loop, not to mention higher-loop examples where our analytic understanding is even more limited.

To show a simple example, let us consider a function of positive variables $u,w>0$,
\begin{equation}
F(u,w) = {\rm Li}_2(1-u) + {\rm Li}_2(1-w) + \log u \log w - \zeta_2 \,.
\label{Fexample}
\end{equation}
This function will appear later in this paper in a particular limit of the NMHV one-loop ratio function, as well as of the BDS-like remainder function.  In the first case the Amplituhedron picture dictates that $F(u,w)<0$ whenever $u+w>1$, while in the second case it requires $F(u,w)>0$ for $u+w<1$. Even in this simple case positivity is not manifest, i.e.~the answer cannot be decomposed into a sum of obviously positive terms (although the positivity proof here is simple, see section~\ref{one_loop_ds_limit}).  Note that for $w=1-u$ we get the famous dilogarithm identity which sets $F(u,1-u)=0$, which also represents a physical vanishing condition on the ratio function in a collinear limit.

In general, positivity relies not only on the sign of transcendental functions like $F(u,w)$, but also on the sign of rational prefactors. For generic kinematics neither has uniform sign on its own. Nevertheless, the sign ambiguities of these individual parts conspire to produce quantities with uniform sign. The statement is even more interesting because not only the bosonic external data, but also the fermionic variables, play a crucial role in establishing this surprising and remarkable property. In the rest of this paper we will flesh out this statement, showcasing numerous regions in which positivity holds.

In this paper, whenever we refer to positivity, we mean that perturbative coefficients in the loop expansion of a given quantity are positive when the expansion parameter is the {\it negative} of the 't Hooft coupling, $-\lambda=-g^2 N_c$.  Or, in terms of a standard, positive 't Hooft coupling (or multiple thereof), we will be testing for strict {\it sign-alternation} with loop order.  That is, one-loop terms should be negative, two-loop terms positive, three-loop terms negative, and so on.  From the point of view of the (dual) Amplituhedron, the overall sign at a given loop order is not dictated; what is really expected is a {\it uniform} sign as a function of the kinematics.  However, we know empirically that the sign alternates for low loop orders, and we also expect it to alternate at very high loop orders.  The reason for the latter statement is that planar ${\cal N}=4$ SYM has no renormalons and no instantons, and so it is expected to have a finite radius of convergence of the perturbation theory.  For some quantities, the radius of convergence is known: it is $\lambda_c=\pi^2$ for the light-like cusp anomalous dimension~\cite{Beisert2006ez}, and $\lambda_c\approx 14.7$ for the Bremsstrahlung function, which is another limit of the velocity-dependent cusp anomalous dimension~\cite{Correa2012at,Henn2013wfa}. These quantities have no singularity on the positive $\lambda$ axis. Hence their finite radius of convergence is controlled by a singularity for negative $\lambda$.  This fact implies sign alternation at very large perturbative orders, with successive perturbative coefficients increasing by a factor that approaches $-1/\lambda_c$.

This paper is organized as follows.  We begin in section \ref{Amplituhedron} by describing the regions in which the Amplituhedron construction leads to positive tree-level amplitudes; these regions are where we wish to test the corresponding loop amplitudes for positivity.  Section \ref{OneLoop} then presents some simple one-loop examples in which this positivity holds for the NMHV ratio function. We also define the double-scaling limit, in which certain monotonicity properties of the amplitude are manifest. In section \ref{DSLimit} we explore this limit at higher loops, both analytically on certain special lines and numerically throughout the full region. We go on in section \ref{Bulk} to present numerical evidence for positivity outside of special limits, in the full space of cross-ratios selected by the Amplituhedron construction.  Section \ref{MHVpos} discusses the positivity properties of the MHV amplitude, and we conclude in section \ref{concl}.

This paper has two appendices. Appendix \ref{appendix_line_u0} provides additional plots on the line $w=0$ within the double-scaling surface, while appendix \ref{c21monotonicity} proves positivity and monotonicity for a quantity, $c^{(2)}_1(u,w)$, relevant at two loops. We also attach ancillary files containing expressions for the quantities we consider on special lines threading the kinematic space.

\section{From the Amplituhedron to positive kinematics}
\label{Amplituhedron}

In this section we review the essential ingredients of the Amplituhedron
construction of the multi-loop integrand for planar ${\cal N}=4$ SYM,
and show how this setup dictates where we should inspect the multi-loop
six-point amplitudes for positivity.

The Amplituhedron space~\cite{ArkaniHamed2013jha,ArkaniHamed2013kca}
${\cal Y}$ is implicitly labeled by $n$, $k$, and $\ell$,
where $n$ is the number of external legs, $k$ is the number of negative
gluon helicities minus 2, and $\ell$ is the loop order.
The formal definition of ${\cal Y}$ is given by the matrix multiplication
\begin{equation}
{\cal Y} = {\cal C}\cdot Z, \label{Amp}
\end{equation}
where ${\cal C}$ is a $(k+2\ell)\times n$ matrix with certain positivity properties, and $Z$ is an $n\times(4+k)$ matrix with all $(4+k)\times(4+k)$ minors positive. The matrix $Z$ corresponds to external data (momentum twistors and Grassmann variables); $Z$ only depends on $k$ while the ${\cal C}$ matrix also depends on $\ell$. The loop integrand $\Omega$ is then a form which behaves logarithmically on the boundaries of ${\cal Y}$.

The conjecture made in ref.~\cite{ArkaniHamed2014dca} is that the form $\Omega$ is positive when the measure is stripped off and it is evaluated inside the Amplituhedron, i.e.~for ${\cal Y}$ satisfying \eqn{Amp} with positive ${\cal C}$ and $Z$ matrices. This property does not follow from the original Amplituhedron proposal.  Rather it would provide evidence for the existence of a ``dual Amplituhedron'' of which $\Omega$ is literally the volume.  This space has not been found yet, but the fact that $\Omega$ is observed to be positive is very encouraging.

Let us now consider the final amplitude rather than the integrand. It has a very complicated branch-cut structure, but no dependence on the loop momenta. If an Amplituhedron-like construction exists for the final amplitude then it is natural to impose the same positivity constraints, but now with $\ell=0$, i.e.
\begin{equation}
Y = C\cdot Z, \label{Amp2}
\end{equation}
where $C$ is the matrix that appears in the definition of the tree-level Amplituhedron. The conjecture now is that a properly-defined amplitude must be positive -- or rather, sign-alternating with loop order -- if evaluated for $Y$ and $Z$ matrices satisfying the positivity conditions. We restrict ourselves to our cases of interest, MHV and NMHV amplitudes ($k=0$ and 1), and review what these conditions imply.

\subsection{MHV positive kinematics}
\label{MHVposkinSubsection}

For MHV amplitudes we have $k=0$ and $l=0$ so there is no $C$ matrix.  That is, the $Y$ space in \eqn{Amp2} becomes trivial and the only conditions come from the positivity of the $(4\times n)$ matrix $Z$. In this case the column vectors composing $Z$ are directly the 4-dimensional momentum twistors $Z_a$ and we have to keep them positive -- in the sense that the following $(4\times 4)$ minors of the $Z$ matrix should be positive: 
\begin{equation}
Z = \left(\begin{array}{cccccc}
\uparrow&\uparrow&\uparrow&\dots&\uparrow&\uparrow\\
Z_1 & Z_2 & Z_3 & \dots & Z_{n-1}& Z_n\\
\downarrow& \downarrow&\downarrow&\dots&\downarrow&\downarrow
\end{array}\right) \qquad \begin{array}{c}\mbox{with}\,\,\, \la abcd\ra \equiv  \mbox{det}(Z_a,Z_b,Z_c,Z_d)>0\\ \mbox{for}\,\,\,a<b<c<d.\end{array}
\end{equation}

Let us now parametrize the positive $Z$ matrix for $n=6$.
Using a $GL(4)$ transformation we fix the first four columns to be the
unit matrix, and parametrize the remaining two columns with eight
positive parameters $x_a>0$, $y_b>0$. One solution that
makes all $(4\times4)$ minors positive is
\begin{equation}
Z = \left(\begin{array}{cccccc} 
1 & 0 & 0 & 0 & -x_1&-y_1 - y_2\frac{x_1}{x_2}-y_3\frac{x_1}{x_3} - y_4\frac{x_1}{x_4}\\
0 & 1 & 0 & 0 & x_2 & y_2 + y_3\frac{x_2}{x_3} + y_4 \frac{x_2}{x_4}\\
0 & 0 & 1 & 0 & -x_3 & -y_3 -y_4\frac{x_3}{x_4}  \\
0 & 0 & 0 & 1 & x_4 & y_4
\end{array}\right) \,.
\label{MHVZposparam}
\end{equation}

We can now build three different dual-conformal cross ratios,
\begin{equation}
u = \frac{\la 6123\ra\la 3456\ra}{\la 6134\ra\la 2356\ra} \,,
\quad v = \frac{\la 1234\ra\la 4561\ra}{\la 1245\ra\la 3461\ra} \,,
\quad w = \frac{\la 2345\ra\la 5612\ra}{\la 2356\ra\la 4512\ra} \,.
\label{ratio1}
\end{equation}
We also consider the combinations
\begin{equation}
  \ve \equiv 1-u-v-w, \qquad \Delta = \ve^2-4uvw.
\label{epsDeltaDef}
\end{equation}
From the positive parametrization~(\ref{MHVZposparam})
of the $Z$ matrix we get,
\begin{align}
&u= \frac{x_2^2x_3^2y_1y_4}{PQ},\quad
v=\frac{x_3x_4y_2}{P}\,,\quad w=\frac{x_1x_2y_3}{Q}\,,\\
&\ve =  \frac{x_2x_3(x_2x_4y_1y_3+x_1x_3y_2y_4)}{PQ}\,,
\quad \Delta =\frac{x_2^2x_3^2(x_1x_3y_2y_4-x_2x_4y_1y_3)^2}{P^2Q^2} \,,
\end{align}
where $P = x_3x_4y_2+x_2x_4y_3+x_2x_3y_4$,
$Q = x_2x_3y_1+x_1x_3y_2+x_1x_2y_3$.
For positive values of $x_a$, $y_b$ the cross ratios $u,v,w$ and
$\ve,\Delta$ are all manifestly positive.  These inequalities combine to define conditions for the MHV positive region,
\be
u,v,w > 0, \quad u+v+w < 1, \quad (1-u-v-w)^2 > 4uvw,
\label{MHVposregion}
\ee
which restrict the cross ratios to be relatively close to the origin,
in contrast to what we will find for the NMHV positive region.
We refer to this region as Region I (see ref.~\cite{Dixon2013eka}
and \eqn{RegionIDef} below).
The only place that $\ve$ can approach zero in Region I, 
given the constraint on $\Delta$, is for $v\to0$, $u+w\to1$,
or cyclic permutations of this line.
In this limit, two gluons become collinear.

Now that we have identified MHV positive kinematics,
we would like to conjecture that a properly-defined IR-finite part of the
MHV amplitude is positive for any positive values $x_a,y_b>0$.  However,
individual on-shell amplitudes are IR divergent, and there is not
a unique way to obtain a finite quantity by removing the IR divergences.
The original way that IR divergences were removed (while preserving
dual conformal symmetry) was to divide by the BDS ansatz~\cite{BDS}.
While this procedure leads to remainder functions
with smooth collinear limits~\cite{Bern2008ap},
it breaks a global analytic property known
as the Steinmann relations~\cite{Steinmann}.
To preserve the Steinmann relations~\cite{CaronHuot2016owq},
at six points (or seven points)
one can divide by a unique ``BDS-like''
ansatz~\cite{Alday2009dv,Dixon2015iva}.
Yet this procedure sacrifices the vanishing in collinear limits of
the six-point BDS remainder function, and the collinear limits form a
boundary of the positive region (e.g.~$v\to0$, $u+w\to1$ makes $\ve$
and $\Delta$ both vanish).
There are also dual-conformal IR regulators based on the Wilson loop
interpretation of the amplitude~\cite{Basso2013vsa}, but they break
a dihedral symmetry.  In short, there is no unique way to define
an IR finite part of the MHV amplitude, nor one that is clearly optimal.
We will discuss the positivity properties of these various choices in
section~\ref{MHVpos}.

\subsection{NMHV positive kinematics}
\label{NMHVPosKinem}

In contrast, when we also consider the NMHV amplitude
there is a natural way to form an IR finite quantity,
the {\it ratio function}, which is defined (at six points)
by dividing the NMHV super-amplitude
by the MHV super-amplitude~\cite{Drummond2008vq}.  IR divergences
are helicity-independent and cancel between numerator and denominator.
We will inspect the ratio function for NMHV positive kinematics.

For the NMHV case, $k=1$, the Amplituhedron lives in a projective space
${\mathbb{P}}^4$. It is defined as all points $Y$ that are linear combinations
of $Z_a$ with positive coefficients,
\begin{equation}
Y = C\cdot Z = c_1 Z_1 + c_2 Z_2 + \dots + c_n Z_n  \quad 
\mbox{with $c_a>0$},\label{Yp}
\end{equation}
where the $Z_a$ are now five-dimensional.  They can be written as 
\begin{equation}
Z_a = \left(\begin{array}{c} z_a\\ \phi\cdot \eta_a\end{array}\right) \,,
\label{Zbosdef}
\end{equation}
where the first four components are momentum twistor variables $z_a$
associated with each particle label, $a=1,2,\ldots,n$ for
$n$-point scattering.  The fifth (last) component is the contraction
$\phi\cdot\eta_a = \epsilon_{IJ}\phi^I\eta^J_a$, $I,J=1,2,3,4$, 
of an auxiliary Grassmann variable $\phi^I$ with the standard
Grassmann variable $\eta_a^J$ of on-shell 
superspace~\cite{Nair1988bq,Georgiou2004by,Drummond2008vq,ArkaniHamed2008gz}.
These bosonic variables then carry all information about the external
particles in the scattering.  The bosonized momentum twistors are projective
variables, defined up to rescaling $Z_a \rightarrow t Z_a$. 

Positivity conditions are then imposed directly on the five-dimensional $Z_a$
rather than the four-dimensional part $z_a$.  The $(n\times 5)$-dimensional
matrix $Z$ has all $(5\times5)$ minors positive; that is,
\begin{equation}
  \la abcde\ra \equiv {\rm det}(Z_a,Z_b,Z_c,Z_d,Z_e) > 0
  \qquad \mbox{for}\,\,\,\,a<b<c<d<e.
\end{equation}
Geometrically, the $Z_a$ form a convex configuration in real projective
space $\mathbb{P}^4$.

In addition to five-brackets containing five $Z_a$,
we can also have five-brackets including the point $Y$ in \eqn{Yp},
which lies inside the Amplituhedron.  The $Y$-containing five-brackets are
given by,
\begin{equation}
\la Yabcd\ra \equiv  {\rm det}(Y,Z_a,Z_b,Z_c,Z_d).
\end{equation}
A subset of these five-brackets is positive when $Y$ is in the Amplituhedron,
specifically those with two pairs of consecutive indices:
$\la Y\,a\,a\pl1\,b\,b\pl1\ra>0$ for all $a,b$.
The three-planes $(Z_a\,Z_{a\pl1}\,Z_b\,Z_{b\pl1})$
are boundaries of the Amplituhedron.
The condition $\la Y\,a\,a\pl1\,b\,b\pl1\ra>0$ puts the point $Y$ on the
correct side of the boundary, inside the Amplituhedron.
From a physics perspective, the term
$\la Y\,a\,a\pl1\,b\,b\pl1\ra \sim s_{a\pl1\ldots b} \equiv (p_{a\pl1}+\dots+p_b)^2$
corresponds to a factorization pole of the tree-level amplitude.

For the six-point case, we redefine the three cross ratios defined in
\eqn{ratio1} by inserting $Y$ into all the four-brackets to
make them five-brackets,
\begin{equation}
u = \frac{\la Y6123\ra\la Y3456\ra}{\la Y6134\ra\la Y2356\ra} \,,
\quad v = \frac{\la Y1234\ra\la Y4561\ra}{\la Y1245\ra\la Y3461\ra} \,,
\quad w = \frac{\la Y2345\ra\la Y5612\ra}{\la Y2356\ra\la Y4512\ra} \,.
\label{ratioY}
\end{equation}
The positive parametrization is now much simpler than in the MHV case
because the matrix $Z$ is $(6\times5)$ rather than $(6\times4)$.
A natural parametrization of $Z$ in terms of five positive parameters $x_a>0$
is,
\begin{equation}
Z = \left(\begin{array}{cccccc} 1&0&0&0&0& x_1\\ 0&1&0&0&0&-x_2\\ 0&0&1&0&0&x_3\\ 0&0&0&1&0&-x_4\\0&0&0&0&1&x_5\end{array}\right)\qquad
\begin{array}{cc} \la 12345\ra = 1,&  \la 23456\ra = x_1, \\ \la 13456\ra = x_2, & \la 12456\ra = x_3, \\  \la 12356\ra = x_4, & \la 12346\ra = x_5.\end{array}\label{posZ}
\end{equation}
Using this parametrization and $Y = C\cdot Z$ from \eqn{Yp},
we can compute all ${6 \choose 2}=15$ five-brackets $\la Yabcd\ra$:
\begin{align}
&\la Y1234\ra = c_5x_6 + c_6x_5,\quad \la Y1235\ra = c_6x_4 -c_4x_6,\quad \la Y6123\ra = c_4x_5+c_5x_4,\nonumber\\
&\la Y1245\ra = c_3x_6 +c_6x_3,\quad \la Y1246\ra = c_3x_5 -
c_5x_3,\quad \la Y1256\ra = c_3x_4 + c_4x_3,\nonumber\\
&\la Y1345\ra = c_6x_2 - c_2x_6,\quad \la Y3461\ra = c_2x_5 +
c_5x_2,\quad \la Y1356\ra = c_4x_2-c_2x_4,\nonumber\\
&\la Y4561\ra = c_2x_3+c_3x_2,\quad\la Y2345\ra = c_1x_6 +
c_6x_1,\quad \la Y2346\ra = c_1x_5 - c_5x_1,\nonumber\\
&\la Y2356\ra = c_1x_4+c_4x_1,\quad \la Y2456\ra = c_1x_3 -
c_3x_1,\quad \la Y3456\ra = c_1x_2+c_2x_1, \label{pos}
\end{align}
where $x_6\equiv1$ is added to make the expressions more uniform.

From \eqn{ratioY}, the cross ratios are now 
\begin{align}
&u = \frac{(c_1x_2+c_2x_1)(c_4x_5+c_5x_4)}{(c_2x_5+c_5x_2)(c_1x_4+c_4x_1)}\,,\qquad v=\frac{(c_2x_3+c_3x_2)(c_5x_6+c_6x_5)}{(c_2x_5+c_5x_2)(c_3x_6+c_6x_3)}\,,\nonumber\\&\hspace{3cm} w=\frac{(c_1x_6+c_6x_1)(c_3x_4+c_4x_3)}{(c_1x_4+c_4x_1)(c_3x_6+c_6x_3)}\,.
\label{ratioxc}
\end{align}
As in the MHV case, the cross ratios are all positive.

Denoting $W=(c_1x_4+c_4x_1)(c_2x_5+c_5x_2)(c_3x_6+c_6x_3)$,
we get for the quantities $\ve$ and $\Delta$ defined in
\eqn{epsDeltaDef},
\begin{equation}
\ve = -\frac{P_1(x_a,c_b)}{W} < 0,\qquad
\Delta = \frac{[P_2(x_a,c_b)]^2}{W^2}>0,
\label{NMHVposconstraints}
\end{equation}
where the $P_j(x_a,c_b)$ are polynomials in $x_a,c_b$ with positive coefficients.
Notice that the sign condition on $\ve$ has flipped from the MHV case,
pushing the cross ratios away from the origin.

The NMHV amplitude also contains $R$-invariants, defined as the following
function of momentum twistors $z_a$ and Grassmann variables $\eta_a$:
\begin{equation}
R[a\,b\,c\,d\,e] = \frac{(\eta_a\la bcde\ra + \eta_b\la cdea\ra + \eta_c\la deab\ra + \eta_d\la eabc\ra + \eta_e\la abcd\ra)^4}{\la abcd\ra\la bcde\ra\la cdea\ra\la deab\ra\la eabc\ra} \,.
\end{equation}
In the bosonized language, the $R$-invariants become functions
of five-brackets, projective in all variables, which we denote as 
\begin{equation}
[a\,b\,c\,d\,e] = \frac{\la Y\,d^4Y\ra \la abcde\ra^4}{\la Yabcd\ra\la Ybcde\ra\la Ycdea\ra\la Ydeab\ra\la Yeabc\ra} \,,
\end{equation}
where $\la Y\,d^4Y\ra$ is the measure in $Y$. For the six-point case,
it is convenient to label this object by the missing index, and to omit
the measure, defining 
\begin{equation}
 (1) \equiv \frac{[23456]}{\la Y\,d^4Y\ra}
 = \frac{\la 23456\ra^4}{\la Y2345\ra\la Y2346\ra\la Y2456\ra\la Y2356\ra\la Y3456\ra}
\end{equation}
and similarly for $(2)=[34561]$, $(3) = [45612]$, etc.

The form for the tree-level NMHV Amplituhedron is then 
\begin{equation}
\Omega_{6,1}^{\rm tree} = (1)+(3)+(5) = (2)+(4)+(6).\label{Tree6}
\end{equation}
This is also the bosonized version of the tree-level NMHV ratio function
${\cal P}_{6,1}^{\rm tree}$, see section~\ref{RatioFunctionSubsection}.

Using the positive parametrization~(\ref{Yp}),
we can rewrite the bosonized $R$-invariants as
\begin{align}
(1) &= \frac{x_1^4}{(c_1x_6+c_6x_1)(c_1x_2+c_2x_1)(c_1x_3-c_3x_1)(c_1x_4+c_4x_1)(c_1x_5-c_5x_1)},\nonumber\\
(2) &= \frac{x_2^4}{(c_1x_2+c_2x_1)(c_2x_3+c_3x_2)(c_2x_4-c_4x_2)(c_2x_5+c_5x_2)(c_2x_6-c_6x_2)},\nonumber\\
(3) &= \frac{x_3^4}{(c_2x_3+c_3x_2)(c_3x_4+c_4x_3)(c_3x_5-c_5x_3)(c_3x_6+c_6x_3)(c_3x_1-c_1x_3)},\nonumber\\
(4) &= \frac{x_4^4}{(c_3x_4+c_4x_3)(c_4x_5+c_5x_4)(c_4x_6-c_6x_4)(c_1x_4+c_4x_1)(c_4x_2-c_2x_4)},\nonumber\\
(5) &=
\frac{x_5^4}{(c_4x_5+c_5x_4)(c_5x_6+c_6x_5)(c_1x_5-c_5x_1)(c_2x_5+c_5x_2)(c_3x_5-c_5x_3)},\nonumber\\
(6) &= \frac{x_6^4}{(c_5x_6+c_6x_5)(c_1x_6+c_6x_1)(c_2x_6-c_6x_2)(c_3x_6+c_6x_3)(c_4x_6-c_6x_4)}. \label{R_inv_parametrization}
\end{align}

Five-brackets corresponding to spurious poles can be identified in
\eqn{pos} as the expressions containing minus signs,
while those corresponding to physical poles are manifestly positive.
Each $R$-invariant $(a)$ contains two spurious poles. For example,
$(1)$ has $\la Y2346\ra$ and $\la Y2456\ra$.  The spurious poles do not have a
fixed sign for all $c_b,x_a>0$, e.g. $\la Y2346\ra = c_1x_5-c_5x_1$.
Therefore, the invariant $(1)$ also does not have a fixed sign and
it is not a manifestly positive object, and similarly for the other $(a)$.
Only in the sum (\ref{Tree6}) do these poles cancel, so that
$\Omega_{6,1}^{\rm tree}$ can be positive in the full positive region.

In fact, we can write the tree amplitude in the form,
\begin{equation}
\Omega_{6,1}^{\rm tree} = \frac{{\cal N}(x_a,c_b)}
{\prod\limits_{|j-k|=1\,{\rm or}\,3}(c_jx_k+c_kx_j)} \,,
\end{equation} 
where ${\cal N}(x_a,c_b)$ is a polynomial in $x_a,c_b$ with all
positive coefficients~\cite{ArkaniHamed2014dca}.

\subsection{The ratio function}
\label{RatioFunctionSubsection}

Scattering amplitudes of massless particles suffer from IR divergences
from both soft and collinear virtual exchange.
It is necessary to introduce a regulator to get a well-defined answer.
In the planar theory, for gauge group $SU(N_c)$ with $N_c\to\infty$,
the IR divergences exponentiate in a relatively simple fashion.
In dimensional regularization with $D=4-2\e$, the poles in $\e$ in
planar ${\cal N}=4$ SYM amplitudes are captured by the BDS ansatz~\cite{BDS},
\begin{equation}
{\cal M}_{n,k} = {\cal M}_{n,k}^{\rm tree}\cdot 
\exp\left[\sum_{\ell=1}^\infty a^\ell
  \left(f^{(\ell)}(\e)\cdot {\cal A}_{n,0}^{\rm 1-loop}(\ell\e)
  + \hbox{finite} \right)\right] \,,
\label{IRdiv}
\end{equation} 
where $a = g^2 N_c/(8\pi^2)$ is the 't Hooft coupling,
$f^{(\ell)}(\e)=f_0^{(\ell)} + f_1^{(\ell)}\e + f_2^{(\ell)}\e^2$
for some constants $f_k^{(\ell)}$, 
and ${\cal A}_{n,0}^{\rm 1-loop}(\e)$ is the regulated one-loop MHV amplitude
${\cal M}_{n,0}^{\rm 1-loop}(\e)$ divided by the tree-level amplitude
${\cal M}_{n,0}^{\rm tree}$.

In the MHV case, $k=0$, the finite part in the exponential in \eqn{IRdiv}
is called the remainder function $R^{(\ell)}_{n}$,
\begin{equation}
{\cal M}_{n,0} = {\cal M}_{n,0}^{\rm tree}\cdot 
\exp\left[\sum_{\ell=1}^\infty a^\ell
  \left(f^{(\ell)}(\e)\cdot {\cal A}_{n,0}^{\rm 1-loop}(\ell\e)
  + R^{(\ell)}_{n}\right)\right] 
\equiv {\cal M}_{n,0}^{\rm BDS}(\e) \cdot \exp[R_n] \,,
\label{IRdivMHV}
\end{equation} 
and it is dual conformally invariant.  However, we can still move finite, dual
conformally invariant terms between the first and second terms in
this expression.  Correspondingly, there are a few possible different
definitions of the remainder function.  In section~\ref{MHVpos} we will 
discuss the possibilities in more detail, and describe one choice
which appears to satisfy MHV positivity properties.

There is a cleaner and less ambiguous way to define an IR-finite object in
the context of scattering amplitudes, simply by taking a ratio of two
amplitudes with different helicities~\cite{Drummond2008vq}.  Because the 
IR divergences~(\ref{IRdiv}) are universal, one can divide any amplitude
${\cal M}_{n,k}$ by the MHV amplitude ${\cal M}_{n,0}$
and get an IR finite ratio function ${\cal P}_{n,k}$.
Expanding the ratio in the coupling constant $a$, we define
the loop expansion coefficients of the ratio function as,
\begin{equation}
{\cal P}_{n,k} = \frac{{\cal M}_{n,k}}{{\cal M}_{n,0}}
= {\cal P}_{n,k}^{\rm tree} + a \cdot {\cal P}_{n,k}^{1-{\rm loop}} + a^2\cdot {\cal P}_{n,k}^{2-{\rm loop}} + \ldots\,, 
\end{equation}
while those of the amplitude normalized by the MHV tree super-amplitude 
(an IR divergent quantity) are denoted by
\begin{equation}
{\cal A}_{n,k} = \frac{{\cal M}_{n,k}}{{\cal M}_{n,0}^{\rm tree}}
= {\cal P}_{n,k}^{\rm tree} + a \cdot {\cal A}_{n,k}^{1-{\rm loop}} + a^2\cdot {\cal A}_{n,k}^{2-{\rm loop}} + \ldots\,.
\end{equation}
The two sets of expansion coefficients are related by,
\begin{align}
{\cal P}_{n,k}^{1-{\rm loop}} &= {\cal A}_{n,k}^{1-{\rm loop}} - {\cal P}_{n,k}^{\rm tree}\cdot {\cal A}_{n,0}^{1-{\rm loop}} \,, \nonumber\\
{\cal P}_{n,k}^{2-{\rm loop}} &= {\cal A}_{n,k}^{2-{\rm loop}} - {\cal P}_{n,k}^{\rm tree}\cdot {\cal A}_{n,0}^{2-{\rm loop}} - {\cal P}_{n,k}^{1-{\rm loop}}\cdot {\cal A}_{n,0}^{1-{\rm loop}} \,,
\end{align}
and so on.

The ratio function ${\cal P}_{n,k}^{\ell-{\rm loop}}$ corresponds to a linear combination of products of amplitudes with different signs. Therefore, it would be quite surprising if it had any positivity properties.  However, numerical checks performed in ref.~\cite{ArkaniHamed2014dca} for the one-loop NMHV $n$-point amplitude for $n\leq12$, and for the one-loop N$^2$MHV amplitude for $n\leq9$ show that this is indeed true!

Let us now focus on the six-point case in more detail. As was pointed out in ref.~\cite{Drummond2008vq}, the ratio function can be expressed in terms of two transcendental functions, $V(u,v,w)$ and $\widetilde{V}(y_u,y_v,y_w)$,
\begin{eqnarray}
{\cal P}_{6,1}&=& \frac{1}{2}
\Bigl([(1)+(4)]V(u,v,w) + [(2)+(5)]V(v,w,u) + [(3)+(6)]V(w,u,v) \nonumber\\ 
&&\hspace{0.2cm}\null + [(1)-(4)]\widetilde{V}(y_u,y_v,y_w) - [(2)-(5)]\widetilde{V}(y_v,y_w,y_u)  + [(3)-(6)]\widetilde{V}(y_w,y_u,y_v)\Bigr) \,, \nonumber\\
{~} \label{ratio_function}
\end{eqnarray}
where the cross ratios $u$, $v$, $w$ are written in terms of our bosonized
variables in \eqn{ratioY}, and the extended cross ratios 
$y_u$, $y_v$, $y_w$~\cite{Dixon2011nj} are also bosonized:
\begin{align}
y_u &= \frac{\la Y1345\ra\la Y2456\ra\la Y1236\ra}{\la Y1235\ra\la Y3456\ra\la Y1246\ra} \,,
\quad y_v = \frac{\la Y1235\ra\la Y2346\ra\la Y1456\ra}{\la Y1234\ra\la Y2456\ra\la Y1356\ra} \,,\nonumber\\ 
&\hspace{3cm} y_w = \frac{\la Y2345\ra\la Y1356\ra\la Y1246\ra}{\la Y1345\ra\la Y2346\ra\la Y1256\ra} \,. \label{ratio2Y}
\end{align}
The function $V$ is even under a parity symmetry that inverts $y_i \lr 1/y_i$,
and leaves $u,v,w$ invariant.
The function $\widetilde{V}$ is parity-odd, changing sign under this inversion.
For this reason, it is better to think of $\widetilde{V}$ as a function
of $y_u,y_v,y_w$ rather than $u,v,w$.

Note that the extended cross ratios do not have any positivity properties due to the presence of spurious poles. Under a cyclic shift $Z_a \to Z_{a+1}$ they transform as
\begin{equation}
y_u \rightarrow \frac{1}{y_v} \,, \qquad y_v\rightarrow
\frac{1}{y_w} \,, \qquad y_w\rightarrow \frac{1}{y_u} \,,\label{cycle_y}
\end{equation}
and the standard cross ratios transform as $u\rightarrow v$, $v\rightarrow w$, $w\rightarrow u$. The ratio function is symmetric under both cyclic shifts and dihedral flips. The combined symmetry group is $D_6$, although acting on the cross ratios $u,v,w$ it reduces to $S_3$, i.e.~all permutations of $u,v,w$.  The individual functions $V$ and $\tilde{V}$ are (anti)symmetric under a $Z_2$ subgroup of $S_3$ that leaves $v$ fixed:
\begin{equation}
V(u,v,w) = V(w,v,u), \qquad \tilde{V}(y_u,y_v,y_w) = - \tilde{V}(y_w,y_v,y_u).
\label{uwsym}
\end{equation}

The transcendental functions $V$ and $\tilde{V}$ have a Euclidean sheet on which they are real, when the cross ratios lie in the positive octant $u,v,w>0$.  We evaluate them on this sheet, with the cross ratios and $R$-invariants further restricted by the positive parametrization $c_b,x_a>0$.  (In some physical scattering regions $V$ and $\tilde{V}$ would acquire imaginary parts, which would make discussing positivity difficult.)

\section{One-loop ratio function}
\label{OneLoop}

At one loop, the parity-odd part vanishes, $\widetilde{V}^{(1)}=0$,
and the full ratio function can be written as 
\begin{equation}
2{\cal P}^{1-{\rm loop}}_{6,1} = [(1)+(4)]V^{(1)}(u,v,w) + [(2)+(5)]V^{(1)}(v,w,u) + [(3)+(6)]V^{(1)}(w,u,v), \label{P1}
\end{equation}
where the one-loop function $V^{(1)}(u,v,w)$ is given by
\bea
V^{(1)}(u,v,w) &=&  \frac12\Bigl[ 
{\rm Li}_2 (1-u) + {\rm Li}_2 (1-v) + {\rm Li}_2 (1-w) \nonumber\\
&&\hskip0.4cm \null + \log u \log v - \log u\log w + \log v\log w 
 - 2\zeta_2 \Bigr] \,.
\eea

Our claim is that \eqn{P1} is negative (because the loop order is odd) within the positive region. Note that the individual pieces in this formula do not have definite signs, neither the $R$-invariants $(a)$, nor the function $V^{(1)}$ which has both plus and minus signs in front of individual terms.  Depending on the values of $u,v,w$, different terms can dominate.

For some purposes it is convenient to separate out the ${\rm Li}_2$ part of the expression. Note that the ${\rm Li}_2$ part is invariant under $S_3$ permutations, and therefore it multiplies all $R$-invariants $(a)$, which can be assembled into the tree-level amplitude,
\begin{align}
2{\cal P}_{6,1}^{1-{\rm loop}} &= {\cal P}_{6,1}^{\rm tree} \cdot \left[ {\rm Li}_2 (1-u) + {\rm Li}_2 (1-v) + {\rm Li}_2 (1-w) - 2\zeta_2\right]\nonumber\\
&\hspace{0.6cm} + [(1)-(2)+(3)]\log u\log v + [(2)-(3)+(4)]\log v \log w\nonumber\\
&\hspace{0.6cm} + [(3)-(4)+(5)]\log w \log u \,, \label{P2}
\end{align}
where we have used the identity $(1)+(3)+(5)=(2)+(4)+(6)$. For some purposes it is more convenient to use~\eqn{P1}, for others~\eqn{P2}. 

\subsection{Simple examples of positivity}
\label{SimpleExampleSubsection}

Let us give a few examples where the overall sign can be easily understood. 

\subsubsection*{Example 1}

Our first case is the point $(u,v,w)=(1,1,1)$, which was studied in detail in ref.~\cite{ArkaniHamed2014dca}. To reach this point, we set $c_3=c_1x_3/x_1$ and $c_5=c_1x_5/x_1$. This preserves positivity of $c_b$, $x_a$, and so it is inside the Amplituhedron. Kinematically, it corresponds to setting $\la Y2456\ra = \la Y 2346\ra = 0$, which is a spurious boundary of the tree-level Amplituhedron, so we are not on the true physical boundary.  Therefore, the tree-level term ${\cal P}_{6,1}^{\rm tree}$ is completely regular and positive here. However, individual $R$-invariants $(a)$ do blow up. In order to approach this point, we first set all cross-ratios to be equal, $u=v=w$, and then take $u\to1$, 
\begin{equation}
{\cal P}_{6,1}^{1-{\rm loop}} \xrightarrow[u=v=w]{}\frac12 {\cal P}_{6,1}^{\rm tree} \cdot \left[3{\rm Li}_2(1-u) + \log^2 u - 2\zeta_2\right] \xrightarrow[u=1]{} -{\cal P}_{6,1}^{\rm tree} \cdot \zeta_2 < 0.
\end{equation}
Thus we obtain the desired negative value.
In section~\ref{HigherLoop111Subsection} we will study the point $(1,1,1)$
at higher loops.

\subsubsection*{Example 2}

Another interesting case is the point $(u,v,w)=(1,0,0)$, which can be reached by setting $c_2=c_3=c_4=0$. Naively, the term $\log v \log w$ dominates, but there is a conspiracy of prefactors which makes the situation more complicated. We can approach this limit by setting $c_2\rightarrow \epsilon c_2$, $c_3\rightarrow \epsilon c_3$, $c_4\rightarrow \epsilon c_4$ and then letting $\epsilon\rightarrow 0$. There are many ways to approach the point $(u,v,w)=(1,0,0)$, but this limit always keeps us in the positive region.

For analyzing the one-loop ratio function in this limit, it is good to use the second representation~(\ref{P2}). The relevant combinations of $R$-invariants behave in this limit as
\begin{align}
{\cal P}_{6,1}^{\rm tree} &= \frac{1}{\epsilon^2}\cdot \frac{x_3}{c_1c_5c_6(c_3x_2+c_2x_3)(c_4x_3+c_3x_4)} \,, \nonumber\\
(1)-(2)+(3) &= - \frac{1}{\epsilon^2}\cdot \frac{x_4}{c_1 c_5 c_6 (c_4 x_2 - c_2 x_4) (c_4 x_3 + c_3 x_4)} \,, \nonumber\\ (3)-(4)+(5) &= \frac{1}{\epsilon^2}\cdot \frac{x_2}{c_1 c_5 c_6 (c_3 x_2 + c_2 x_3) (c_4 x_2 - c_2 x_4)} \,.
\end{align}
while the term $(2)-(3)+(4) = {\cal O}(1)$ is subleading. 
Combining these limits with those of the polylog parts,
the individual pieces in \eqn{P2} behave like
\begin{align}
{\cal P}_{6,1}^{\rm tree}\cdot (\dots) &= \frac{\log\epsilon}{\epsilon}
\cdot \frac{X}{c_1^2c_5^2c_6^2x_2x_4(c_3x_2+c_2x_3)(c_4x_3+c_3x_4)
} \,, \label{limterm1}\\
[(1)-(2)+(3)]\cdot (\dots) &= - \frac{\log\epsilon}{\epsilon}\cdot \frac{c_1x_5-c_5x_1}{c_1^2c_5^2c_6x_2(c_4x_3+c_3x_4)} \,, \label{limterm2}\\
[(3)-(4)+(5)]\cdot (\dots) &= \frac{\log\epsilon}{\epsilon}\cdot \frac{c_1x_5-c_5x_1}{c_1^2c_5^2c_6x_4(c_3x_2+c_2x_3)} \,, \label{limterm3}
\end{align}
where
\be
X = c_4c_5x_2x_3(c_6x_1+c_1x_6)+c_1c_2x_3x_4(c_6x_5+c_5x_6)
   +c_3x_2x_4(c_5c_6x_1+c_1c_6x_5+2c_1c_5x_6),
\ee
while the last term is subleading in this limit, $[(2)-(3)+(4)]\cdot (\dots) = {\cal O}(\log^2\epsilon)$.  This suppression may be counter-intuitive (as that term had the dominant logarithms), but the rational prefactor is regular in this limit, while the prefactors of other terms diverge.  We see that the terms~(\ref{limterm2}) and (\ref{limterm3}) do not have fixed sign, but if we combine all three pieces together we get
\begin{equation}
{\cal P}_{6,1}^{1-{\rm loop}} =
\frac{\log\epsilon}{\epsilon}\cdot \frac{Y}{2c_1^2c_5^2c_6^2x_2x_4(c_3x_2+c_2x_3)(c_4x_3+c_3x_4)} \,, \label{lim2}
      \end{equation}
where
\be
Y = c_5 c_6 x_1x_4  (c_3 x_2 + c_2 x_3) + c_1 c_6 x_2 x_5(c_4 x_3 + c_3 x_4)
  + c_1 c_5 x_6(c_4 x_2 x_3 + 2 c_3 x_2 x_4 +  c_2 x_3 x_4),
\ee 
which is manifestly negative for $\epsilon\rightarrow 0$ while keeping $c_a,x_b>0$. The negativity of the final expression requires a conspiracy between the rational prefactors and the polylog part, as well as between different parts of the answer in \eqn{P2}. We can also start with representation~(\ref{P1}), but in this case the cancellation is even more complicated. Individual pieces would also contain logs of $c_a,x_b$ as prefactors of $\frac{\log\epsilon}{\epsilon}$. 
These logs would all cancel when taking the sum, leaving us with the rational expression~(\ref{lim2}). 

\subsection{Double-scaling limit}
\label{one_loop_ds_limit}

In the previous examples the rational prefactors played a central role in proving positivity. Let us now discuss an example where positivity relies on a relation between polylogarithms. Such a case can be found near the boundary $\la Y1234\ra =0$, which we can approach by setting $c_5= \epsilon \hat{c}_5$, $c_6= \epsilon \hat{c}_6$ and taking the limit $\epsilon \rightarrow 0$ with $ \hat{c}_5, \hat{c}_6$ fixed. As can be seen from \eqn{R_inv_parametrization}, the two dominant $R$-invariants are equal to each other in this limit,
\begin{equation}
(5) = (6) = \frac{1}{\epsilon}\cdot \frac{1}{c_1c_2c_3c_4(\hat{c}_6 x_5+ \hat{c}_5 x_6)},
\end{equation} 
while the $R$-invariants $(1),(2),(3)$ and $(4)$ remain finite. Similarly, the cross ratios become
\begin{equation}
u = \frac{c_4(c_2x_1+c_1x_2)}{c_2(c_4x_1+c_1x_4)}, \quad v = \mathcal{O}(\epsilon),\quad w = \frac{c_1(c_4 x_3+c_3x_4)}{c_3(c_4x_1+c_1x_4)}\label{uw}
\end{equation}
in this limit.

Thus this limit sends the cross ratio $v\to0$, but leaves $u,w$ fixed.  This limit has been studied in the context of the operator product expansion (OPE), where it is referred to as the double-scaling limit and corresponds to contributions with the maximum number of gluonic flux-tube excitations~\cite{Gaiotto2011dt,Basso2014nra,Drummond2015jea}.  While the conventional OPE addresses configurations near the collinear limit $v\to0$, $u+w\to1$, the double-scaling limit allows $u$ and $w$ to be generic.  

For NMHV positive kinematics, $u$ and $w$ are not totally generic, because
we have
\begin{equation}
\left[ u+w \right]_{c_5, c_6 \rightarrow 0} = 1 + \frac{c_1 c_4 (c_2 x_3 +c_3 x_2)}{c_2 c_3 (c_1 x_4 + c_4 x_1)} > 1.
\end{equation}
This turns out to be the only additional constraint; that is, the correct
NMHV positive region within the double-scaling limit is the semi-infinite plane
\begin{align}
u > 0, \quad w > 0,\quad u+w > 1. \label{ds_positive_region}
\end{align}
In order to show that the entire region~(\ref{ds_positive_region}) corresponds to positive kinematics, we use the fact that the lines $u=1$ and $w=1$ divide the region~(\ref{ds_positive_region}) into four subregions.  Each of the four subregions corresponds to solving \eqn{uw} for two of the $c_b$, $b=1,2,3,4$, in terms of $u,w$ and the remaining $c_b,x_a$, in a manifestly positive manner.  There are six possible pairs of $c_b$, but the pairs $\{c_1,c_3\}$ and $\{c_2,c_4\}$ do not work.  For example, solving \eqn{uw} for $c_2,c_3$ gives
\be
c_2 = \frac{c_1c_4x_2}{uc_1x_4+(u-1)c_4x_1} \,, \qquad
c_3 = \frac{c_1c_4x_3}{wc_4x_1+(w-1)c_1x_4} \,,
\label{c2c3soln}
\ee
which is manifestly positive in the subregion $u>1$, $w>1$.  This solution shows that this entire subregion is covered.  The other subregions work in the same way.

Since polylogarithms can generate at most $\log\epsilon$ behavior, the one-loop ratio function in the double-scaling limit becomes dominated by terms involving the singular (and equal) $R$-invariants $(5)$ and $(6)$:
\be
{\cal P}_{6,1}^{1-{\rm loop}} \Big|_{c_5,c_6 \rightarrow 0} = \frac{1}{2\epsilon} \cdot  \frac{1}{c_1c_2c_3c_4(\hat{c}_6x_5+\hat{c}_5x_6)}
\cdot C^{(1)}(u,w), \label{Poneloopds}
\ee
where
\be
C^{(1)}(u,w) = {\rm Li}_2 (1-u) +  {\rm Li}_2 (1-w) + \log u \log w - \zeta_2 \,.
\label{dsoneloop}
\ee
While the rational prefactor in this expression is manifestly positive for all positive values of the $c_a$, it's not yet obvious what can be said about the sign of the polylogarithmic part $C^{(1)}(u,w)$ in region~(\ref{ds_positive_region}).  In fact, ${\cal P}_{6,1}^{1-{\rm loop}}|_{c_5, c_6 \rightarrow 0}$, and hence also $C^{(1)}(u,w)$, are required to vanish on the boundary $u+w=1$, because this line corresponds to a limit in which two adjacent particles become collinear. In general, this would mean that the six-point ratio function should match onto the five-point ratio function -- but the five-point ratio function receives no loop-level corrections~\cite{BDS}.  The vanishing boundary condition holds to all loop orders.  At one loop, it is a trivial dilog identity, ${\rm Li}_2(1-u) = \zeta_2 - \log u \log(1-u) - {\rm Li}_2(u)$.

Given a vanishing boundary condition at the boundary $u+w=1$, we can learn about the sign of the one-loop ratio function throughout the NMHV positive region by looking instead at the radial derivative of $C^{(1)}(u,w)$,
\begin{align}
\big (u \partial_u + w \partial_w \big) C^{(1)}(u,w) =
\frac{\log u}{1-u} + \frac{\log w}{1-w} \,.
\end{align}
This derivative is manifestly negative for all $u,w > 0$.  Also, radial flow can be used to reach any point $(u,w)$ starting from some point on the boundary, namely the point $(\frac{u}{u+w},\frac{w}{u+w})$.  Thus $C^{(1)}(u,w)$ and ${\cal P}_{6,1}^{1-{\rm loop}}|_{c_5, c_6 \rightarrow 0}$ must be negative throughout region~(\ref{ds_positive_region}).

\section{Positivity in the double-scaling limit}
\label{DSLimit}

We now begin to extend our investigation of positivity from one loop to higher loop orders.  In this section, we focus on the double-scaling limit just discussed in section~\ref{one_loop_ds_limit}.  Because the $R$-invariants are independent of loop order, the only difference in going to higher loops is that the transcendental function $C^{(1)}(u,w)$ in \eqn{dsoneloop} is replaced by the sum of the coefficients of the $R$-invariants $(5)$ and $(6)$, in \eqn{ratio_function} for ${\cal P}_{6,1}$.  Up to a factor of $1/2$, we denote this sum by $C(u,v,w)$. In terms of the functions $V$ and $\widetilde V$, it is given by
\begin{align}
C(u,v,w) = V(v,w,u) + V(w,u,v)+  \widetilde{V}(y_v,y_w,y_u) - \widetilde{V}(y_w,y_u,y_v) \,. \label{DS_comb}
\end{align}

The limit $v\to 0$ with $u,w$ held fixed (or $c_5, c_6 \rightarrow 0$ in the positive parametrization) acts on the extended cross ratios $y_i$ by sending
\be
y_u \rightarrow \frac{1-w}{u} \,, \quad
y_v \rightarrow \frac{(1-u-w)^2}{v (1-u) (1-w)} \,, \quad
y_w \rightarrow \frac{1-u}{w} \,.
\label{yuyvywToDS}
\ee
(Because $u,v,w$ remain stationary under parity, while $y_u,y_v,y_w$ invert, one might think that one could send the $y_i$ variables instead to the reciprocal of the three values chosen in \eqn{yuyvywToDS}.  However, this choice is inconsistent with the positive parametrization~(\ref{ratio2Y}).)

In general, the functions $V$ and $\widetilde{V}$ diverge logarithmically in this limit, because the amplitude has a physical branch cut at $v=0$, where the Mandelstam variables $s_{23}$ and $s_{56}$ vanish.  We therefore parametrize the limiting behavior of $C(u,v,w)$ as an expansion in powers of $\log(1/v)$ as well as loop order,
\begin{equation}
C(u,v\to0,w)  = \sum_{\ell=0}^\infty \sum_{n=0}^{\ell-1} (-a)^\ell c^{(\ell)}_n(u,w)
               \log^n (1/v),
\label{C_expansion}
\end{equation}
up to power-suppressed terms. The upper limit on the sum over $n$ reflects the empirical observation that the leading-logarithmic contribution is
$\log^{\ell-1}(1/v)$ at $\ell$ loops. We expect that this observation should have a OPE-based explanation.

The one-loop case studied in the previous section is the only one with no logarithmic divergence:
\be
C^{(1)}(u,v\to0,w) = C^{(1)}(u,w) = -c_0^{(1)}(u,w).
\label{C1example}
\ee
The use of $(-a)$ in \eqn{C_expansion} ensures that all the coefficients $c^{(\ell)}_n(u,w)$ will be empirically positive, given the overall sign alternation with loop order discussed in the introduction.  The boundary condition discussed in the previous subsection, that the ratio function vanishes in the collinear limit, tells us that
\be
c^{(\ell)}_n(u,1-u) = 0,
\label{vanishbc}
\ee
for all $\ell$ and $n$.

The limiting values~(\ref{yuyvywToDS}) for the $y_i$ imply that the coefficient functions $c^{(\ell)}_n (u,w)$ in \eqn{C_expansion} can be expressed as multiple polylogarithms~\cite{FBThesis,Gonch} of weight $2\ell-n$
with symbol letters drawn from the set~\cite{Dixon2014iba,Drummond2015jea} 
\begin{equation}
\mathcal{S}_\text{DS} = \{ u, w, 1-u, 1-w, 1-u -w \}, \label{DS_alphabet}
\end{equation}
and branch cuts only in the letters $u$ and $w$.  This ``double-scaling'' function space is a subspace of the 2dHPL function space introduced by Gehrmann and Remiddi~\cite{GehrmannRemiddi} for four-point scattering with one massive leg
and three massless legs.  

The $c^{(\ell)}_n(u,w)$ can be computed from $V$ and $\widetilde V$ by expressing them as multiple polylogarithms and taking the double scaling limit directly using the replacements (\ref{yuyvywToDS}) for the $y_i$ variables.  In this process, one can also extract the $\log(1/v)$ dependence.  Alternatively, one can construct the double-scaling function space more abstractly at first, using the set of relations between derivatives and coproducts implied by the symbol alphabet $\mathcal{S}_\text{DS}$.  These relations are limiting versions of the coproduct relations used in the hexagon function bootstrap.  Then one can find matching conditions between these functions and the $v\to0$ limit of one's basis of hexagon functions. For an example of the latter procedure see Appendix D of ref.~\cite{Dixon2015iva}. 

In the latter approach, at high loop order it may be preferable to perform intermediate steps using the BDS-like normalized MHV and NMHV amplitudes that satisfy the Steinmann relations, because the space of Steinmann-satisfying hexagon functions is much smaller~\cite{CaronHuot2016owq}.  The limiting behavior of the (non-Steinmann) functions $V$ and $\tilde{V}$ can then be computed from the limiting values of the Steinmann functions.

In section~\ref{FullDoubleScalingSubsection} we will show plots for the coefficient functions $c_n^{(\ell)}(u,w)$ on the full two-dimensional double-scaling surface~(\ref{ds_positive_region}).  First, however, we would like to examine their behavior on three one-dimensional lines that trace through this surface.

\subsection{Positivity along lines in the double-scaling limit}

The space of functions relevant for six-gluon scattering amplitudes simplifies further in three one-dimensional subspaces of the double-scaling limit, where everything can be expressed in terms of harmonic polylogarithms (HPLs) of a single variable~\cite{Remiddi1999ew}.  On these lines, we can evaluate the ratio function numerically in Mathematica using the HPL package~\cite{Maitre2005uu}. Correspondingly, we first explore the behavior of the functions $c_n^{(\ell)}(u,w)$ in these special kinematic regions, before enlarging the scope of our study to the full double-scaling limit. As we will see later, these lines turn out to capture most of the interesting information about the ratio function in the double-scaling limit.

\subsubsection{The line \texorpdfstring{$w=1$}{w=1}}

\begin{figure*}[t]
\centering
        \begin{subfigure}[b]{0.325\textwidth}
            \includegraphics[width=\textwidth]{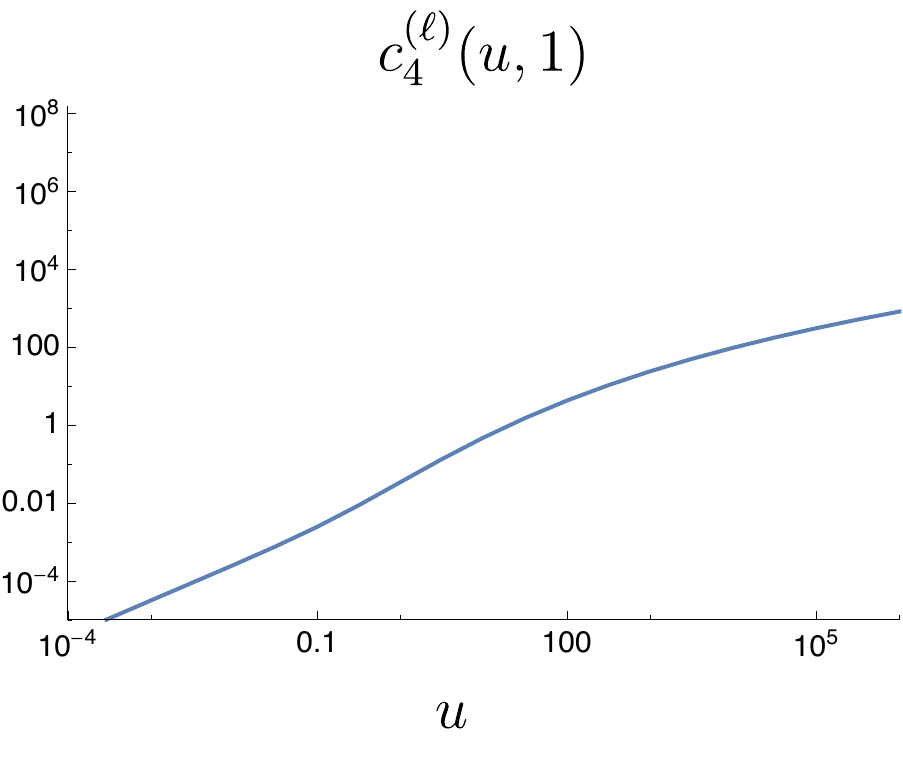}
        \end{subfigure}
        \begin{subfigure}[b]{0.325\textwidth}
            \includegraphics[width=\textwidth]{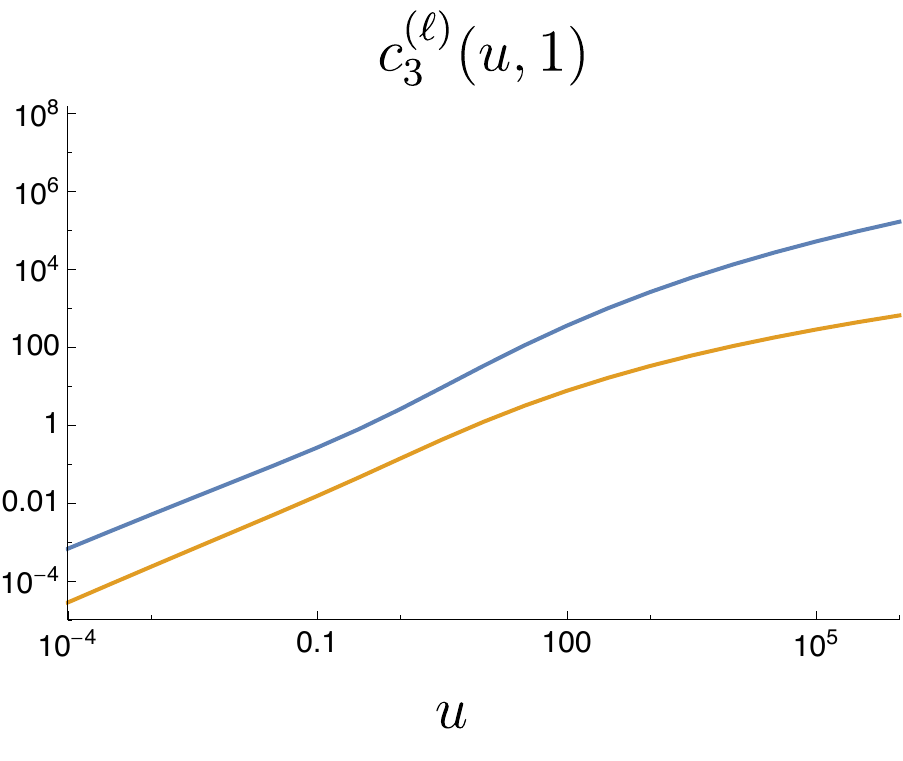}
        \end{subfigure}
        \begin{subfigure}[b]{0.325\textwidth}  
            \includegraphics[width=\textwidth]{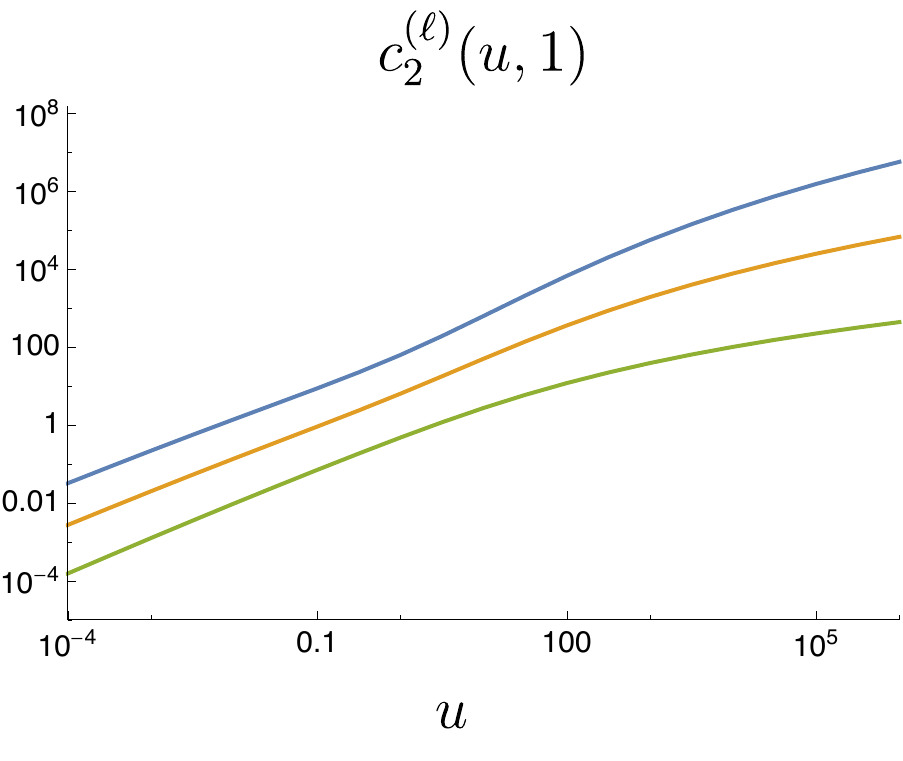}
        \end{subfigure}
    \vskip\baselineskip \vspace*{-.5cm}
        \begin{subfigure}[b]{0.325\textwidth}
            \includegraphics[width=\textwidth]{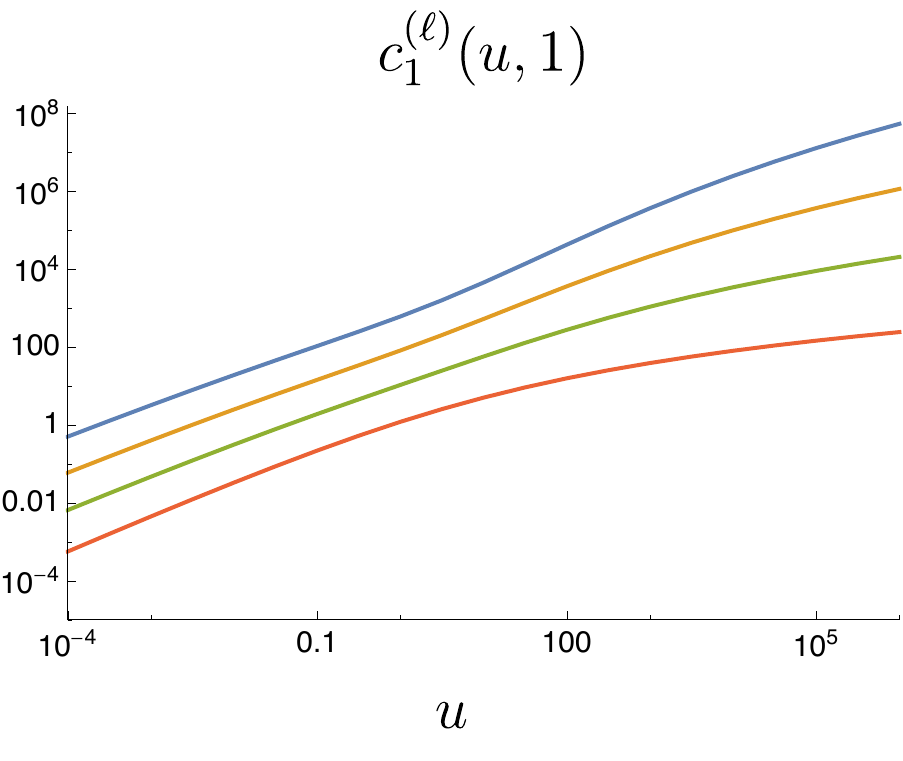}
        \end{subfigure}
        \hspace{.3cm}
        \begin{subfigure}[b]{0.325\textwidth}  
            \includegraphics[width=\textwidth]{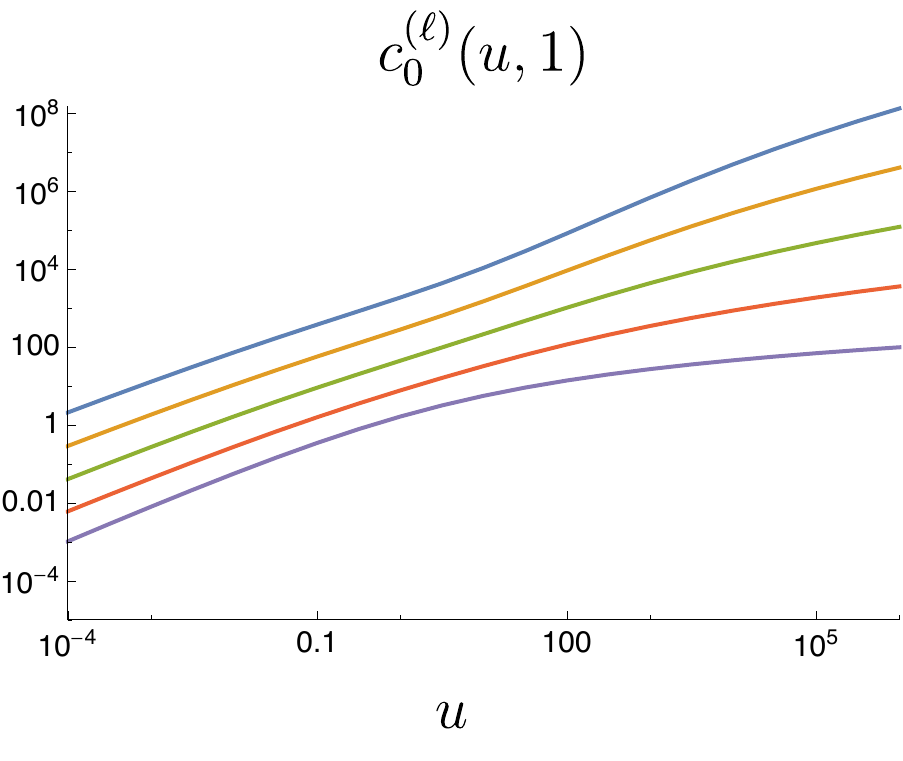}
        \end{subfigure}
        \caption[]
        {The coefficient functions $c_n^{(\ell)}(u,1)$ that multiply $\log^n(1/v)$ in the double-scaling limit at $\ell$ loops. Five loops is shown in blue, four loops in yellow, three loops in green, two loops in red, and one loop in purple.}
        \label{line_u1}
\end{figure*}

The first simple line in the double-scaling limit corresponds to setting $w=1$. This collapses $\mathcal{S}_\text{DS}$ to the simpler set of letters $\{u,1-u\}$, which implies that the functions $c^{(\ell)}_n(u,1)$ can be written as a sum of HPLs with argument $1-u$. This representation can be built up through iterative integrations, using the fact that the $u$ derivative of a generic hexagon function $F$ collapses to
\begin{align}
\frac{\partial F}{\partial u} \bigg|_{v\rightarrow 0; w = 1}  = \frac{F^{u}-F^{y_u}+2F^{y_v}}{u} - \frac{F^{1-u} - F^{y_v} + F^{y_w}}{1-u}
\end{align}
along this line. To carry out this integration on a generic hexagon function, one must also set the integration constant at each weight. This can be done by integrating from the point $(u,v,w)=(1,1,1)$, where the additive constants of hexagon functions are usually defined, to the point $(1,0,1)$ along the line $(1,v,1)$. Hexagon functions all collapse to HPLs with argument $1-v$ along the line $(1,v,1)$, so this integration is also simple~\cite{Dixon2013eka}. Using this procedure, we have computed the functions $c_n^{(\ell)}(u,1)$ through five loops, which we plot in figure~\ref{line_u1}. We also provide their HPL expressions in an ancillary file. 

The vanishing of the ratio function along the collinear line $u+w=1$, \eqn{vanishbc}, requires that the $c_n^{(\ell)}(u,1)$ all vanish at the point $u=0$. We can also check the behavior of these functions as $u\rightarrow\infty$, where they reduce to polynomials in $\log u$. For instance, the coefficient functions $c_0^{(\ell)}(u\rightarrow \infty, 1)$ become
\begin{align}
c_0^{(1)}(u\rightarrow \infty, 1) &= \frac{1}{2} \log^2 u  + 2 \zeta_2 \,, \\
c_0^{(2)}(u\rightarrow \infty, 1) &= \frac{1}{12} \log^4 u + \frac{7}{4} \zeta_2 \log^2 u + \frac{1}{2} \zeta_3 \log u +\frac{59}{4} \zeta_4 \,, \\
c_0^{(3)}(u\rightarrow \infty, 1) &= \frac{1}{80} \log^6 u + \frac{25}{48} \zeta_2 \log^4 u + \frac{1}{24} \zeta_3 \log^3 u + \frac{287}{16} \zeta_4 \log^2 u \nonumber \\
                                                 &\hspace{1cm} + \frac{7}{4} \zeta_5 \log u + \frac{3}{2} \zeta_3^2 + \frac{6303}{64} \zeta_6  \,, \\
c_0^{(4)}(u\rightarrow \infty, 1) &= \frac{37}{20160} \log^8 u+ \frac{11}{96} \zeta_2 \log^6 u - \frac{1}{480} \zeta_3 \log^5 u + \frac{459}{64} \zeta_4 \log^4 u \nonumber \\
                                                 &\hspace{1cm}  - \bigg( \frac{1}{2} \zeta_2 \zeta_3 + \frac{19}{48} \zeta_5 \bigg) \log^3 u + \bigg( \frac{3}{2} \zeta_3^2 + \frac{108763}{768} \zeta_6 \bigg) \log^2 u  \nonumber \\
                                                 &\hspace{1cm} + \bigg( \frac{381}{128} \zeta_7 - \frac{443}{32} \zeta_4 \zeta_3 - \frac{107}{16} \zeta_5 \zeta_2 \bigg) \log u \nonumber \\
                                                 &\hspace{1cm} - \frac{1}{4} \zeta_{5,3} + \frac{3299555}{4608} \zeta_8 + \frac{63}{4} \zeta_5 \zeta_3 + \frac{85}{16} \zeta_3^2 \zeta_2 \,, \\
c_0^{(5)}(u\rightarrow \infty, 1) &= \frac{13}{48384} \log^{10} u + \frac{899}{40320} \zeta_2 \log^8 u - \frac{7}{5760} \zeta_3 \log^7 u +\frac{2559}{1280} \zeta_4 \log^6 u \nonumber \\
                                                 &\hspace{1cm} - \bigg( \frac{223}{960} \zeta_3 \zeta_2 + \frac{71}{320} \zeta_5 \bigg) \log^5 u + \bigg( \frac{103}{192} \zeta_3^2 + \frac{105113}{1536} \zeta_6 \bigg) \log^4 u  \nonumber \\
                                                 &\hspace{1cm} - \bigg( \frac{1613}{96} \zeta_4 \zeta_3 + \frac{1769}{192} \zeta_2 \zeta_5 + \frac{1913}{256} \zeta_7 \bigg) \log^3 u \nonumber \\
                                                 &\hspace{1cm} + \bigg( \frac{691}{64} \zeta_2 \zeta_3^2 + \frac{659}{32} \zeta_5 \zeta_3 - \frac{3}{8} \zeta_{5,3} + \frac{21436813}{18432} \zeta_8 \bigg) \log^2 u \nonumber \\
                                                 &\hspace{1cm}  - \bigg(  \frac{79}{48} \zeta_3^3 + \frac{60801}{256} \zeta_6 \zeta_3  + \frac{3209}{16} \zeta_4 \zeta_5 + \frac{6913}{64} \zeta_7 \zeta_2 + \frac{66545}{1152} \zeta_9 \bigg) \log u \nonumber \\
                                                 &\hspace{1cm} - \frac{101}{160} \zeta_2 \zeta_{5,3} - \frac{543}{512} \zeta_{7,3} + \frac{10267}{128} \zeta_4 \zeta_3^2 + \frac{2707}{32} \zeta_2 \zeta_5 \zeta_3 \nonumber \\
                                                 &\hspace{1cm} + \frac{1717}{16} \zeta_7 \zeta_3 + \frac{28635}{512} \zeta_5^2 + \frac{592519707}{102400} \zeta_{10} \,,
 \end{align}
which all approach positive infinity, as expected. More generally, we have checked that $c_n^{(\ell)}(u\rightarrow \infty, 1) \rightarrow +\infty$ for all $\ell\leq5$ and for all $n$ between $0$ and $\ell-1$.

Since $v$ is very small, positivity strictly requires only the leading-log coefficients $c_{\ell-1}^{(\ell)}(u,1)$ to be positive.  However, we find a much stronger result: The coefficients $c_n^{(\ell)}(u,1)$ are all positive for $u>0$ and for any $n$ between 0 and $\ell-1$.  Furthermore, figure~\ref{line_u1} shows that they all increase monotonically with $u$.
 
\subsubsection{The line \texorpdfstring{$w=0$}{w=0}}
\label{NMHVweq0SubSubsection}

\begin{figure*}[t]    
\centering 
        \begin{subfigure}[b]{0.325\textwidth}
            \includegraphics[width=\textwidth]{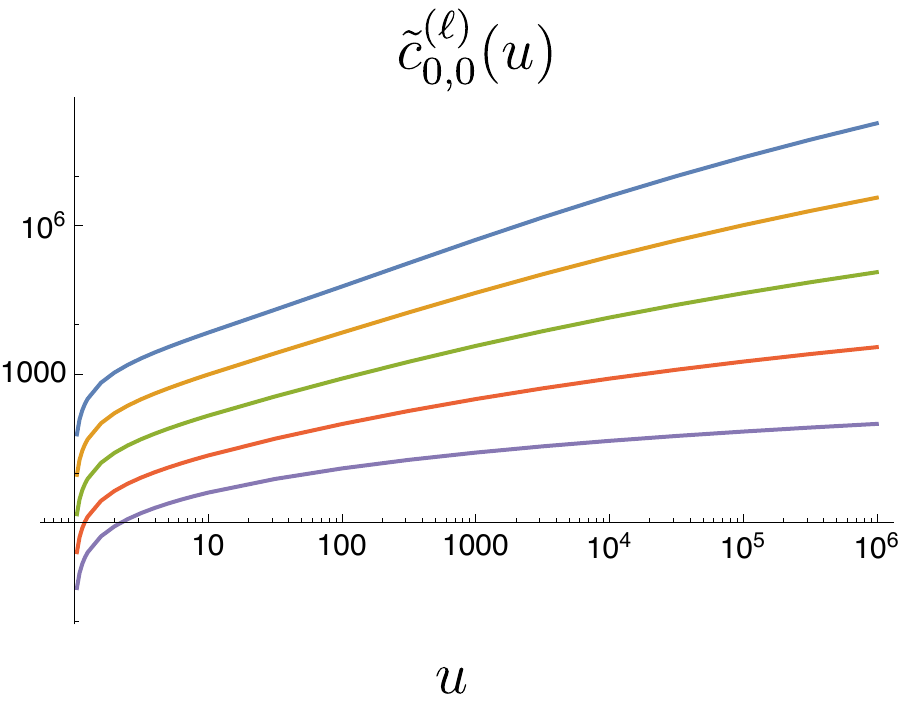}
        \end{subfigure}
        \begin{subfigure}[b]{0.325\textwidth}  
            \includegraphics[width=\textwidth]{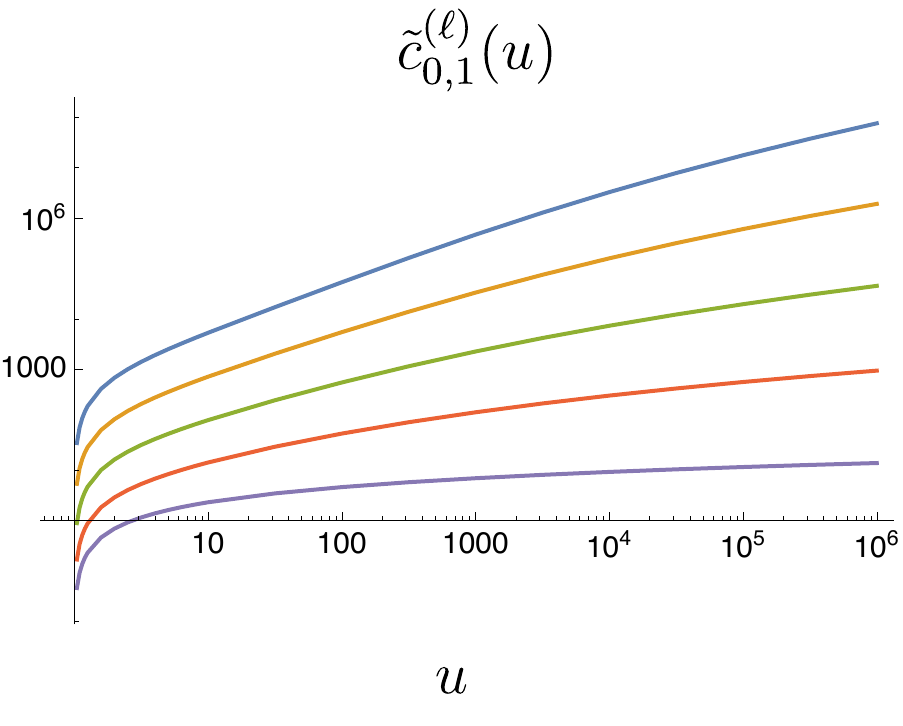}
        \end{subfigure}
        \begin{subfigure}[b]{0.325\textwidth}
            \includegraphics[width=\textwidth]{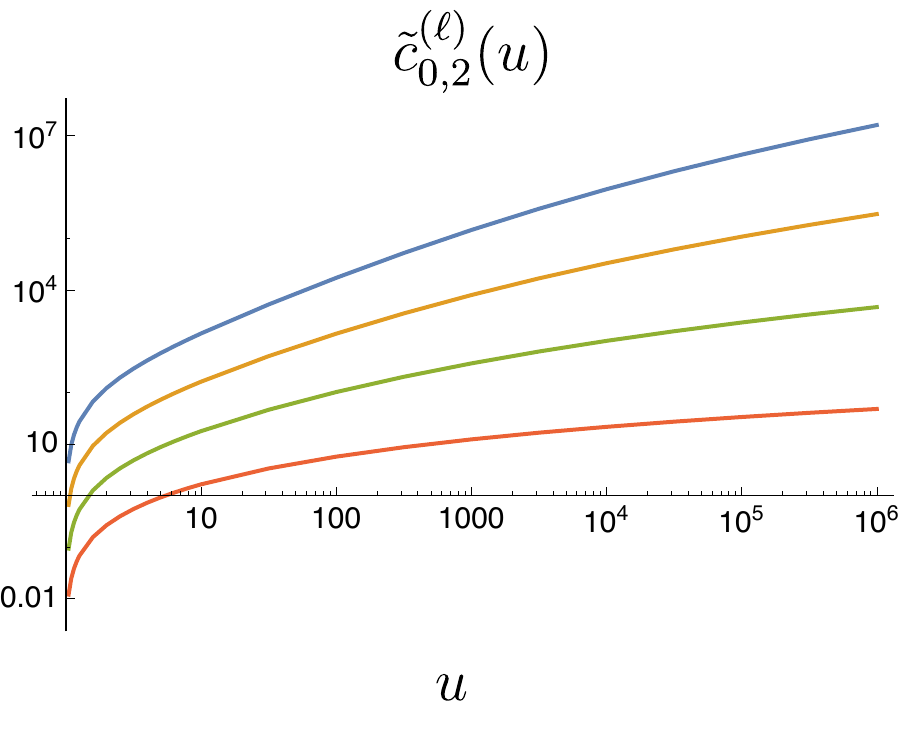}
        \end{subfigure}
    \vskip\baselineskip \vspace*{-.3cm}
        \begin{subfigure}[b]{0.325\textwidth}  
            \includegraphics[width=\textwidth]{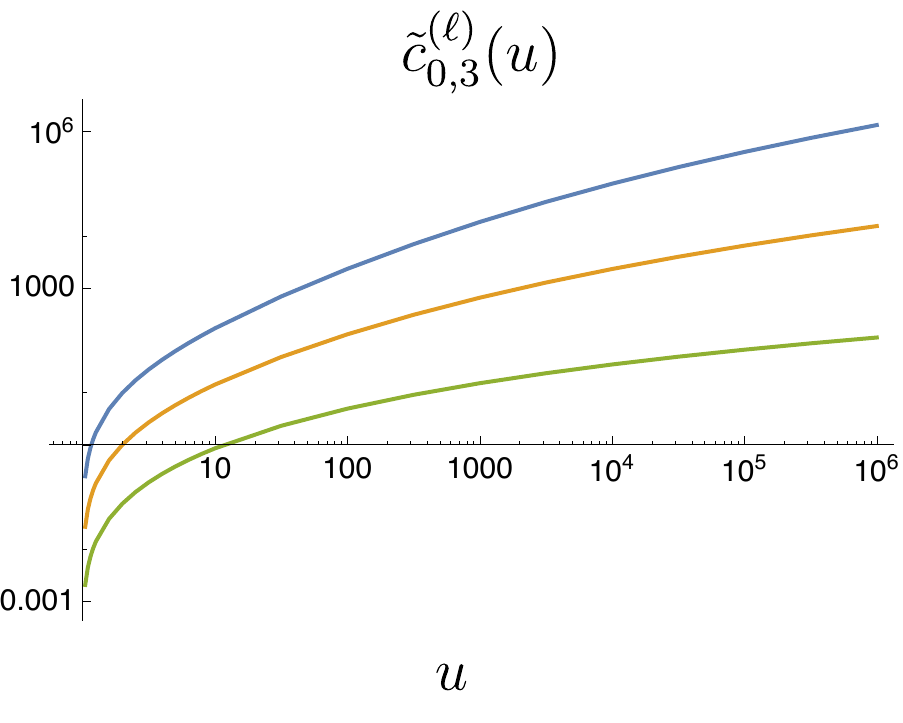}
        \end{subfigure}
        \begin{subfigure}[b]{0.325\textwidth}   
            \includegraphics[width=\textwidth]{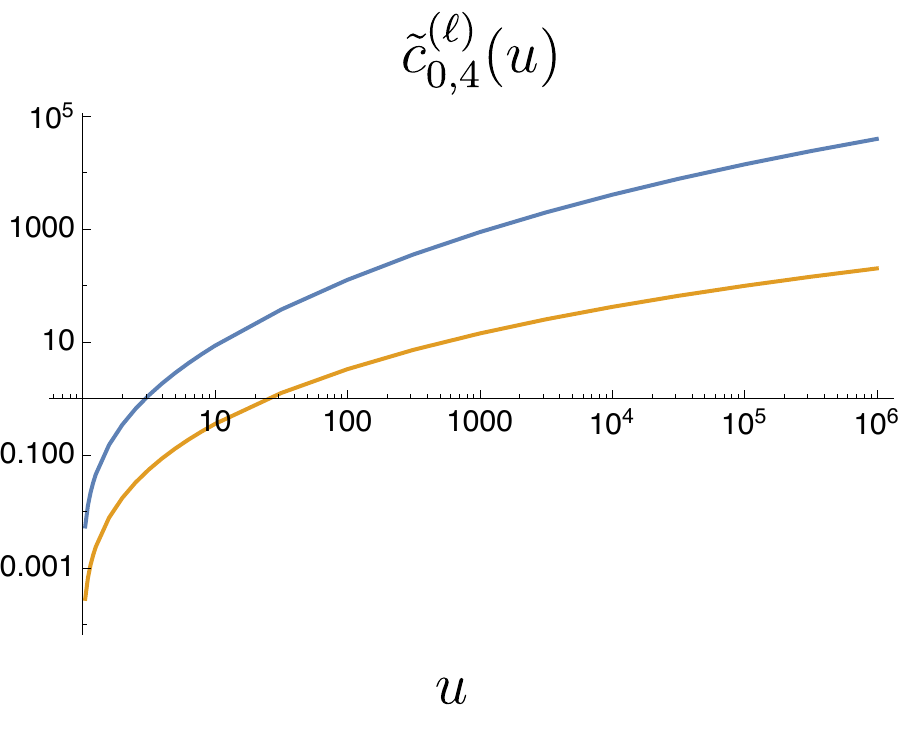}
        \end{subfigure}
        \begin{subfigure}[b]{0.325\textwidth}   
            \includegraphics[width=\textwidth]{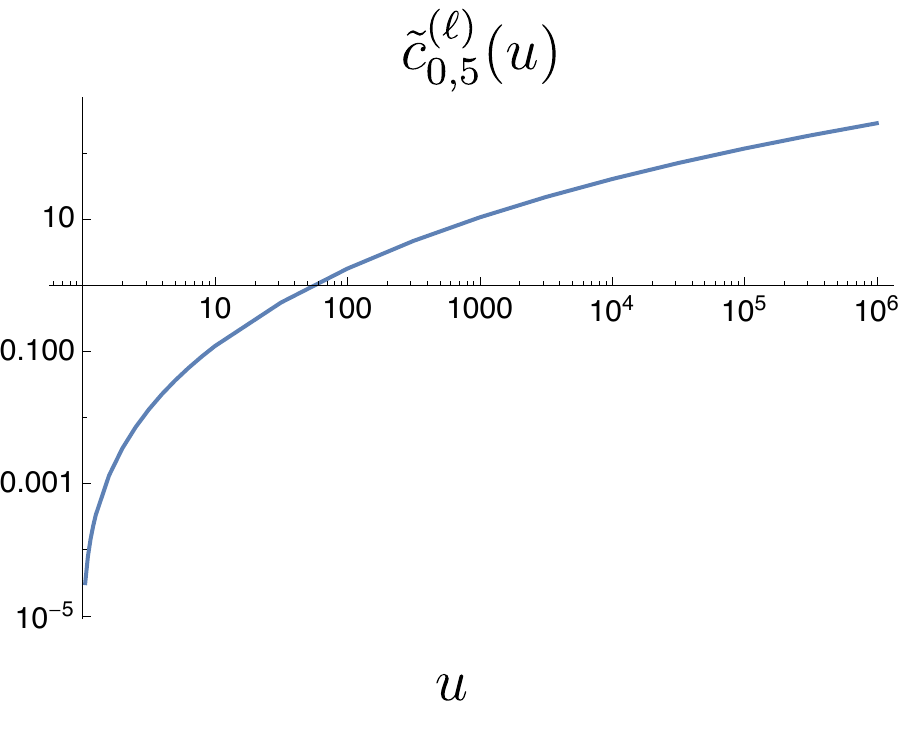}
        \end{subfigure}
        \caption[]
        {The coefficient functions $\tilde{c}^{(\ell)}_{0,k}(u)$ for the $w\to0$ edge of the double-scaling limit at $\ell$ loops. Five loops is shown in blue, four loops in yellow, three loops in green, two loops in red, and one loop in purple.}
        \label{line_u0_c0}
\end{figure*}

The second simple line we will look at is $w=0$.  It forms an edge of the positive double-scaling region~(\ref{ds_positive_region}).  As was the case for the $w=1$ line, $\mathcal{S}_\text{DS}$ collapses to $\{u,1-u\}$. However, $c^{(\ell)}_n (u, w \rightarrow 0)$ diverges logarithmically in $w$ due to a physical branch cut analogous to the branch cut in $v$. The functions $c^{(\ell)}_n (u, w \rightarrow 0)$ are therefore expressible as an expansion in powers of $\log(1/w)$,
\be
c^{(\ell)}_n (u, w \rightarrow 0) = \sum_{k=0}^{\ell-n} \tilde{c}_{n,k}^{(\ell)}(u) \log^k(1/w).
\label{cu0coeffsdef}
\ee
The coefficients $\tilde{c}_{n,k}^{(\ell)}(u)$ are drawn from the space of HPLs with argument $1-u$, and empirically they vanish unless $k$ is between 0 and $\ell-n$, where we recall that $n$ is the power of $\log(1/v)$ in the expansion~(\ref{C_expansion}).  

The derivative of a generic hexagon function $F$ along the line $(u,w\to0)$ is given by
\begin{align}
\frac{\partial F}{\partial u} \bigg|_{v, w \rightarrow 0} = \frac{F^u - F^{y_u}}{u} - \frac{F^{1-u} + F^{y_v} + F^{y_w}}{1-u} \,.
\end{align}
The integration constant can be set at $u=1$, using the $v\to0$ endpoint of the line $(u,v,w) = (1,v,0)$, which is just an $S_3$ permutation of the line $(u,0,1)$ considered in the previous subsection.

We have carried out the corresponding integration through five loops and we include HPL representations of all the $\tilde{c}_{n,k}^{(\ell)}(u)$ in an ancillary file. The functions $\tilde{c}_{0,k}^{(\ell)}(u)$, which multiply different powers of $\log(1/w)$ in the non-$\log(1/v)$ part, are plotted in figure~\ref{line_u0_c0}. Due to the large number of independent functions multiplying different powers of large logs on this line, we have relegated plots of the other $\tilde{c}_{n,k}^{(\ell)}(u)$ functions to appendix~\ref{appendix_line_u0}. 

The vanishing of the ratio function along the collinear line $u+w=1$, \eqn{vanishbc}, requires these coefficient functions to become zero at $u=1$. We have also checked analytically that each of these functions approaches positive infinity in the limit $u\to\infty$.  Once again, we observe that all the coefficient functions -- not just the leading-log ones -- are positive, and furthermore that they are monotonically increasing with $u$.

Interestingly, there is an HPL representation in which the positivity and monotonicity of the $\tilde{c}_{n,k}^{(\ell)}(u)$ is {\it almost} manifest.  We let the argument of the HPLs be $z=1-1/u$.  As $u$ increases from 1 to $\infty$, $z$ increases from 0 to 1. In this range of $z$, the HPLs with trailing 1's in their weight vectors are manifestly positive and monotonic, simply from their integral definition,
\be
H_{0,\vec{w}}(z) = \int_0^z \frac{dt}{t} H_{\vec{w}}(t), \quad
H_{1,\vec{w}}(u) = \int_0^z \frac{dt}{1-t} H_{\vec{w}}(t), 
\label{Hdef}
\ee
because the integrand is a lower-weight HPL of the same form, $H_{\vec{w}}(t)$, multiplied by a kernel that is positive for $0 < t < 1$.  Hence if the $\tilde{c}_{n,k}^{(\ell)}(u)$ could be written in terms of such HPLs with only positive coefficients, positivity and monotonicity would both be manifest. 

At one and two loops, this is the case; the non-vanishing coefficients are
\bea
\tilde{c}^{(1)}_{0,1} &=& H_1 \,, \nonumber\\
\tilde{c}^{(1)}_{0,0} &=& H_{0,1} + H_{1,1} \,, \nonumber\\
\tilde{c}^{(2)}_{1,1} &=& \frac{1}{2} H_{0,1} + \frac{1}{2} H_{1,1}  \,, \nonumber\\
\tilde{c}^{(2)}_{1,0} &=&  H_{0,0,1} + H_{0,1,1}
             + \frac{1}{2} H_{1,0,1} + \frac{1}{2} H_{1,1,1} \,, \nonumber\\
\tilde{c}^{(2)}_{0,2} &=& \frac{1}{4} H_{0,1} + \frac{1}{2} H_{1,1} \,, \nonumber\\
\tilde{c}^{(2)}_{0,1} &=&  2 H_{0,0,1} + \frac{5}{2} H_{0,1,1}
    + \frac{3}{2} H_{1,0,1} + 2 H_{1,1,1} + \zeta_2 H_{1} \,, \nonumber\\
\tilde{c}^{(2)}_{0,0} &=& \frac{9}{2} H_{0,0,0,1} + 5 H_{0,0,1,1} + 3 H_{0,1,0,1}
 + \frac{7}{2} H_{0,1,1,1} + 2 H_{1,0,0,1} + \frac{5}{2} H_{1,0,1,1} \nonumber\\
&&\hskip0cm\null
 + \frac{3}{2} H_{1,1,0,1} + 2 H_{1,1,1,1}
 + \zeta_2 \Bigl( \frac{1}{2} H_{0,1} + H_{1,1} \Bigr) \,, 
\label{u0HPL12loop}
\eea
where we have suppressed the argument $z=1-1/u$ of the HPLs $H_{\vec{w}}(z)$, displaying only their weight vector $\vec{w}$.

Since all the coefficients in \eqn{u0HPL12loop} are positive, positivity and monotonicity on the line $w=0$ is manifest through two loops.  However, the plot thickens at three loops.  All 9 nonzero coefficient functions $\tilde{c}^{(3)}_{n,k}$ have positive coefficients in their representations, except for $\tilde{c}^{(3)}_{1,0}$ and $\tilde{c}^{(3)}_{0,0}$.  The only negative coefficients in these functions are those in terms containing $\zeta_3$ -- for example,
\bea
\tilde{c}^{(3)}_{1,0} &=& 6 H_{0,0,0,0,1} + \frac{45}{4} H_{0,0,0,1,1}
 + 6 H_{0,0,1,0,1} + \frac{45}{4} H_{0,0,1,1,1} + 4 H_{0,1,0,0,1}
 + \frac{31}{4} H_{0,1,0,1,1} \nonumber\\
&&\hskip0cm\null
 + 4 H_{0,1,1,0,1} + \frac{23}{4} H_{1,0,1,1,1}
 + 2 H_{1,1,0,0,1} + 4 H_{1,1,0,1,1} + 2 H_{1,1,1,0,1} + 4 H_{1,1,1,1,1}  \nonumber\\
&&\hskip0cm\null
+ \frac{31}{4} H_{0,1,1,1,1} + 3 H_{1,0,0,0,1} + \frac{23}{4} H_{1,0,0,1,1}
 + 3 H_{1,0,1,0,1}   \nonumber\\
&&\hskip0cm\null
 + \zeta_2 \Bigl( \frac{3}{2} H_{0,0,1} + \frac{7}{4} H_{0,1,1}
                + \frac{3}{4} H_{1,0,1} + H_{1,1,1} \Bigr)
 - \frac{1}{2} \zeta_3 H_{0,1} \,.
\label{c3_10_problem}
\eea
Because the numerical coefficient in front of the $\zeta_3$ is relatively small, it doesn't change the actual positivity or monotonicity properties; it just makes them less manifest.

Continuing on to four and five loops, there are 14 and 20 nonzero coefficient functions, respectively, with weights that range from 4 up to 10.  The sign in front of each HPL in each coefficient function is completely predictable:  positive, unless the term has an odd number of odd zeta values, in which case it is negative.  The (mostly) consistent signs for the HPL coefficients are reminiscent of the behavior found for the velocity-dependent cusp anomalous dimension $\Omega_0(x)$ in ref.~\cite{Henn2012qz}.

\subsubsection{The line \texorpdfstring{$u=w$}{u=w}}

The final simple line in the double-scaling limit is given by setting $u=w$. Here, the symbol letters in $\mathcal{S}_\text{DS}$ collapse to the set $\{ u, 1-u, 1-2u \}$. This makes the functions $c^{(\ell)}_n(u,u)$ expressible as HPLs of argument $x \equiv 1-2u$ with weight vectors involving $-1$, 0, and 1. The derivative of a generic hexagon function $F$ along this line takes the form
\begin{align}
\frac{\partial F}{\partial x}\bigg|_{v\rightarrow 0; u,w= (1-x)/2} &= \frac{2 F^{y_v}}{x} + \frac{F^{1-u} + F^{1-w} + F^{y_u} + F^{y_w} - 2 F^{y_v}}{1+x} \nonumber \\
 &\hspace{0.5cm} - \frac{F^{u} + F^{w} - F^{y_u} - F^{y_w} }{1-x} \,,
\end{align}
while the integration constant can be set by matching to the $v \rightarrow 0$ endpoint of the line $(u,v,w) = (1,v,1)$. This requires setting the argument $x = -1$, which introduces transcendental constants beyond the multiple zeta values $\zeta_m$ and $\zeta_{m,n}$. At low weights, there are identities relating these new constants to multiple zeta values, $\log 2$, and ${\rm Li}_n(1/2)$ with $n\ge4$, but starting at weight 6 new alternating sums alt$_{\vec w} \equiv (-1)^{|{\vec w}|} H_{\vec w}(-1)$ are needed~\cite{Blumlein2009}, where $|{\vec w}|$ is the depth of $\vec w$. The numerical values of these constants can be calculated using the HPL package.  

\begin{figure*}[t]    
\centering 
        \begin{subfigure}[b]{0.47\textwidth}
            \includegraphics[width=\textwidth]{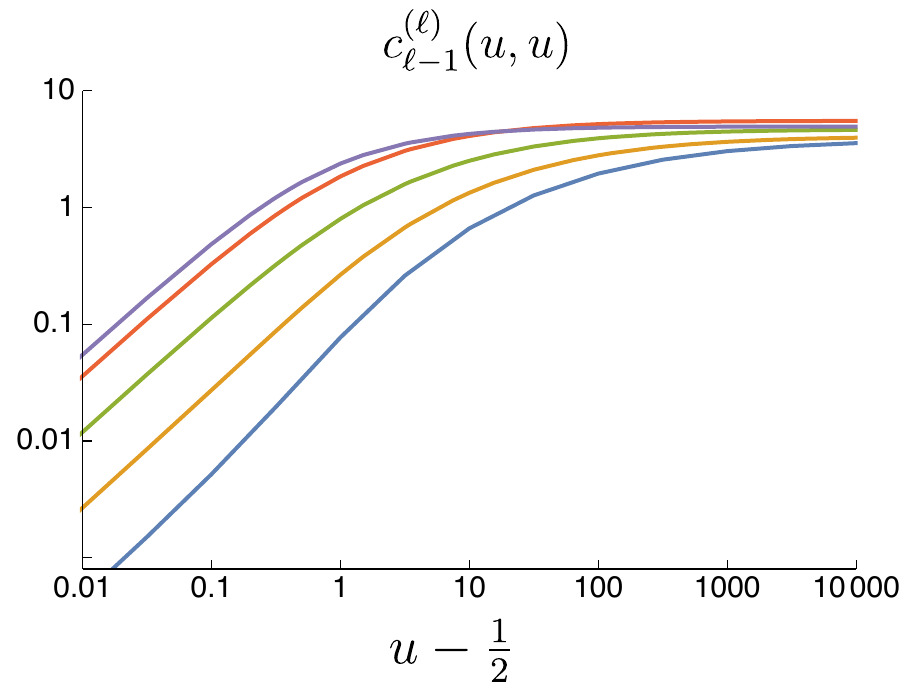}
        \end{subfigure}
        \begin{subfigure}[b]{0.47\textwidth}  
            \includegraphics[width=\textwidth]{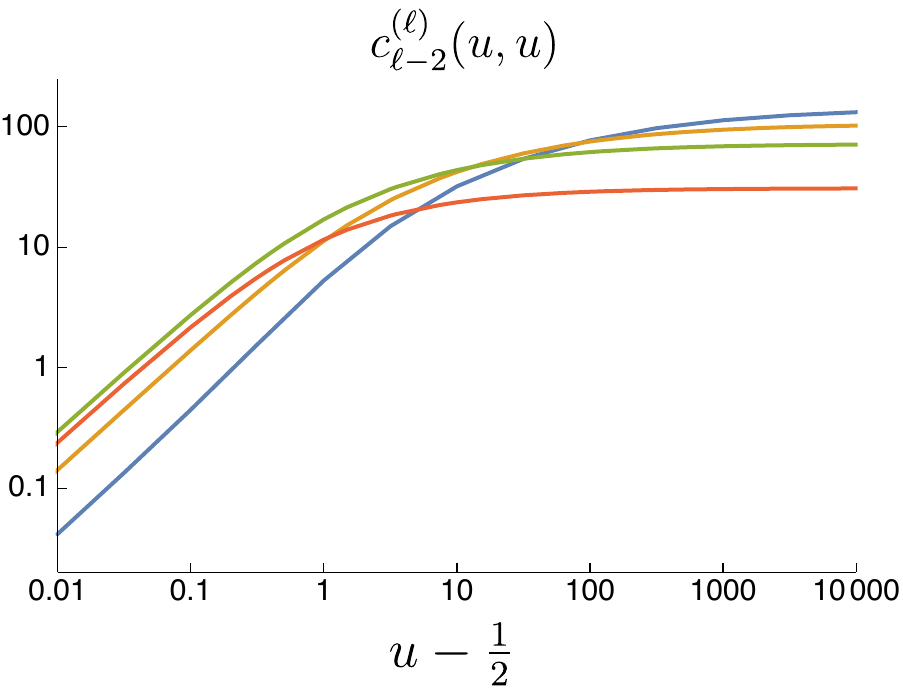}
        \end{subfigure}
        \caption[]
        {The functions $c^{(\ell)}_{\ell-1}(u,u)$ and $c^{(\ell)}_{\ell-2}(u,u)$ governing the leading-log and next-to-leading-log behavior of the ratio function at $\ell$ loops in the double scaling limit. The variable $u$ has been shifted by $\tfrac{1}{2}$ to make it possible to plot on a log scale. Five loops is shown in blue, four loops in yellow, three loops in green, two loops in red, and one loop in purple.}
        \label{line_uu}
\end{figure*}

We have computed the functions $c^{(\ell)}_{n}(u,u)$ through five loops and include their HPL representations in an ancillary file. The functions governing the leading-log and next-to-leading-log contributions in $1/v$ are plotted in figure~\ref{line_uu}. These functions must vanish at $u=\frac{1}{2}$ where they intersect the collinear line $u+w=1$.  While $c_n^{(\ell)}(u,w)$ diverges at large $u$ along the $w=1$ and $w=0$ lines, it has a finite large $u$ limit along the line $u=w$.  That is, figure~\ref{line_uu} shows that the coefficient functions $c^{(\ell)}_{n}(u,u)$ all asymptote to a constant as $u \rightarrow \infty$. This constant can be computed analytically using our HPL representation; for instance, the constants for $n=0$ are given through four loops by
\begin{align}
c_0^{(1)}(u,u)|_{u\rightarrow \infty} &= 3 \zeta_2  \,, \nonumber \\
c_0^{(2)}(u,u)|_{u\rightarrow \infty} &= 27 \zeta_4 + 6 \zeta_2 \log^2 2 - 6 \text{Li}_4(1/2) - \frac{1}{4} \log^4 2 \,, \nonumber \\
c_0^{(3)}(u,u)|_{u\rightarrow \infty} &= 213 \zeta_6 + \frac{55}{16} \zeta_3^2 + \frac{341}{64} \zeta_5 \log 2 + \frac{2835}{32} \zeta_4 \log^2 2 + \frac{23}{16} \zeta_2 \log^4 2 \nonumber \\
                                                        &\hspace{1cm} - \frac{51}{2} \zeta_2 \text{Li}_4(1/2) - 30 \text{Li}_6(1/2) - \frac{1}{24} \log^6 2 - \frac{11}{4} \text{alt}_{5,1} \,, \nonumber \\
c_0^{(4)}(u,u)|_{u\rightarrow \infty} &= \frac{2714608937}{1474560} \zeta_8 + \frac{6793}{512} \zeta_2 \zeta_3^2 + \frac{10285}{4096}  \zeta_3 \zeta_5 - \frac{11683}{20480} \zeta_{5,3}  + \frac{20489}{512} \zeta_3 \zeta_4 \log 2 \nonumber \\
                                                       &\hspace{1cm} + \frac{2871}{64} \zeta_2 \zeta_5 \log 2 + \frac{354801}{16384} \zeta_7 \log 2  - \frac{729}{512} \zeta_3^2 \log^2 2 + \frac{477873}{512} \zeta_6 \log^2 2 \nonumber \\
                                                       &\hspace{1cm} + \frac{787}{192} \zeta_2 \zeta_3 \log^3 2  + \frac{2015}{384} \zeta_5 \log^3 2  + \frac{7423}{128} \zeta_4 \log^4 2 - \frac{221}{960} \zeta_3 \log^5 2 \nonumber \\
                                                       &\hspace{1cm} - \frac{457}{720} \zeta_2 \log^6 2 + \frac{11}{768}  \log^8 2 - \frac{5231}{16} \text{Li}_4 (1/2) \zeta_4 - \frac{43}{2} \text{Li}_4(1/2) \zeta_2 \log^2 2 \nonumber \\
                                                       &\hspace{1cm} + \frac{43}{48} \text{Li}_4(1/2) \log^4 2 + \frac{43}{4} \text{Li}_4(1/2){}^2  + \frac{221}{8} \text{Li}_5(1/2) \zeta_3 \nonumber \\
                                                       &\hspace{1cm} + \frac{9}{2} \text{Li}_5(1/2) \zeta_2 \log 2 - 135 \text{Li}_6(1/2) \zeta_2 -175 \text{Li}_8(1/2) - \frac{67}{16} \text{alt}_{5,1,1,1} \nonumber \\
                                                       &\hspace{1cm} +\frac{193}{64} \text{alt}_{4,2,1,1} + \frac{5281}{256} \text{alt}_{7,1} - \frac{327}{16} \text{alt}_{5,1} \zeta_2  + \frac{67}{16} \text{alt}_{5,1,1} \log 2 \nonumber \\
                                                       &\hspace{1cm} - \frac{193}{64} \text{alt}_{4,2,1} \log 2 -\frac{65}{8} \text{alt}_{5,1} \log^2 2 \,,
\label{c0uu_to_inf_analytic}
\end{align}
while the five loop expression $c_0^{(5)}(u,u)|_{u\rightarrow \infty}$ proves too unwieldy to present.  At one loop this constant is manifestly positive. Evaluating the higher-loop expressions numerically confirms that they are positive as well:
\bea
c_0^{(1)}(u,u)|_{u\rightarrow \infty} &=& 4.93480220054\ldots, \nonumber\\
c_0^{(2)}(u,u)|_{u\rightarrow \infty} &=& 30.8020253462\ldots, \nonumber\\
c_0^{(3)}(u,u)|_{u\rightarrow \infty} &=& 235.199512804\ldots, \nonumber\\
c_0^{(4)}(u,u)|_{u\rightarrow \infty} &=& 2091.54312703\ldots, \nonumber\\
c_0^{(5)}(u,u)|_{u\rightarrow \infty} &=& 22406.9101345\ldots.
\label{c0uu_to_inf_num}
\eea
Indeed, numerical checks reveal that the functions $c_n^{(\ell)}(u,u)$ are positive throughout the positive region, and increase monotonically with $u$. This has been checked exhaustively through four loops and for $n > 1$ at five loops. The higher-weight expressions $c_1^{(5)}(u,u)$ and $c_0^{(5)}(u,u)$ are more computationally challenging to check at finite $u$, and have only been checked in the limit $u \rightarrow \infty$.

\subsection{The full double-scaling surface}
\label{FullDoubleScalingSubsection}

Figures~\ref{line_u1}, \ref{line_u0_c0} and \ref{line_uu},
as well as those in appendix~\ref{appendix_line_u0},
exhibit a remarkable feature -- the functions $c_n^{(\ell)}(u,w)$ are not only positive along these lines, but increase monotonically as they move away from the $u+w=1$ line.  We proved this radial monotonicity at one loop, for $c_0^{(1)}(u,w)$, in section~\ref{one_loop_ds_limit}.  In appendix~\ref{c21monotonicity} we show it for the next simplest case, $c_1^{(2)}(u,w)$, a weight-3 function.
These results make it natural to conjecture that the monotonicity of $c_n^{(\ell)}(u,w)$ holds to all loop orders. 

In the rest of this section we check the monotonicity of the $c_n^{(\ell)}(u,w)$ numerically throughout the double-scaling surface. This can be done by expressing the functions in terms of Goncharov polylogarithms, which can be numerically evaluated using the program {\sc GiNaC}~\cite{Bauer2000cp, Vollinga2004sn} wherever these functions admit a convergent series expansion. The convergence condition for a Goncharov polylogarithm $G(\vec{a}, z)$ is that $|z| \le |a_i|$ for all nonzero $a_i$. This condition is satisfied in the triangle subregion $u+w>1$, $u<1$, $w<1$ if we work in the following basis of Goncharov polylogarithms:
\begin{equation}
G_\text{DS} = \biggl\{ G(\vec{a};1-w) \Big| a_i \in (0,u,1) \biggr\} \cup \biggl\{ G(\vec{a};1-u) \Big| a_i \in (0,1) \biggr\} \,. \label{DS_polylogs}
\end{equation}
This basis can also be used in the remainder of the NMHV positive region, where $u$ and/or $w$ is larger than 1, because {\sc GiNaC} automatically employs identities to relate functions outside their region of convergence to ones that do admit a convergent expansion. This procedure can generate imaginary parts for individual $G$ functions, but the imaginary parts cancel out in the final result.

\begin{figure*}[ht]    
\centering 
        \begin{subfigure}[b]{0.47\textwidth}
            \includegraphics[width=\textwidth]{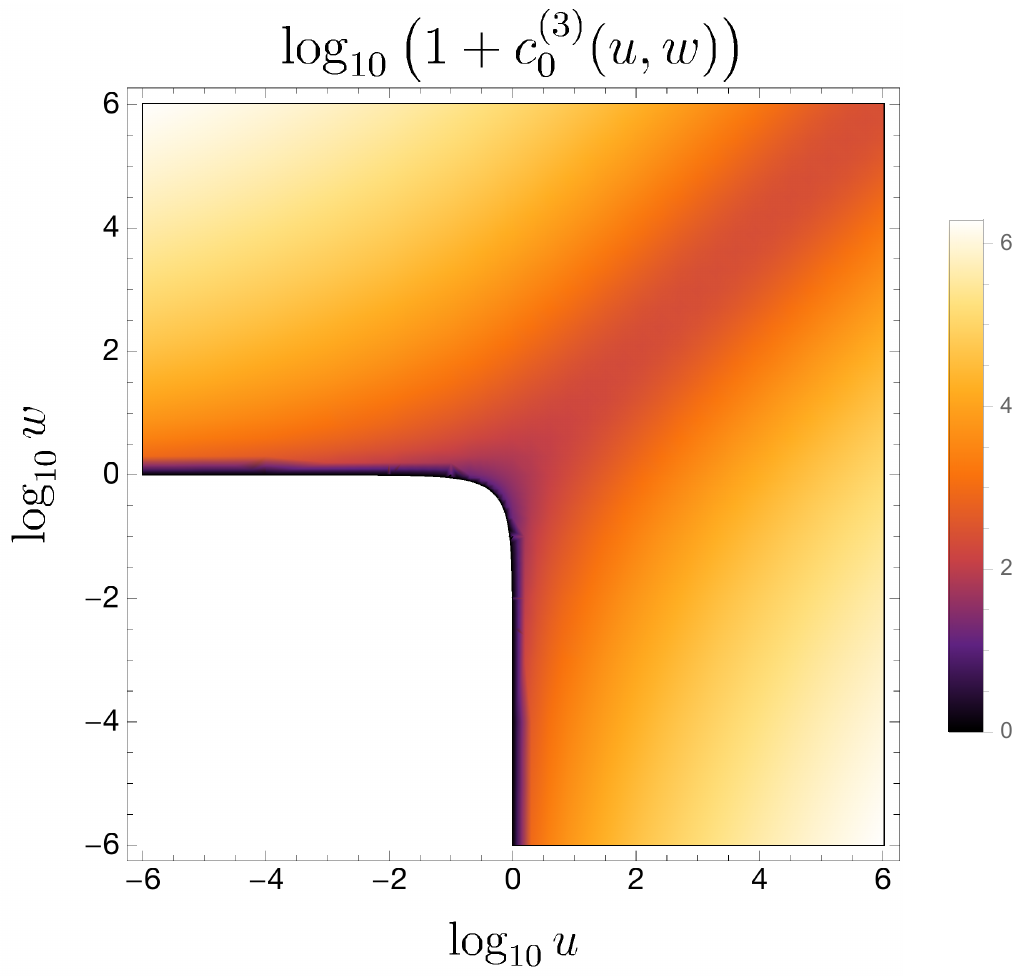}
        \end{subfigure}
    \vskip\baselineskip \vspace*{-.3cm}
        \begin{subfigure}[b]{0.47\textwidth}  
            \includegraphics[width=\textwidth]{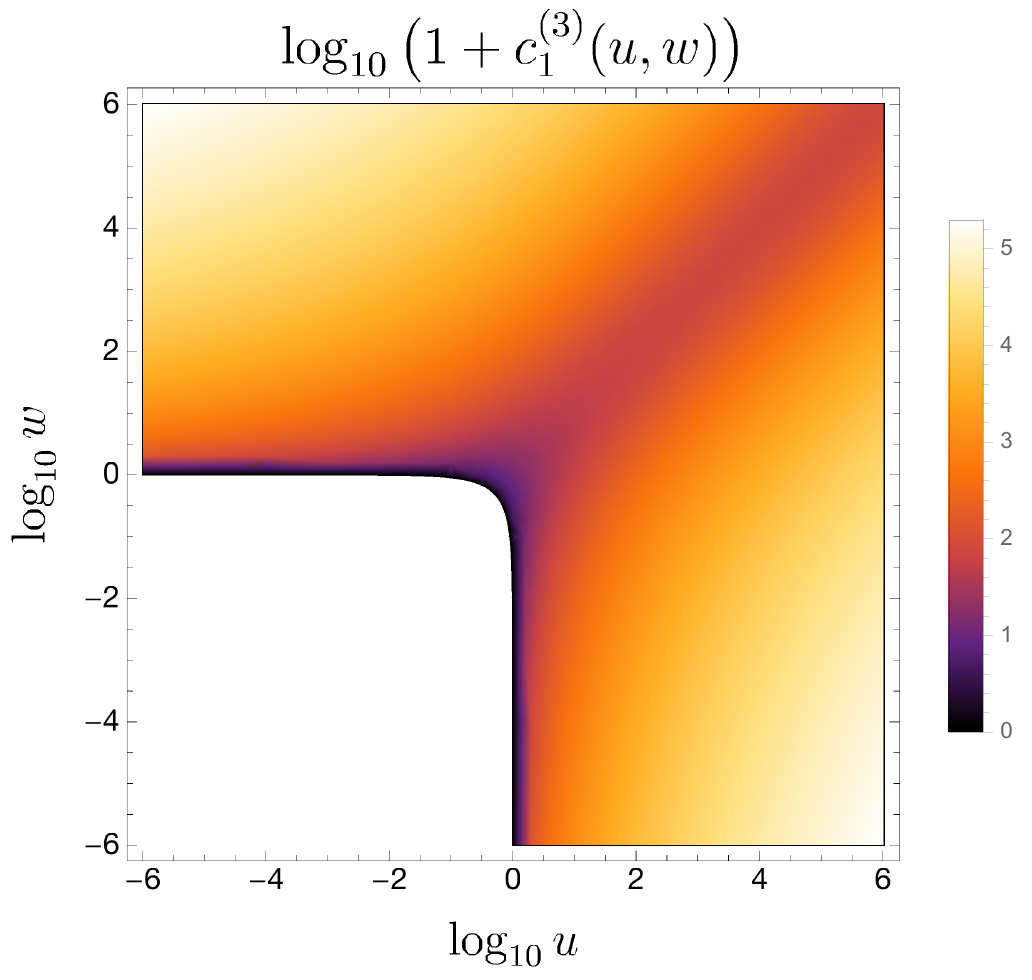}
        \end{subfigure}
        \begin{subfigure}[b]{0.47\textwidth}
            \includegraphics[width=\textwidth]{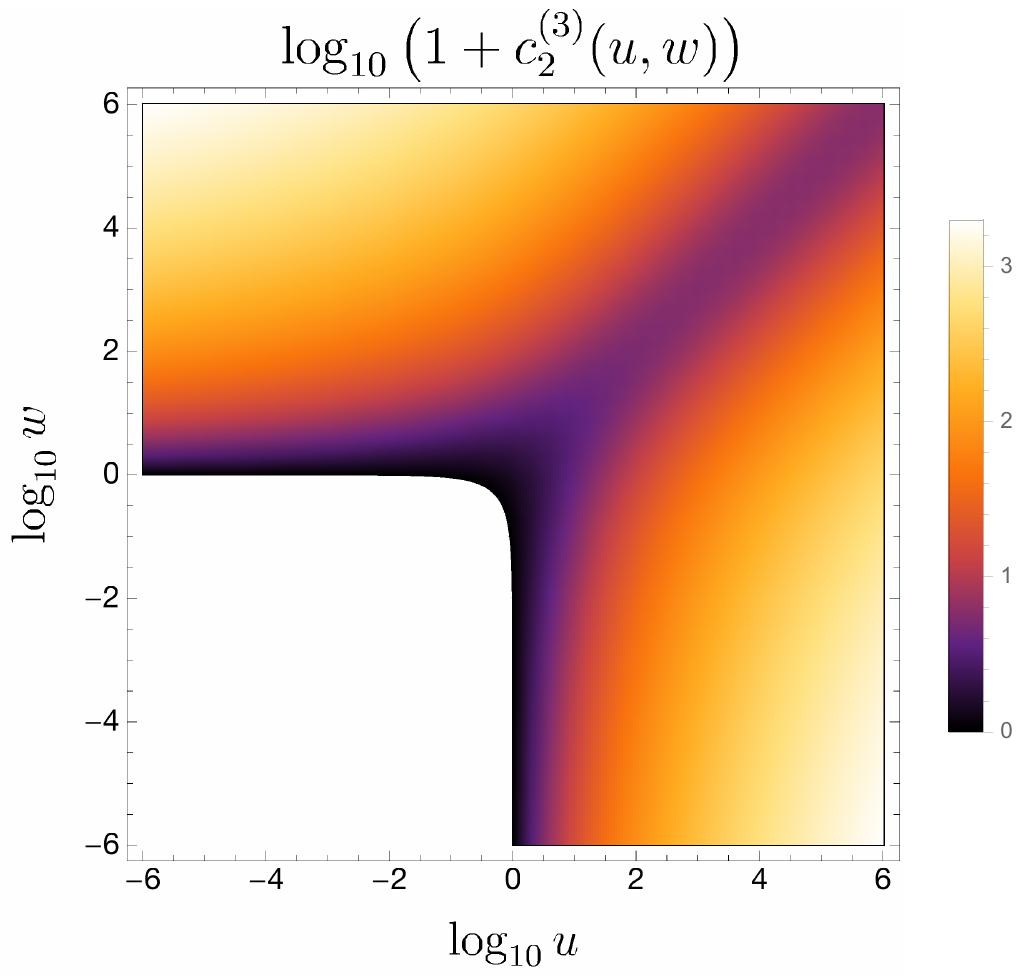}
        \end{subfigure}
        \caption[]
        {The three-loop coefficient functions $c_n^{(3)}(u,w)$ in the double-scaling limit, shifted to make it possible to plot them on a log scale. By plotting these functions against $\log u$ and $\log w$ we deform the $u+w=1$ line to the concave boundary seen in each plot.}
        \label{ds_surface_three_loops}
\end{figure*}

All the numerical checks we have performed on the double-scaling surface support both positivity and monotonic radial growth for every function $c_n^{(\ell)}(u,w)$.  We plot the functions, rather than their radial derivatives, in order to make interpretation of the magnitudes appearing in these plots more clear. In particular, we provide two sequences of plots that illustrate the trends the functions $c_n^{(\ell)}(u,w)$ exhibit as $n$ and $\ell$ are varied. The first sequence, in figure~\ref{ds_surface_three_loops}, shows how the three-loop result $c_n^{(3)}(u,w)$ changes as we move from the coefficient of the next-to-next-to-leading log in $1/v$ ($n=0$) to the leading log in $1/v$ ($n=2$) in the expansion~(\ref{C_expansion}). The plots all display the $u\lr w$ symmetry of $C(u,v,w)$, which is manifest from its definition~(\ref{DS_comb}) and the (anti)symmetry properties of $V$ and $\tilde{V}$, \eqn{uwsym}.  More interestingly, the coefficient of the leading log term grows the most slowly in the radial direction at a given loop order, particularly near the line of symmetry, $u=w$, where it asymptotes to a constant.  This result holds at least through four loops. (The five loop expressions proved too computationally taxing to explore exhaustively.) 

\begin{figure*}[t]    
\centering 
        \begin{subfigure}[b]{0.47\textwidth}
            \includegraphics[width=\textwidth]{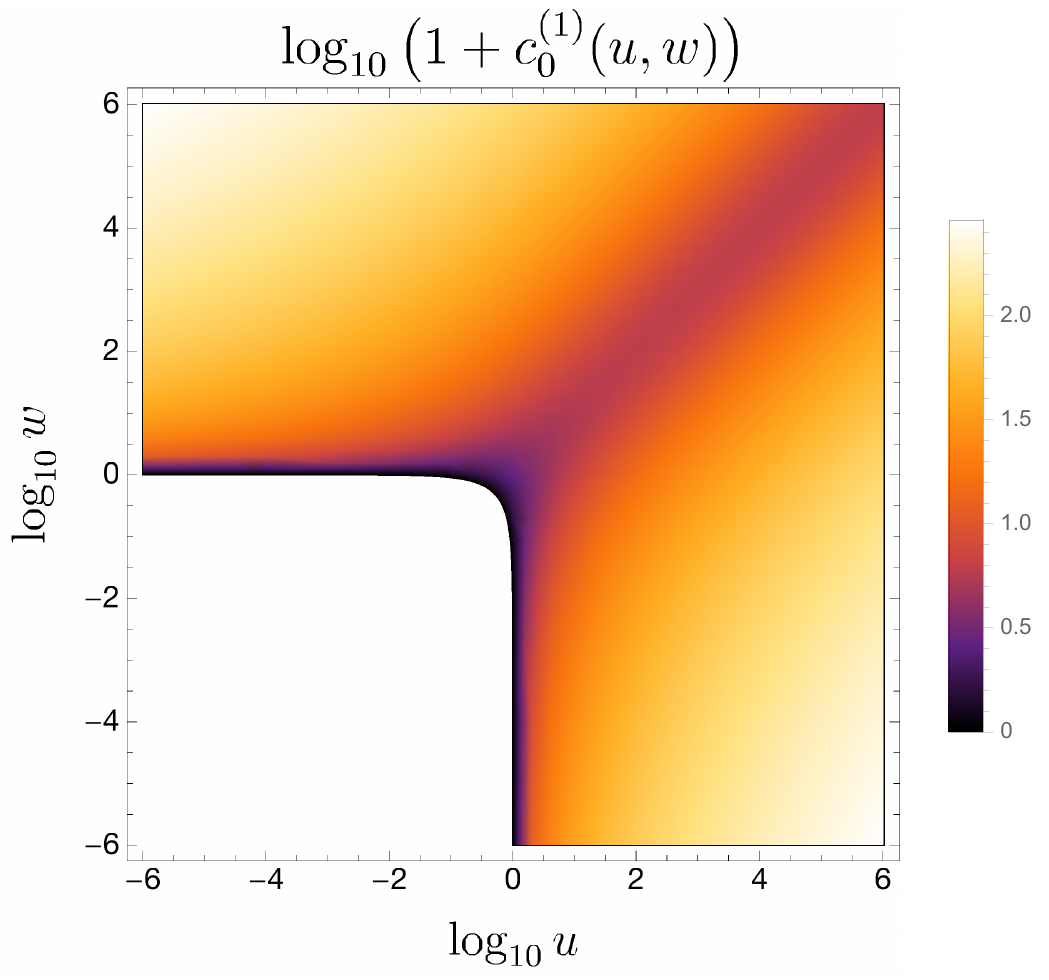}
        \end{subfigure}
        \begin{subfigure}[b]{0.47\textwidth}  
            \includegraphics[width=\textwidth]{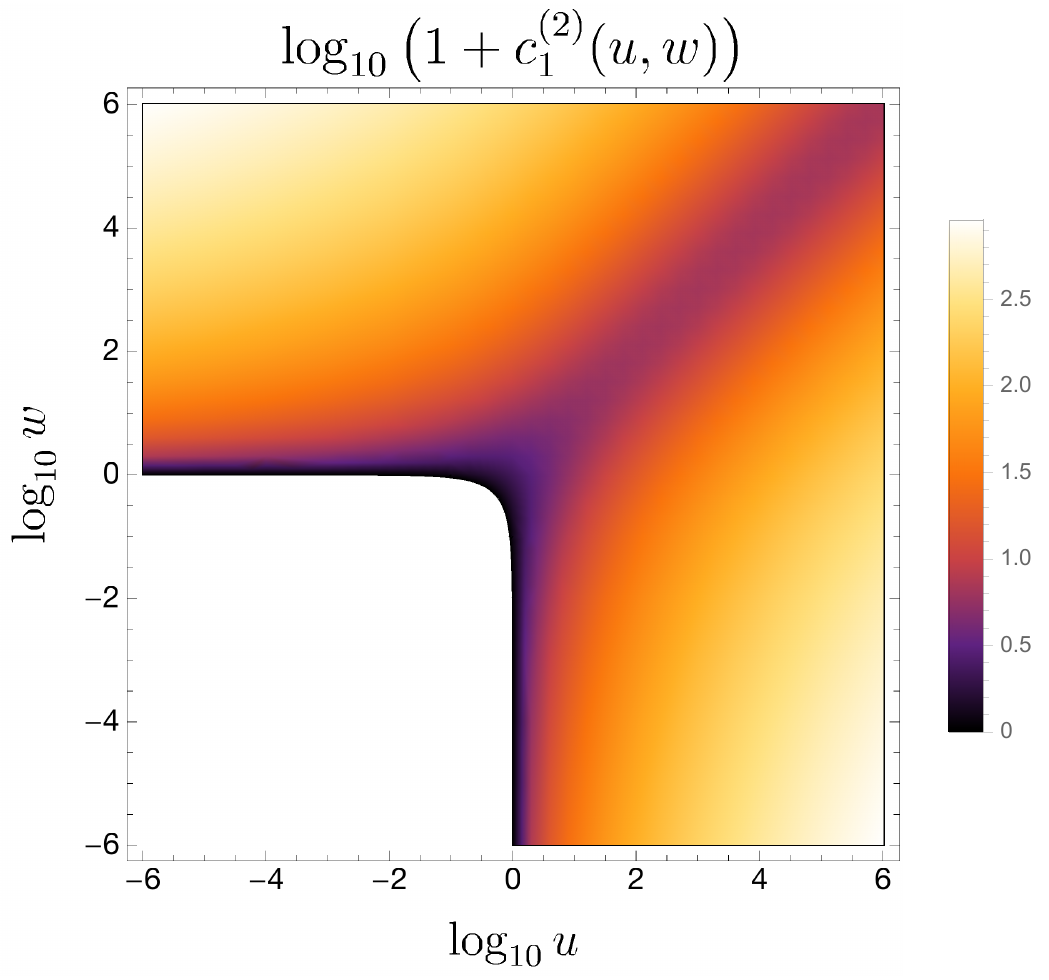}
        \end{subfigure}
    \vskip\baselineskip \vspace*{-.3cm}
        \begin{subfigure}[b]{0.47\textwidth}
            \includegraphics[width=\textwidth]{C3DSln2v.pdf}
        \end{subfigure}
        \begin{subfigure}[b]{0.47\textwidth}
            \includegraphics[width=\textwidth]{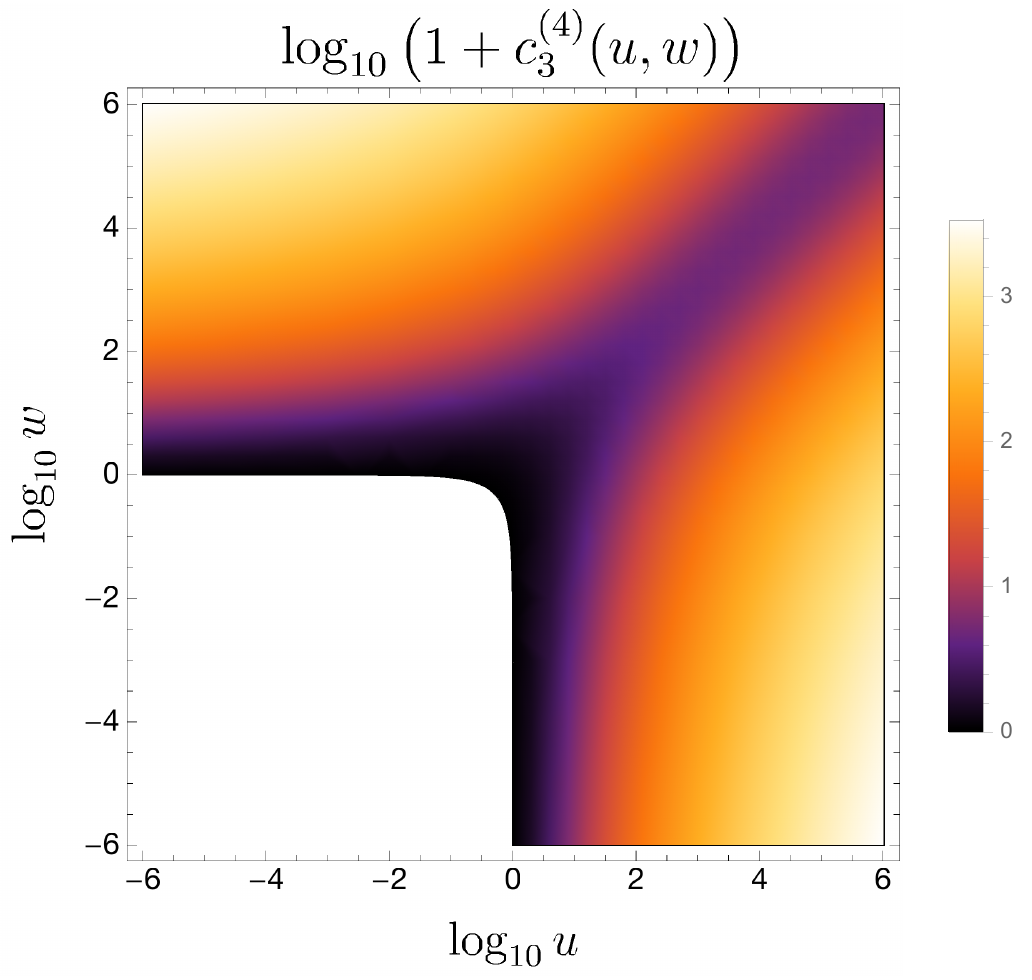}
        \end{subfigure}
        \caption[]
        {The leading-log coefficient functions $c_{\ell-1}^{(\ell)}(u,w)$ in the double-scaling limit from one to four loops, shifted to make it possible to plot them on a log scale. By plotting these functions against $\log u$ and $\log w$ we deform the $u+w=1$ line to the concave boundary seen in each plot.}
        \label{ds_surface_leading_log}
\end{figure*}

In figure \ref{ds_surface_leading_log} we plot the slowest-growing, leading-log coefficient functions $c_{\ell-1}^{(\ell)}(u,w)$ from one to four loops.  As the loop order increases, the functions experience slower radial growth. Moreover, the functions $c_n^{(\ell)}(u,w)$ interpolate smoothly between the lines $u=w$ and $w=0$, implying that most of the interesting information about these functions in present on these two lines.  In particular, the functions always grow the most slowly along the line $u=w$.

\section{Bulk positivity at higher loops}
\label{Bulk}

The previous sections verified the positivity of the ratio function in various limits, nearly all of which were on the boundary of the positive octant, i.e.~the double-scaling limit.  In this section, we check the positivity of the ratio function in the bulk, where all three cross ratios are bounded away from zero.
Except for the point $(u,v,w)=(1,1,1)$, the topic of the next subsection, 
our investigations will be numerical.  After a brief review of our procedure for numerically evaluating hexagon functions, we outline the checks performed. Positivity appears to continue to hold in the bulk through at least four loops, after which it gets too computationally taxing to check.

\subsection{The point \texorpdfstring{$(u,v,w)=(1,1,1)$}{(u,v,w)=(1,1,1)}}
\label{HigherLoop111Subsection}

The parity-odd functions $\tilde{V}^{(\ell)}$ all vanish at the point $(1,1,1)$,
because they are odd about the surface $\Delta(u,v,w)=0$, which includes this
point.  Thus we can repeat the analysis from Example 1 in
section~\ref{SimpleExampleSubsection}, obtaining
\begin{equation}
{\cal P}_{6,1}^{\ell-{\rm loop}} \xrightarrow[u=v=w]{}
\ {\cal P}_{6,1}^{\rm tree} \times V^{(\ell)}(1,1,1).
\label{lloop111}
\end{equation}

So all we need to do is check that the sign of $V^{(\ell)}(1,1,1)$
alternates with loop order $\ell$.
The value of the functions $V^{(\ell)}(1,1,1)$ were supplied
through four loops in ref.~\cite{Dixon2015iva}, and we have extracted
the five-loop value from ref.~\cite{CaronHuot2016owq}:
\bea
V^{(1)}(1,1,1) &=& - \zeta_2 \,, \nonumber\\
V^{(2)}(1,1,1) &=& 9 \, \zeta_4 \,, \nonumber\\
V^{(3)}(1,1,1) &=& - \frac{243}{4} \, \zeta_6 \,, \nonumber\\
V^{(4)}(1,1,1) &=& \frac{5051}{12} \, \zeta_8 + 3 \, \zeta_2 \, (\zeta_3)^2 
- 15 \, \zeta_3 \, \zeta_5 - 3 \, \zeta_{5,3} \,, \nonumber\\
V^{(5)}(1,1,1) &=& - \frac{244257}{80} \, \zeta_{10}
- \frac{93}{2} \, \zeta_4 \, (\zeta_3)^2 - 21 \, \zeta_2 \, \zeta_3 \, \zeta_5
+ \frac{399}{2} \, \zeta_3 \, \zeta_7 + \frac{777}{8} \, (\zeta_5)^2
\nonumber\\
&&\hskip0cm\null
+ \frac{9}{2} \, \zeta_2 \, \zeta_{5,3} + \frac{57}{4} \, \zeta_{7,3}  \,.
\label{V1V2V3V4V5_111}
\eea
The desired sign alternation is manifest from \eqn{V1V2V3V4V5_111}
through three loops; after that it relies on the numerical values
of the multiple zeta values:
\bea
V^{(1)}(1,1,1) &=& -1.64493406684\ldots, \nonumber\\
V^{(2)}(1,1,1) &=& +9.74090910340\ldots, \nonumber\\
V^{(3)}(1,1,1) &=& -61.8035910155\ldots, \nonumber\\
V^{(4)}(1,1,1) &=& +410.9535753669\ldots, \nonumber\\
V^{(5)}(1,1,1) &=& -2825.3845732862\ldots.
\label{V1V2V3V4V5_111_num}
\eea
We remark that the numerical result for $V^{(\ell)}(1,1,1)$
is dominated by the $\zeta_{2\ell}$ term through five loops (it gives
the correct value to within 10\%).

\subsection{Method for obtaining bulk numerics and positivity tests}
\label{BulkNumericsSubsection}

Next we turn to numerical evaluation of the ratio function at
random points in the bulk of the NMHV positive region.
To evaluate the ratio function numerically at higher loops,
we followed the procedure pioneered in ref.~\cite{Dixon2013eka}. 

Representing the ratio function in terms of multiple polylogarithms allows us to evaluate them using powerful existing code like {\sc GiNaC}~\cite{Bauer2000cp, Vollinga2004sn}. In order to do this, we choose a representation in which the multiple polylogarithms have convergent series expansions. We also prefer our representations to be manifestly real to reduce the potential for numerical error.

These conditions lead to two conditions on our multiple polylogarithms. For a multiple polylogarithm $G(w_1,\ldots,w_n;z)$, we obtain a convergent series expansion when $|z| \le |w_i|$ for all nonzero $w_i$, and our result is manifestly real if $z$ and all $w_i$ are real and positive.

In order to avoid square roots and their attendant branch-cut ambiguities, we work in the variables $(y_u,y_v,y_w)$. Following ref.~\cite{Dixon2013eka}, we find four different multiple polylog representations, corresponding to four different kinematic regions. In particular, for MHV studies we use
\begin{equation}
\begin{split}
\mathcal{G}_I^L =& \biggl\{ G(\vec{w};y_u) | w_i \in (0,1) \biggr\} 
\cup \biggl\{ G(\vec{w};y_v) \Big| 
w_i \in\Bigl(0,1,\frac{1}{y_u} \Bigr)
\biggr\} \\ 
&\quad\quad\cup \biggl\{ G(\vec{w};y_w) \Big| w_i 
\in \Bigl(0,1,\frac{1}{y_u},\frac{1}{y_v},\frac{1}{y_u y_v} \Bigr) 
\biggr\}
\label{GL_I}
\end{split}
\end{equation}
which is manifestly convergent for points in Region I, the MHV positive
kinematic region defined by
\begin{equation}
\textrm{Region I}:\quad
\left\{
\begin{array}{l}
\Delta > 0\,,\quad 0<u_i<1\,, \quad~\textrm{and}~\quad u+v+w<1,\\
0< y_i < 1 \,.
\end{array}
\right.
\label{RegionIDef}
\end{equation}

For studying the ratio function in NMHV positive kinematics, we use
\begin{equation}
\begin{split}
\mathcal{G}_{II}^L =& \biggl\{ G\Bigl(\vec{w};\frac{1}{y_u}\Bigr) \Big| 
w_i \in (0,1) \biggr\} 
\cup \biggl\{ G\Bigl(\vec{w};\frac{1}{y_v}\Bigr) \Big| 
w_i \in(0,1,y_u)\biggr\} \\
&\quad\quad
\cup \biggl\{ G(\vec{w};y_w) \Big| w_i 
\in \Bigl(0,1,\frac{1}{y_u},\frac{1}{y_v},\frac{1}{y_u y_v} \Bigr) 
\biggr\}
\label{GL_II}
\end{split}
\end{equation}
for points in Region II:
\begin{equation}
\textrm{Region II}:\quad
\left\{
\begin{array}{l}
\Delta > 0\,,\quad 0<u_i<1\,, \quad~\textrm{and}~\quad u+v-w>1,\\
0< y_w < \frac{1}{y_u y_v} < \frac{1}{y_u} , \frac{1}{y_v} < 1\,.
\end{array}
\right.
\label{RegionIIDef}
\end{equation}
Cycling the $y_i$ in Region II lets us define two other regions, Region III and Region IV, where we have multiple polylog representations in the bulk. 
Because the bosonized ratio function is $S_3$ symmetric, Regions III and IV do not add any new information.  The NMHV positive region always has $\Delta>0$ (see \eqn{NMHVposconstraints}). However, Region II lies entirely within the unit cube in $(u,v,w)$, and the bulk NMHV positive region extends well beyond it (as is clear from the double-scaling plots in the previous section).  So our bulk positivity tests will be confined to points inside the unit cube.

In order to perform this test, we randomly generate a phase-space point in the NMHV positive region by picking eleven random values of the positive parameters $(c_b,x_a)$, each between 0 and 100 ($x_6$ is set to 1, as discussed in section~\ref{NMHVPosKinem}).  For each set of values we use eqs.~(\ref{pos}) and (\ref{ratioY}) to compute the three cross ratios $u,v,w$.  If the point $(u,v,w)$ is not inside the unit cube, we stop and generate a new point.  If it is inside the unit cube, we use eqs.~(\ref{R_inv_parametrization}) and (\ref{ratio2Y}) to compute the $R$-invariants and extended cross ratios $y_u,y_v,y_w$.  We plug the latter into the arguments of the multiple polylogarithms in our Region II (or III or IV) representation of the ratio function, performing the numerical evaluation with {\sc GiNaC}.  We examined 585 points at loop orders from one through four, and the ratio function always has the expected sign, alternating with loop order.


\section{MHV positivity}
\label{MHVpos}

Having found strong evidence that the NMHV ratio function is positive through
five loops in the NMHV positive region, we now return to studying various
IR-finite versions of the MHV amplitude in the MHV positive region.

\subsection{The remainder function fails}
\label{RemainderFunctionFailsSubsection}

As mentioned in section~\ref{MHVposkinSubsection}, there are a variety
of possibilities.  They are all fairly simply related to each other analytically, but they still can have different positivity properties.  First we consider
the six-point remainder function $R_6$, which is defined as the logarithm of the MHV amplitude divided by the BDS ansatz, as in \eqn{IRdivMHV},
\be
\exp[R_6]\ =\ \frac{{\cal M}_{6,0}}{{\cal M}_{6,0}^{\rm BDS}} \,.
\label{R6Def}
\ee
The remainder function vanishes at one loop by construction.
Its positivity in the MHV
positive region~(\ref{MHVposregion}) was investigated at two
loops~\cite{ArkaniHamed2014dca}, three loops~\cite{Dixon2013eka},
and four loops~\cite{Dixon2014voa}.  All points investigated
numerically were found to have the correct sign. 

However, it turns out that there are regions
close to the origin in $(u,v,w)$ that have the wrong sign for $R_6^{(4)}$.
To exhibit such points, we consider the same line $v=0$, $w=0$ on
which the ratio function was studied for $u>1$
in section~\ref{NMHVweq0SubSubsection},
but now we take $0<u<1$ in order to be in the MHV positive region.
As was true for the ratio function, the remainder function develops
logarithmic singularities in both $v$ and $w$ as they approach zero,
\be
R_6(u,v\to0,w\to0) = \sum_{\ell=2}^\infty \sum_{n,k=0}^{\ell-1}
(-a)^\ell \, r_{n,k}^{(\ell)}(u) \, \log^n(1/v) \log^k(1/w),
\label{R6u00expansion}
\ee
up to power-suppressed terms in $v$ and $w$.  Since $R_6$ is $S_3$ permutation
symmetric, $r_{k,n}(u) = r_{n,k}(u)$.  Also, the coefficient functions vanish
unless $n + k \leq \ell$.

At two and three loops, there are no problems in this region.
The independent nonzero coefficient functions are given by,
\bea
r_{1,1}^{(2)} &=& \frac{1}{4} H_{0,1} \,, \nonumber\\
r_{1,0}^{(2)} &=& \frac{1}{4} \Bigl[ 2 H_{0,0,1} + H_{1,0,1} \Bigr] \,, \nonumber\\
r_{0,0}^{(2)} &=& \frac{1}{4} \Bigl[ 6 H_{0,0,0,1} + 3 H_{0,1,0,1}
     + 4 H_{1,0,0,1} + 2 H_{1,1,0,1} -  2 \zeta_2 (H_{0,1} + H_{1,1}) \Bigr]  \,,
\label{R62_u00}
\eea
and
\bea
r_{2,1}^{(3)} &=& \frac{1}{16} \Bigl[ H_{0,0,1}  -  H_{0,1,1} \Bigr] \,, \nonumber\\
r_{2,0}^{(3)} &=& \frac{1}{16} \Bigl[ 3 H_{0,0,0,1} - 2 H_{0,0,1,1}
 + H_{0,1,0,1} + H_{1,0,0,1} - H_{1,0,1,1} \Bigr] \,, \nonumber\\
r_{1,1}^{(3)} &=& \frac{1}{4}   \Bigl[ 3 H_{0,0,0,1} - 2 H_{0,0,1,1}
 + H_{1,0,0,1} - H_{1,0,1,1} + 2 \zeta_2 H_{0,1} \Bigr] \,, \nonumber\\
r_{1,0}^{(3)} &=& \frac{1}{8}  \Bigl[ 18 H_{0,0,0,0,1} - 9 H_{0,0,0,1,1}
 + 3 H_{0,0,1,0,1} + 7 H_{0,1,0,0,1} - 4 H_{0,1,0,1,1}
 + H_{0,1,1,0,1} \nonumber\\
&&\hskip0.3cm\null
 + 9 H_{1,0,0,0,1} - 6 H_{1,0,0,1,1}
 + H_{1,0,1,0,1} + 3 H_{1,1,0,0,1} - 3 H_{1,1,0,1,1} \nonumber\\
&&\hskip0.3cm\null
 + \zeta_2 (5 H_{0,0,1} - H_{0,1,1} + 2 H_{1,0,1}) \Bigr] \,, \nonumber\\
r_{0,0}^{(3)} &=& \frac{1}{4}   \Bigl[ 30 H_{0,0,0,0,0,1}
 - 12 H_{0,0,0,0,1,1} + 6 H_{0,0,0,1,0,1}
 + 12 H_{0,0,1,0,0,1} - 5 H_{0,0,1,0,1,1} \nonumber\\
&&\hskip0.3cm\null
 + 2 H_{0,0,1,1,0,1} + 15 H_{0,1,0,0,0,1} - 8 H_{0,1,0,0,1,1} + 2 H_{0,1,0,1,0,1}
 + 5 H_{0,1,1,0,0,1}  \nonumber\\
&&\hskip0.3cm\null
 - 4 H_{0,1,1,0,1,1} + 18 H_{1,0,0,0,0,1} - 9 H_{1,0,0,0,1,1} + 3 H_{1,0,0,1,0,1}
 + 7 H_{1,0,1,0,0,1}  \nonumber\\
&&\hskip0.3cm\null
 - 4 H_{1,0,1,0,1,1} + H_{1,0,1,1,0,1}
 + 9 H_{1,1,0,0,0,1} - 6 H_{1,1,0,0,1,1} + H_{1,1,0,1,0,1} \nonumber\\
&&\hskip0.3cm\null
 + 3 H_{1,1,1,0,0,1} - 3 H_{1,1,1,0,1,1} \nonumber\\
&&\hskip0.3cm\null
 + \zeta_2 (3 H_{0,0,0,1} - 2 H_{0,0,1,1} + H_{0,1,0,1}
      + H_{1,0,0,1} - H_{1,0,1,1}) \nonumber\\
&&\hskip0.3cm\null
 -  2 \zeta_3 (H_{0,0,1}+H_{0,1,1})
 - 11 \zeta_4 (H_{0,1}+H_{1,1}) \Bigr] \,,
\label{R63_u00}
\eea
where the suppressed HPL argument is $1-u$.  It can be checked that
they are all positive for $0<u<1$.  

The problem starts at four loops with the leading log coefficients,
\bea
r_{3,1}^{(4)}(u) &=& \frac{1}{96} \Bigl[
 H_{0,0,0,1} - 2 H_{0,0,1,1} - 2 H_{0,1,0,1} + H_{0,1,1,1} \Bigr] \,, \nonumber\\
r_{2,2}^{(4)}(u) &=& \frac{1}{32} \Bigl[
 H_{0,0,0,1} - 5 H_{0,0,1,1} - H_{0,1,0,1} + H_{0,1,1,1} \Bigr] \,,
\label{R64_u00_ll}
\eea
which turn negative for $u < 0.15$ and $u < 0.2$, respectively,
and stay negative as $u\to0$. The leading terms in their expansions around
$u=0$ are clearly negative:
\bea
r_{3,1}^{(4)}(u) &\sim& - \frac{u}{96} \Bigl[
\frac{1}{6} \log^3(1/u) + \frac{1}{2} \log^2(1/u)
- ( 2 \zeta_2 - 1 ) \log(1/u) + 3 \zeta_3 - 2 \zeta_2 + 1 \Bigr] \,,
\nonumber\\
r_{2,2}^{(4)}(u) &\sim& - \frac{u}{32} \Bigl[
\frac{1}{6} \log^3(1/u) + \frac{1}{2} \log^2(1/u)
- ( \zeta_2 - 1 ) \log(1/u) - 2 \zeta_3 - \zeta_2 + 1 \Bigr] \,,
\label{R64_u00_ll_smallu}
\eea
Thus $R_6^{(4)}(u,v,w)$ is negative for very small $v$ and $w$ and $u < 0.14$.

\subsection{Logarithmic fixes fail}
\label{LogFixesFailSubsection}

One might first try to fix the problem with $R_6^{(4)}$ at the logarithmic level.
Consider the logarithm of the BDS-like normalized amplitude,
\be
{\cal E}\ =\ \frac{{\cal M}_{6,0}}{{\cal M}_{6,0}^{\rm BDS-like}}
 = \exp\biggl[ R_6 - \frac{\gamma_K}{8} Y \biggr] \,,
\label{EMHVDef}
\ee
where $\gamma_K$ is the cusp anomalous dimension and 
\be
Y(u,v,w)\ =\ {\rm Li}_2(1-u) + {\rm Li}_2(1-v) + {\rm Li}_2(1-w)
+ \frac{1}{2} \Bigl( \log^2 u + \log^2 v + \log^2 w \Bigr),
\label{Ydef}
\ee
so that
\be
\log{\cal E}(u,v,w) = R_6(u,v,w) - \frac{\gamma_K}{8} Y(u,v,w).
\label{logE}
\ee
This attempt immediately runs into trouble, because the limiting
behavior of $Y$,
\be
Y(u,v\to0,w\to0) \sim \frac{1}{2} \log^2 v + \frac{1}{2} \log^2 w 
+ \frac{1}{2} \log^2 u + {\rm Li}_2(1-u) + 2 \zeta_2 \,,
\label{Y_u0}
\ee
like that of any one-loop function,
does not have enough logarithms of $v$ or $w$ to compete with the four
powers of logs in the problematic terms in $R_6^{(4)}$.

One can also consider the logarithm of the hexagonal Wilson loop
framed by two pentagons and a box~\cite{Alday2010ku,Basso2013vsa},
\be
W_{\rm ratio} = \frac{\la W_{\rm hex} \ra \la W_{\rm box} \ra}
{\la W_{\rm pent} \ra \la W_{\rm pent'} \ra}
= \exp\biggl[ R_6 + \frac{\gamma_K}{8} X \biggr] \,,
\label{framedWdef}
\ee
where
\bea
X(u,v,w) &=& - {\rm Li}_2(1-u) - {\rm Li}_2(1-v) - {\rm Li}_2(1-w)
\nonumber\\
&&\hskip0cm\null
- \log\biggl(\frac{uv}{w(1-v)}\biggr)\log (1-v)
- \log u \log w + 2 \zeta_2 \,.
\label{Xuvw}
\eea
Since $X$ is a one-loop function, it cannot produce enough logs
in the limit to compete with $R_6^{(4)}$, and thus $\log W_{\rm ratio}$
cannot be strictly positive either by four loops.

\subsection{Other fixes fail}
\label{OtherFixesFailSubsection}

Next we turn to functions that are defined at the level of the MHV
amplitude, rather than its logarithm.  First we consider the
BDS-normalized amplitude $\exp[R_6]$.  At one and two loops,
it is the same as $R_6$, while its four-loop coefficient receives
an extra positive contribution:
\be
\Bigl[ \exp[R_6] \Bigr]^{(4)} = R_6^{(4)}
 + \frac{1}{2} \Bigl[ R_6^{(2)} \Bigr]^2 \,.
\label{expR64}
\ee
Taking into account \eqn{R62_u00}, the leading-log $[r_{1,1}^{(2)}]^2$ part
of $[ R_6^{(2)} ]^2$ can and does flip the sign of the $\log^2(1/v) \log^2(1/w)$
coefficient function to positive.  But it clearly leaves the 
$\log^3(1/v) \log(1/w)$ term unaltered.
So the addition of $[ R_6^{(2)} ]^2$
cannot cancel the negative behavior of $R_6^{(4)}$ for kinematics with
$0 < v \ll w \ll u < 0.14$, for which
$\log^3(1/v) \log(1/w) \gg \log^2(1/v) \log^2(1/w)$.

Can the negative behavior be fixed by the framed Wilson
loop $W_{\rm ratio}$ defined in \eqn{framedWdef}?
Now $X$ is not $S_3$ symmetric, and the three cyclically-related
line segments all belong to the MHV positive regions:
$v,w\to0$, $0<u<1$; $w,u\to0$, $0<v<1$; $u,v\to0$, $0<w<1$.
We need to ensure positivity along all three lines and for both
orderings of the two vanishing cross ratios.
Equivalently, since $R_6$ is $S_3$ symmetric, we should consider the
$v,w\to0$, $0<u<1$ limits of all six permutations of $X$.
The original orientation $X(u,v,w)$ already reveals a problem:
\be
X(u,v\to0,w\to0) \sim - \log(1/w) \log(1/u) - {\rm Li}_2(1-u) \,.
\label{Xuvw_u0}
\ee
Because there are no $\log(1/v)$'s in this expression, powers
of $X$ cannot fix the sign problem that $\exp[R_6]$ still has
in the region $0 < v \ll w \ll u < 0.14$.

\subsection{BDS-like normalized amplitude works}
\label{BDSlikeWorksSubsection}

Finally, we consider the BDS-like normalized amplitude itself,
${\cal E}(u,v,w)$ defined in \eqn{EMHVDef}.  Since the limiting
behavior of $Y$ in \eqn{Y_u0} contains both $\log^2(1/v)$ and $\log^2(1/w)$,
it can potentially fix the negative behavior. 
Indeed it does fix the problem through five loops,
at least for $v,w\to0$, $0<u<1$,
or (by symmetry) on cyclic permutations of this line segment.
It also leads to monotonically increasing behavior as $u$ decreases from 1.
The expansion on this line segment now contains many higher powers
of the singular logs, all the way up to $2\ell$,
\be
{\cal E}(u,v\to0,w\to0) = \sum_{\ell=0}^\infty \sum_{n,k=0}^{2\ell}
(-a)^\ell \, \tilde{e}_{n,k}^{(\ell)}(u) \, \log^n(1/v) \log^k(1/w),
\label{EMHVu00expansion}
\ee
up to power-suppressed terms.  Here $\tilde{e}_{k,n}^{(\ell)} = \tilde{e}_{n,k}^{(\ell)}$ and $n+k\leq2\ell$ for a nonzero coefficient.

As was the case for the NMHV ratio function on the continuation of
this line to $u>1$, discussed in section~\ref{NMHVweq0SubSubsection},
there is an HPL representation which {\it almost} makes manifest
the positivity and monotonicity.  In this case we use the argument 
$1-u$ rather than $1-1/u$, since the argument $1-u$ runs from 0 to 1
as $u$ runs from the collinear point $u=1$ down to the origin.
Positivity is manifest from the signs in front of the HPLs
at one and two loops:
\be
\tilde{e}_{2,0}^{(1)} = \frac{1}{4} \,, \qquad
\tilde{e}_{1,1}^{(1)} = 0 \,, \qquad
\tilde{e}_{1,0}^{(1)} = 0 \,, \qquad
\tilde{e}_{0,0}^{(1)} = \frac{1}{2} \Bigl[ H_{0,1} + H_{1,1} + 2 \zeta_2 \Bigr]  \,,
\label{EMHV1_u00}
\ee
\bea
\tilde{e}_{4,0}^{(2)} &=& \frac{1}{32} \,, \qquad
\tilde{e}_{3,1}^{(2)} = 0 \,, \qquad
\tilde{e}_{2,2}^{(2)} = \frac{1}{16} \,, \qquad
\tilde{e}_{3,0}^{(2)} = 0 \,, \qquad
\tilde{e}_{2,1}^{(2)} = 0 \,, \nonumber\\
\tilde{e}_{2,0}^{(2)} &=& \frac{1}{8} \Bigl[ H_{0,1} + H_{1,1} + 4 \zeta_2 \Bigr] \,,
\qquad
\tilde{e}_{1,1}^{(2)} = \frac{1}{4} H_{0,1} \,, \qquad
\tilde{e}_{1,0}^{(2)} = \frac{1}{4} \Bigl[ 2 H_{0,0,1} + H_{1,0,1} \Bigr] \,, \nonumber\\
\tilde{e}_{0,0}^{(2)} &=& \frac{1}{4} \Bigl[ 6 H_{0,0,0,1} + 2 H_{0,0,1,1}
 + 4 H_{0,1,0,1} + 3 H_{0,1,1,1} + 4 H_{1,0,0,1} + 2 H_{1,0,1,1} \nonumber\\
&&\hskip0.3cm\null
 + 3 H_{1,1,0,1} + 3 H_{1,1,1,1} + 2 \zeta_2 (H_{0,1} + H_{1,1})
 + 15 \zeta_4 \Bigr] \,.
\label{EMHV2_u00}
\eea

At three loops the HPL representation no longer makes manifest
the positivity of all terms; for example,
\bea
\tilde{e}_{2,1}^{(3)} &=& \frac{1}{16} \Bigl[ 3 H_{0,0,1} + H_{1,0,1} - H_{0,1,1} \Bigr] \,,
\nonumber\\
\tilde{e}_{1,0}^{(3)} &=& \frac{1}{8} \Bigl[ 
   18 H_{0,0,0,0,1} + 3 H_{0,0,0,1,1} + 9 H_{0,0,1,0,1}
 + 6 H_{0,0,1,1,1} + 9 H_{0,1,0,0,1} + 2 H_{0,1,0,1,1}  \nonumber\\
&&\hskip0.3cm\null
 + 5 H_{0,1,1,0,1} + 9 H_{1,0,0,0,1} + 2 H_{1,0,0,1,1}
 + 5 H_{1,0,1,0,1} + 3 H_{1,0,1,1,1} + 5 H_{1,1,0,0,1}  \nonumber\\
&&\hskip0.3cm\null
 + H_{1,1,0,1,1} + 3 H_{1,1,1,0,1}
 + \zeta_2 ( 9 H_{0,0,1} + 4 H_{1,0,1} - H_{0,1,1} ) \Bigr] \,.
\label{EMHV3_u00_some}
\eea
In both of these cases, it is easy to see that the terms with a minus sign 
are overpowered by the previous term.  At higher-loop orders, positivity
and monotonicity of the coefficient functions
becomes tricky to prove analytically, but we have verified it numerically
for all $\tilde{e}_{n,k}^{(\ell)}$ coefficients through five loops.

What about positivity of ${\cal E}$ in other parts of the MHV positive region?
The double-scaling limit intersects this region in the triangle,
\begin{align}
u > 0, \quad w > 0,\quad u+w < 1. \label{mhv_ds_positive_region}
\end{align}
which is the complement of the NMHV double-scaling positive
region~(\ref{ds_positive_region}) in the positive quadrant.
The expansion of ${\cal E}$ in this limit is
\be
{\cal E}(u,v\to0,w) = \sum_{\ell=0}^\infty \sum_{n=0}^{2\ell}
(-a)^\ell \, e_{n}^{(\ell)}(u,w) \, \log^n(1/v).
\label{EMHVu0wexpansion}
\ee
The one-loop coefficient functions are,
\bea
e_2^{(1)}(u,w) &=& \frac{1}{4} \,, \nonumber\\
e_1^{(1)}(u,w) &=& 0 \,, \nonumber\\
e_0^{(1)}(u,w) &=& \frac{1}{4} \log^2(u/w) + \zeta_2 + \frac{1}{2} C^{(1)}(u,w).
\label{EMHV1_ds}
\eea
Now $C^{(1)}(u,w)$ is negative in the NMHV positive region,
but the same radial-derivative argument shows that it flips sign
around the collinear boundary, where it vanishes.  So $C^{(1)}(u,w)$ is
positive in the MHV positive region, and the representation~(\ref{EMHV1_ds})
makes manifest the desired sign (and monotonicity)
for ${\cal E}^{(1)}(u,v,w)$ in the double-scaling limit of the
MHV positive region.

Similarly at two loops we have,
\bea
e_4^{(2)}(u,w) &=& \frac{1}{32} \,, \nonumber\\
e_3^{(2)}(u,w) &=& 0 \,, \nonumber\\
e_2^{(2)}(u,w) &=& \frac{1}{4} \Bigl[ e_0^{(1)}(u,w) + \zeta_2 \Bigr]
\,, \nonumber\\
e_1^{(2)}(u,w) &=& - \frac{1}{2} c_1^{(2)}(u,w),
\label{EMHV2_ds}
\eea
where $e_0^{(1)}(u,w)$ was just argued to be positive.
The positivity of $c_1^{(2)}(u,w)$ is proved in the NMHV positive region
in appendix~\ref{c21monotonicity}.  But again the argument does not
rely on $u+w>1$ -- except for the overall sign, which flips when crossing
the collinear boundary dividing the MHV and NMHV positive regions.
Hence $c_1^{(2)}(u,w)$ is negative in the MHV positive region, implying that
$e_1^{(2)}(u,w)$ is positive. 

\begin{figure*}[t]    
\centering 
        \begin{subfigure}[b]{0.47\textwidth}  
            \includegraphics[width=\textwidth]{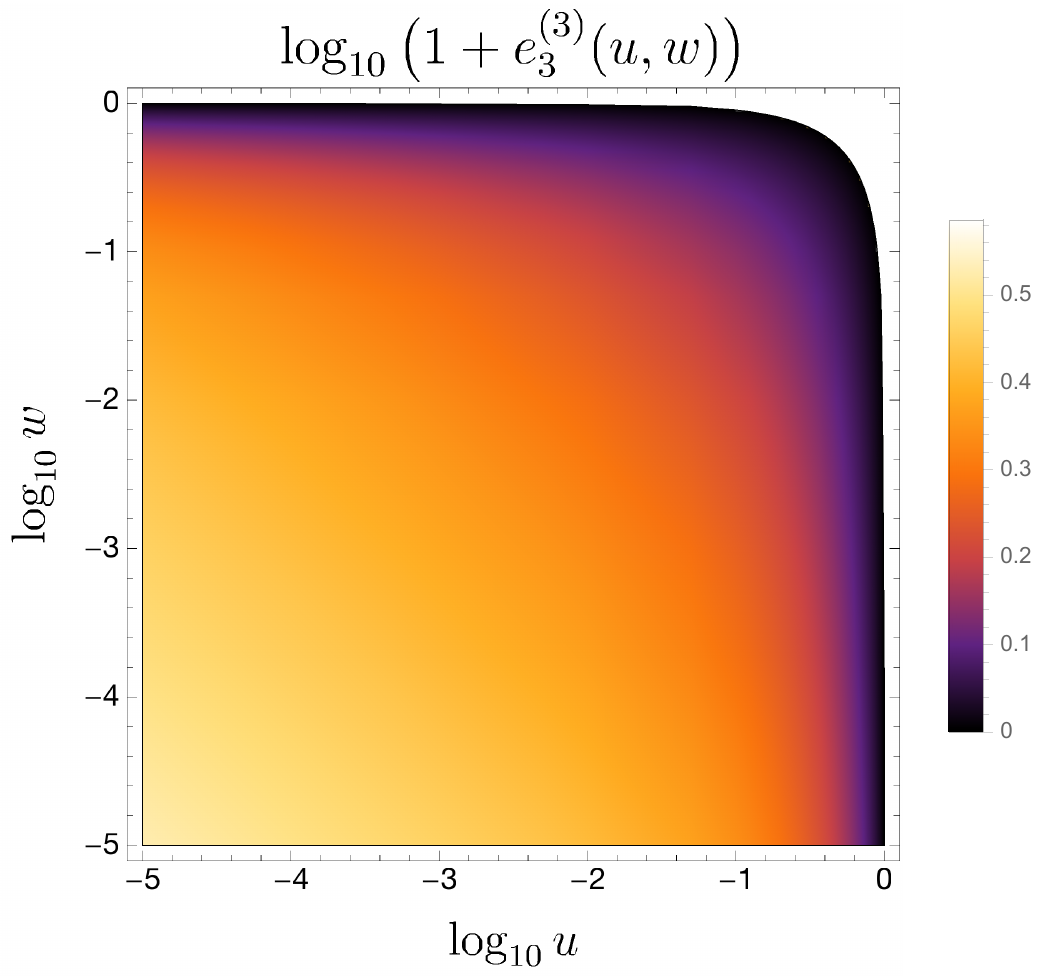}
        \end{subfigure}
        \begin{subfigure}[b]{0.47\textwidth}
            \includegraphics[width=\textwidth]{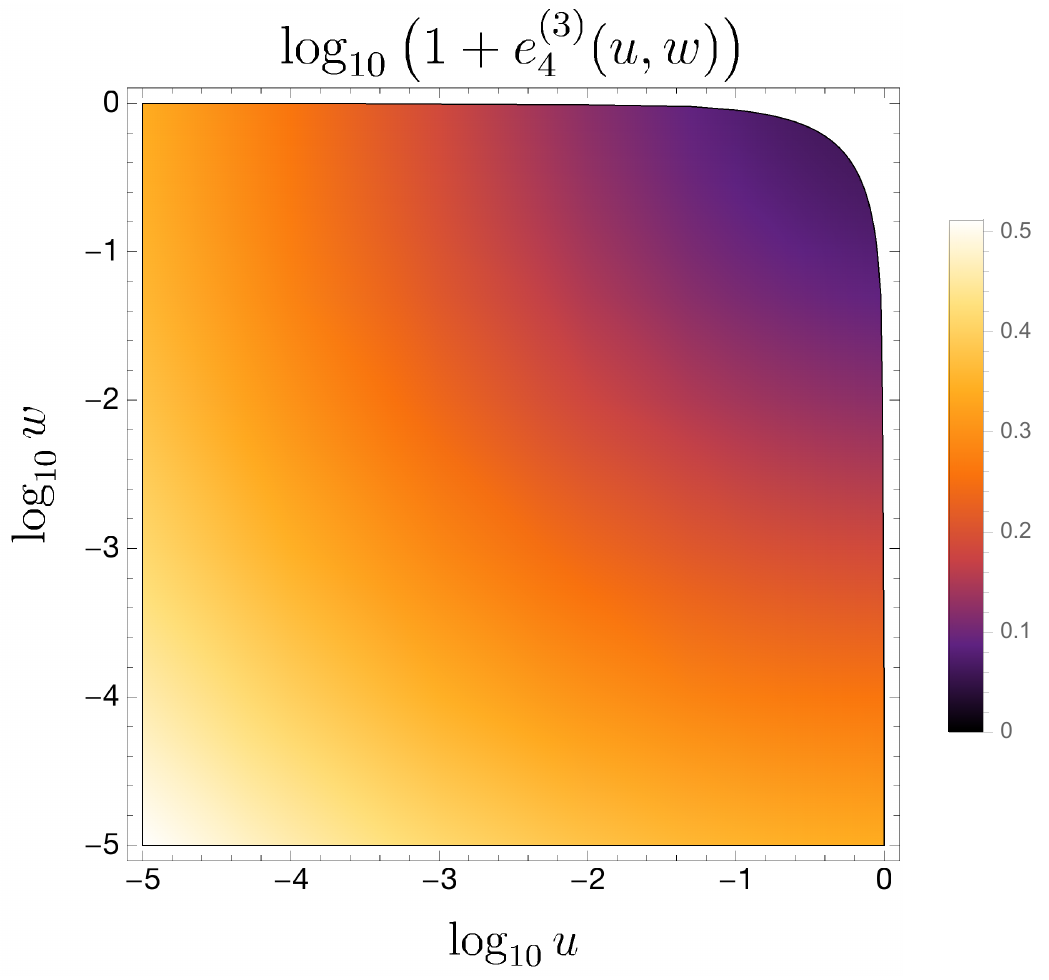}
        \end{subfigure}
        \caption[]
        {The three-loop coefficient functions $e_3^{(3)}(u,w)$ and $e_4^{(3)}(u,w)$ in the double-scaling limit, shifted to make it possible to plot them on a log scale. By plotting these functions against $\log u$ and $\log w$ we deform the $u+w=1$ line to the convex boundary seen in each plot.}
        \label{mhv_ds_surface_three_loops}
\end{figure*}

The positivity and monotonicity of the last two-loop coefficient, 
$e_0^{(2)}(u,w)$, is not as simple to prove, but has been confirmed numerically with {\sc GiNaC} using the basis of multiple polylogarithms
given in \eqn{DS_polylogs}. Similar numerical checks confirm the positivity 
and monotonicity of all the three loop coefficient functions $e_{n}^{(3)}(u,w)$; we plot the functions governing the leading-log and next-to-leading log behavior in figure~\ref{mhv_ds_surface_three_loops}. As can be seen in these plots, $\cal E$ is not generically required to vanish on the line $u+w=1$. However, the collinear vanishing of $R_6$ on this line is inherited by the coefficient functions $e_{n}^{(\ell)}(u,w)$ that multiply odd 
powers of logs. This is due to the fact that the function $Y$ that converts between $\cal E$ and $R_6$ in~\eqn{EMHVDef} can only provide even powers of logs, as can be seen from its definition in~\eqn{Ydef}. Correspondingly, $e_3^{(3)}(u,w)$ vanishes along the line $u+w=1$ while $e_4^{(3)}(u,w)$ does not. These plots also exhibit the $u\lr w$ symmetry that the functions $e_{n}^{(\ell)}(u,w)$ inherit from the total symmetry of $\cal E$. 

Finally, we examined the values for ${\cal E}^{(\ell)}(u,v,w)$ in
the bulk MHV positive region (Region I), from one to four loops, using
the representations for ${\cal E}^{(\ell)}$ in terms of
multiple polylogarithms referred to in section~\ref{BulkNumericsSubsection}.
After randomly generating 1608 points in this region, we found that ${\cal E}^{(\ell)}$ had the correct sign through four loops for every point examined.

\section{Conclusion}
\label{concl}

In this paper we have demonstrated that the positivity properties of the Amplituhedron persist after momentum integration, at least in some cases. In particular, the ratio function (the IR-finite ratio of the NMHV and MHV amplitudes) has uniform sign in the same region in which the Amplituhedron is positive. The MHV amplitude also has uniform sign provided that we normalize by a ``BDS-like'' ansatz. In both cases, it appears that the Minkowski contour of integration preserves positivity more completely than would have been expected.

While we have not provided a general proof, we do provide analytic evidence on a variety of lines, as well as numerical checks through the bulk of kinematic space, all of which support positivity. In doing so, we have observed that the ratio function and ${\cal E}$ both appear to be not just of uniform sign but, at least in the double-scaling limit, they are monotonic in a radial direction away from the collinear limit. This property appears to be quite robust, and falls in line with older observations of ratio function numerics, all of which suggest that the ratio function is significantly simpler than the complicated expressions used to represent it might imply.

In the future, it would be interesting to explore whether a more general proof of positivity can be devised.
It seems possible that one could find rules for which positive integrands result in positive amplitudes, and such rules would likely be useful in much broader contexts. This would likely involve finding some contour of integration that, unlike the usual Minkowski contour, manifestly preserves positivity. Understanding such a contour could also shed new light on the Amplituhedron, suggesting that there could be an Amplituhedron-like construction of finite quantities such as ratio functions or BDS-like normalized MHV amplitudes, both for the integrands and the final results.

\vfill\eject

\vskip0.5cm
\noindent {\large\bf Acknowledgments}
\vskip0.3cm

We are grateful to John Schwarz for helping to construct ${\cal N}=4$ super-Yang-Mills theory forty years ago and for his positive influence since then.  We thank Nima Arkani-Hamed, Jake Bourjaily, Simon Caron-Huot, Claude Duhr, Erik Panzer and Georgios Papathanasiou for illuminating discussions. This research was supported by the US Department of Energy under contract DE--AC02--76SF00515, and by Perimeter Institute for Theoretical Physics. Research at Perimeter Institute is supported by the Government of Canada through the Department of Innovation, Science and Economic Development Canada and by the Province of Ontario through the Ministry of Research, Innovation and Science.  LD is grateful to the Walter Burke Institute at Caltech and the Kavli Institute for Theoretical Physics (National Science Foundation grant NSF PHY11-25915) for hospitality.  AM thanks the Higgs Centre of the University of Edinburgh for its hospitality.

\vfill\eject


\appendix

\section{More results for the double-scaling line \texorpdfstring{$w=0$}{w=0}} \label{appendix_line_u0}

This appendix provides additional plots of the coefficient functions $\tilde{c}^{(1)}_{n,k}(u)$ describing the behavior of the ratio function on the $w\to0$ edge of the double-scaling limit, beyond the case $n=0$ already plotted in figure~\ref{line_u0_c0}.  Figure~\ref{ci(u,0)} gives the remaining cases $n=1,2,3,4$.  Again all coefficient functions are positive and monotonically increasing for the $u>1$ region of NMHV positive kinematics.

\begin{figure*}[ht]
\centering
        \begin{subfigure}[b]{0.24\textwidth}
            \includegraphics[width=\textwidth]{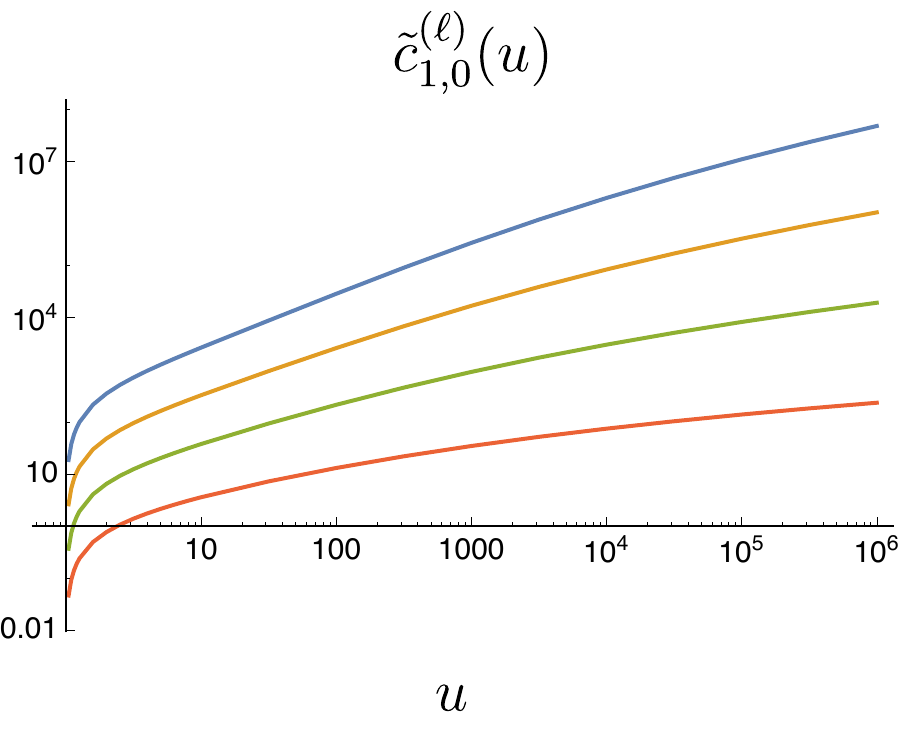}
        \end{subfigure}
        \begin{subfigure}[b]{0.24\textwidth}
            \includegraphics[width=\textwidth]{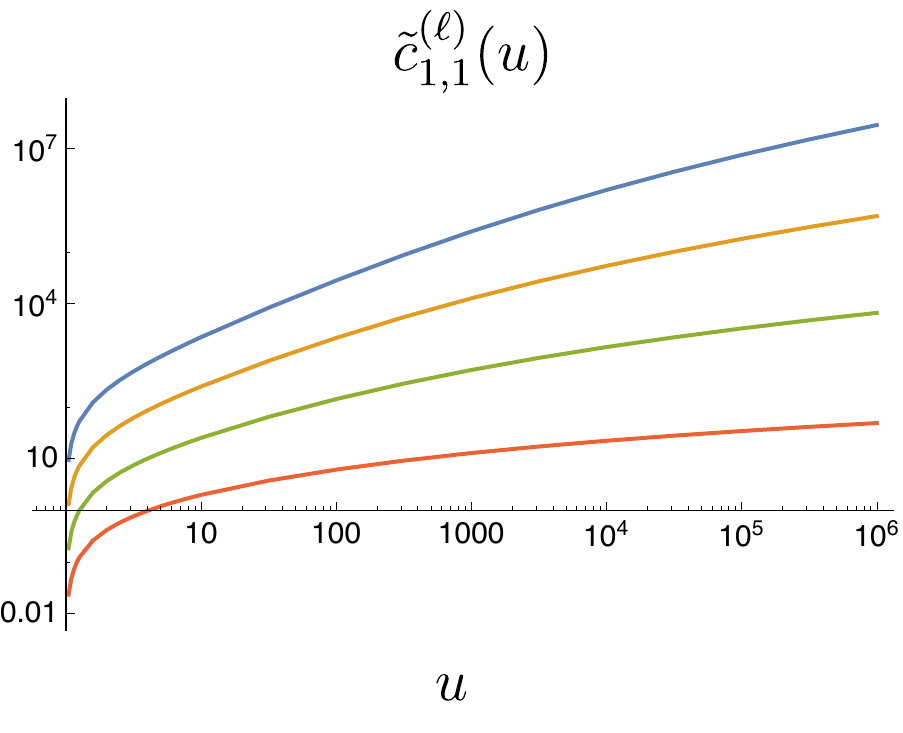}
        \end{subfigure}
        \begin{subfigure}[b]{0.24\textwidth}  
            \includegraphics[width=\textwidth]{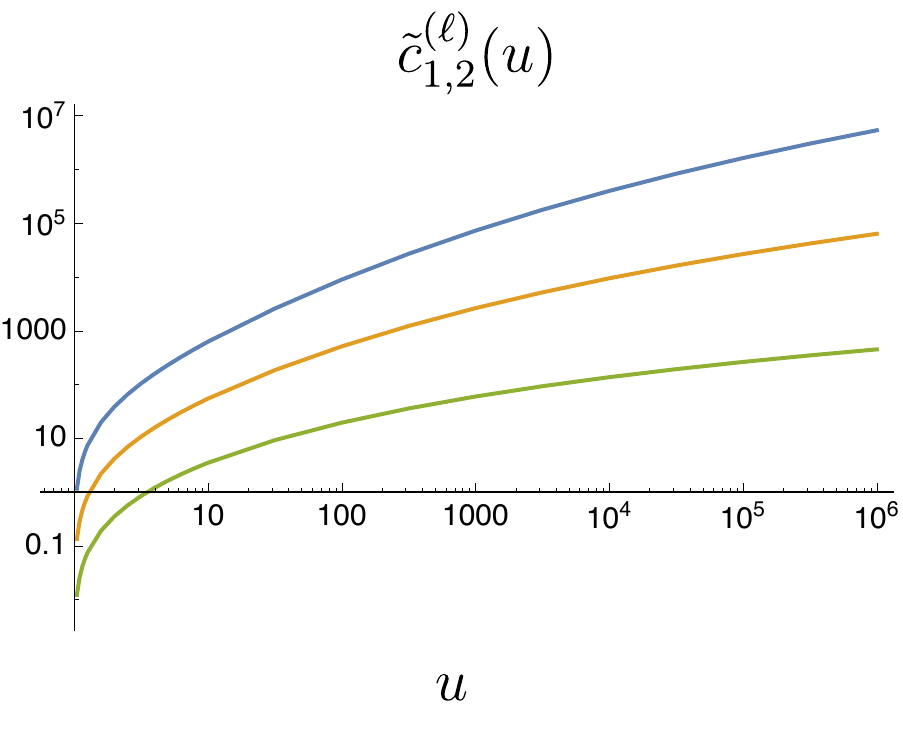}
        \end{subfigure}
        \begin{subfigure}[b]{0.24\textwidth}   
            \includegraphics[width=\textwidth]{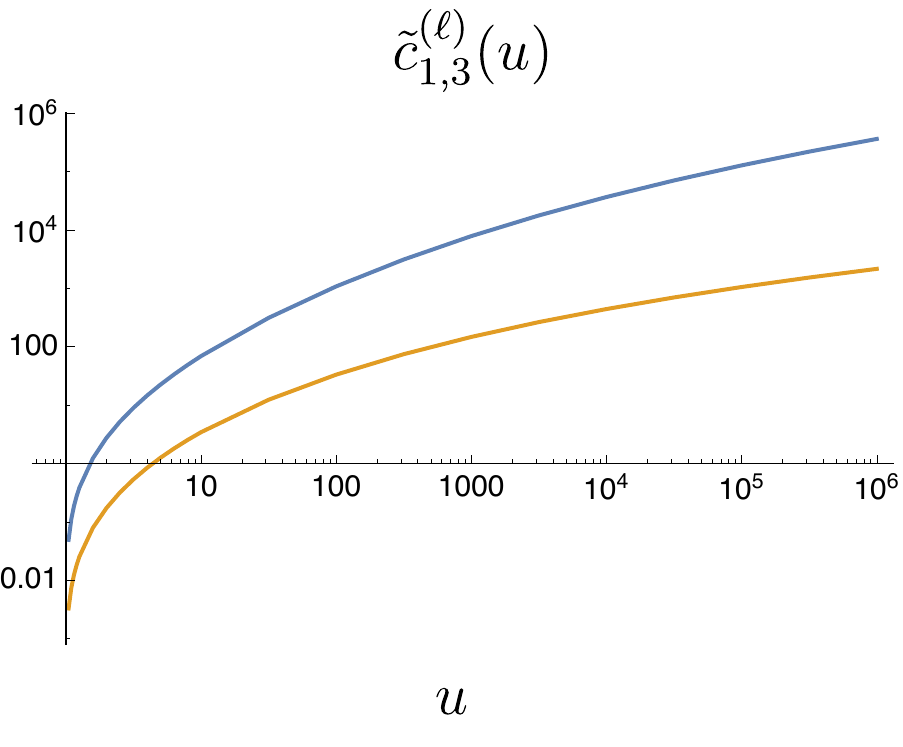}
        \end{subfigure}
  \vskip\baselineskip \vspace*{-.3cm}
        \begin{subfigure}[b]{0.24\textwidth}   
            \includegraphics[width=\textwidth]{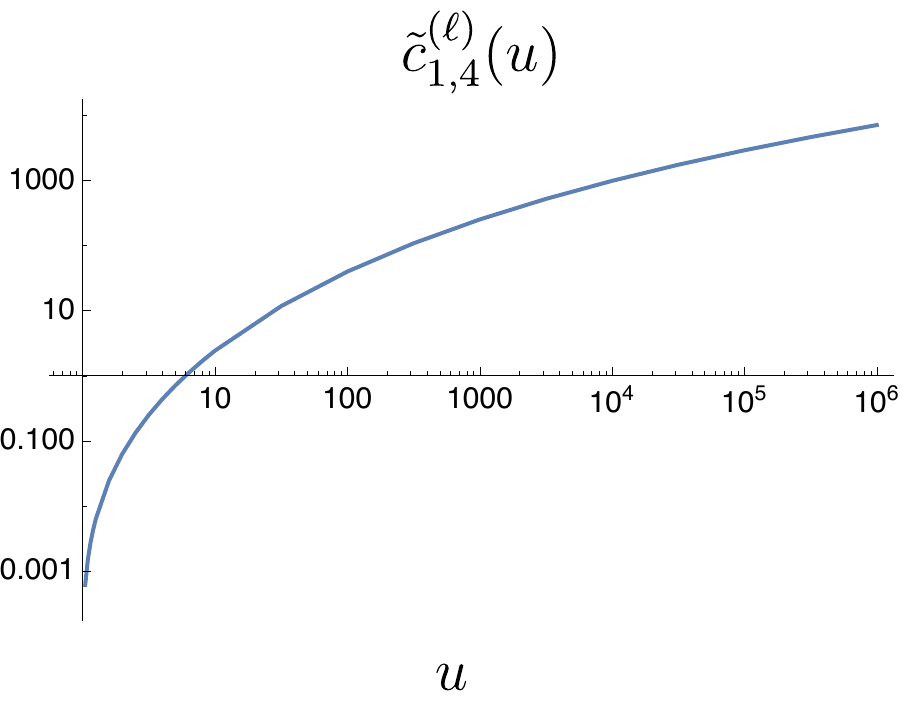}
        \end{subfigure}
        \begin{subfigure}[b]{0.24\textwidth}
            \includegraphics[width=\textwidth]{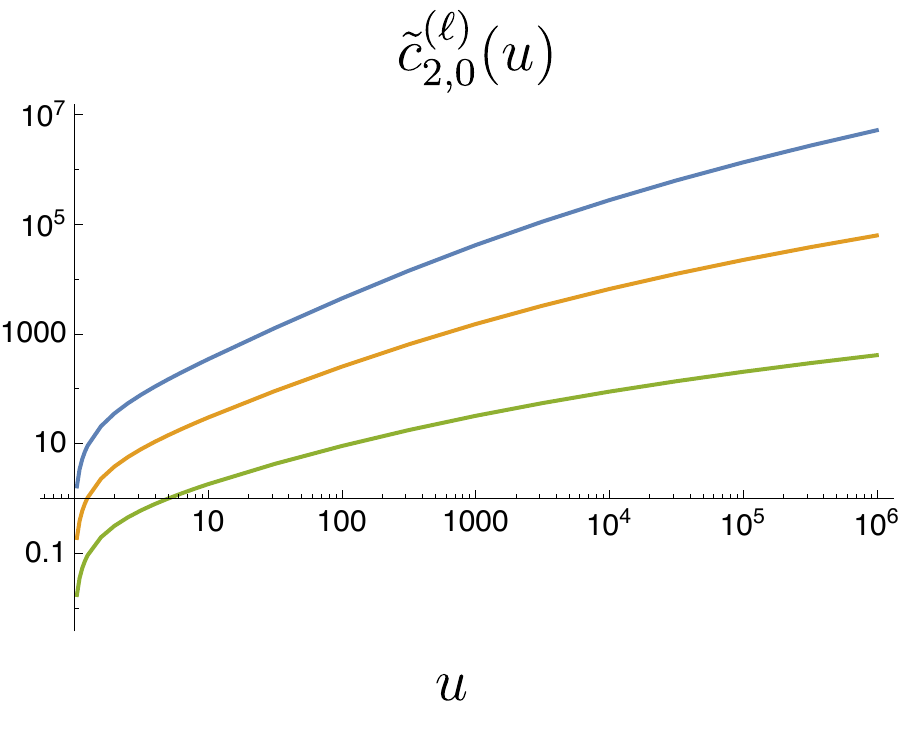}
        \end{subfigure}
        \begin{subfigure}[b]{0.24\textwidth}  
            \includegraphics[width=\textwidth]{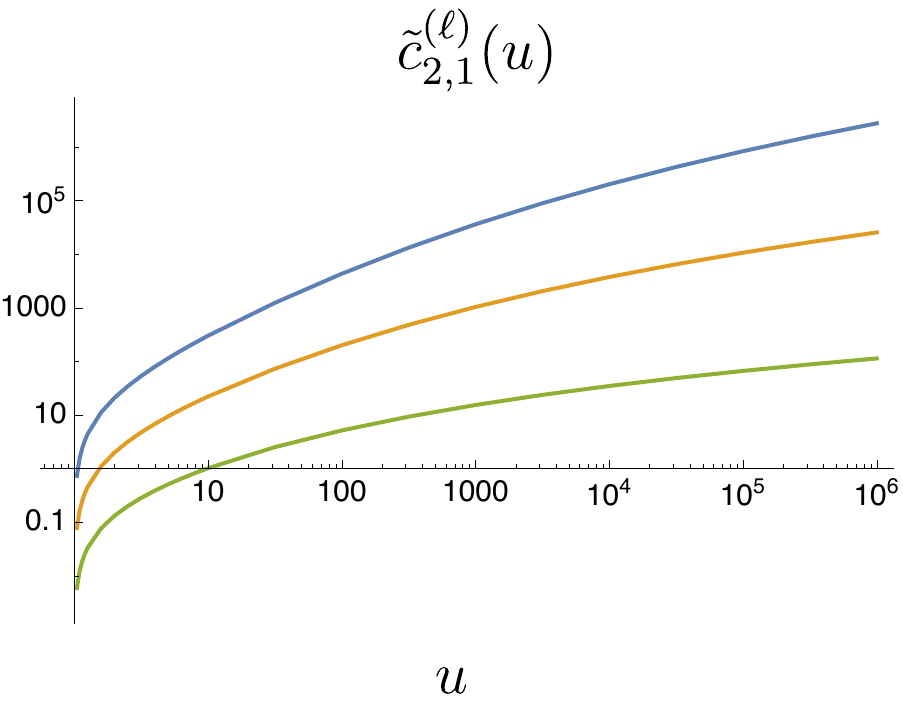}
        \end{subfigure}
        \begin{subfigure}[b]{0.24\textwidth}   
            \includegraphics[width=\textwidth]{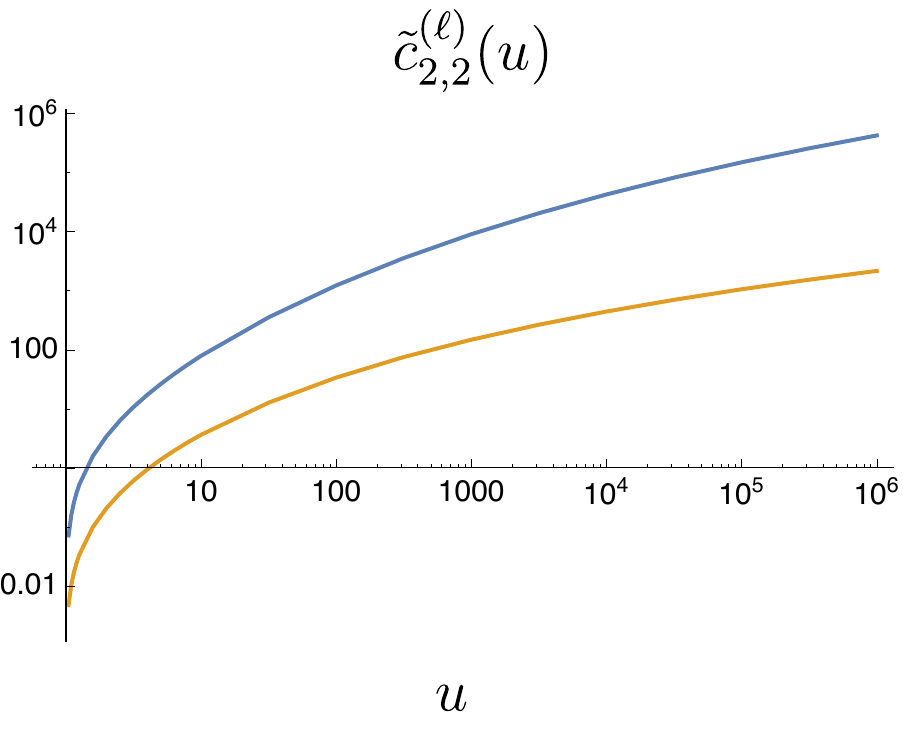}
        \end{subfigure}
  \vskip\baselineskip \vspace*{-.3cm}
        \begin{subfigure}[b]{0.24\textwidth}   
            \includegraphics[width=\textwidth]{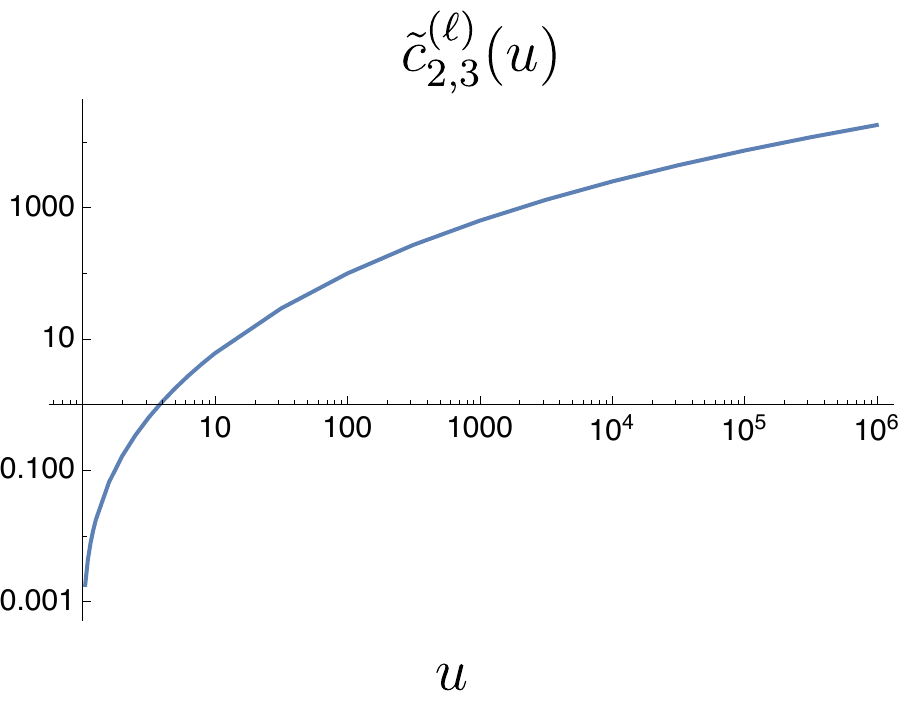}
        \end{subfigure}
        \begin{subfigure}[b]{0.24\textwidth}
            \includegraphics[width=\textwidth]{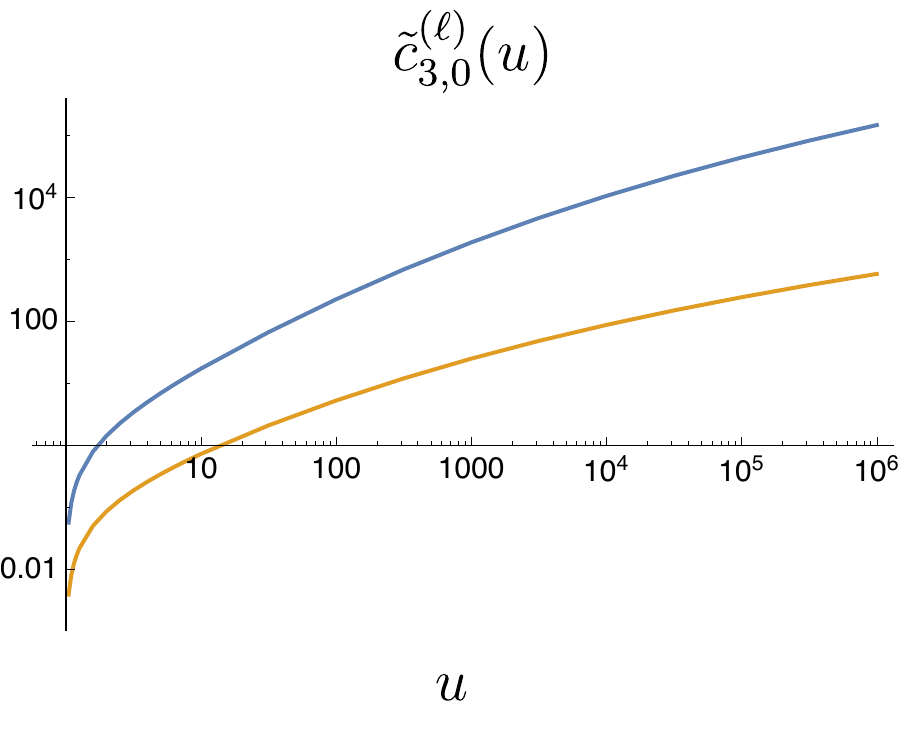}
        \end{subfigure}
        \begin{subfigure}[b]{0.24\textwidth}   
            \includegraphics[width=\textwidth]{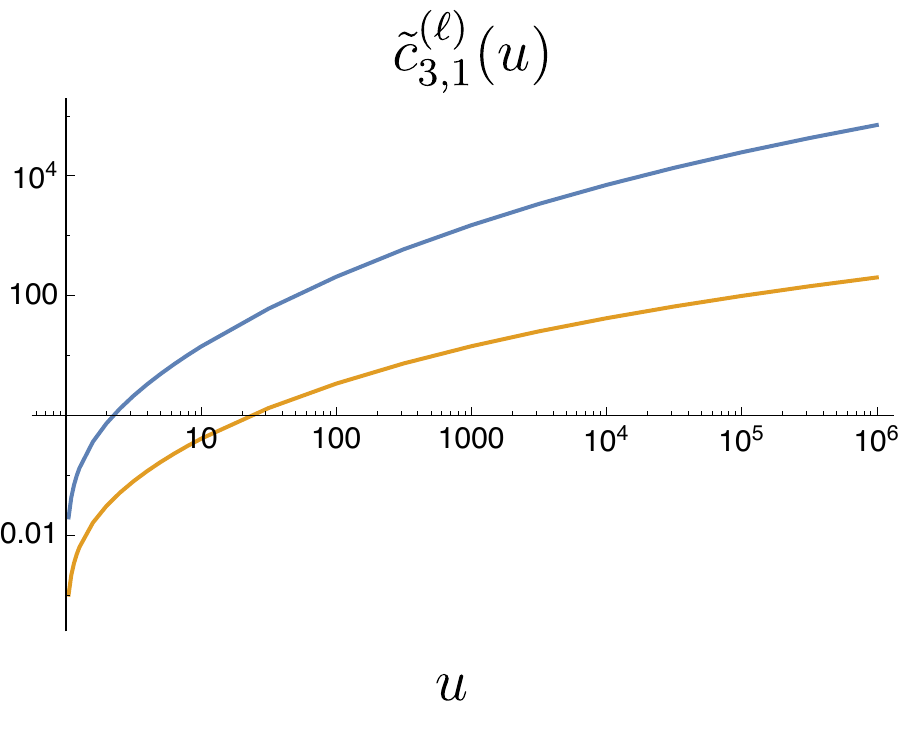}
        \end{subfigure}
        \begin{subfigure}[b]{0.24\textwidth}   
            \includegraphics[width=\textwidth]{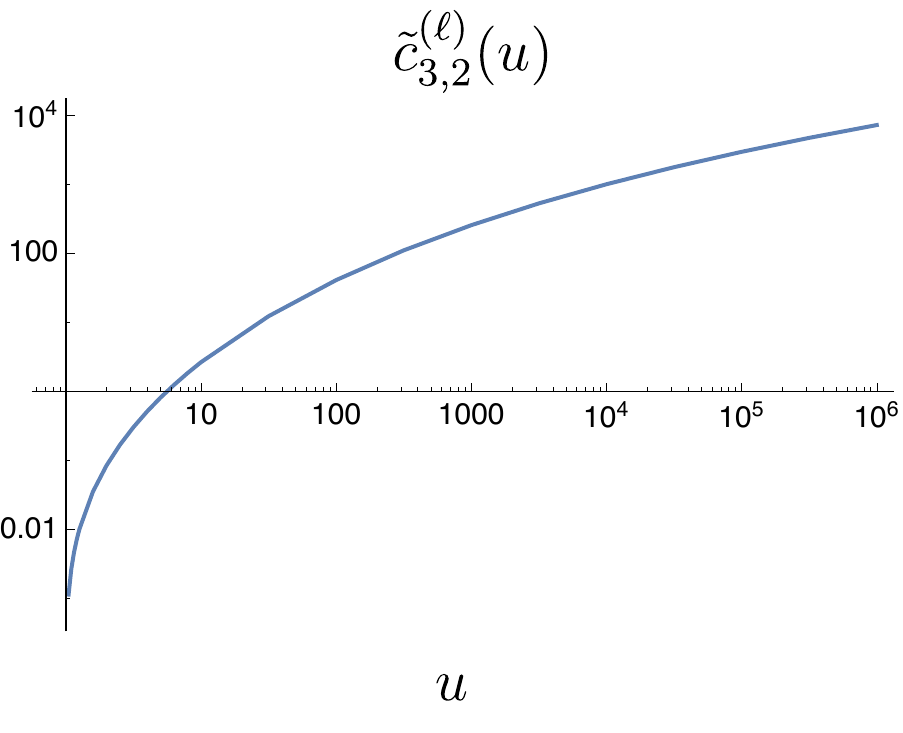}
        \end{subfigure}
  \vskip\baselineskip \vspace*{-.3cm}
        \begin{subfigure}[b]{0.24\textwidth}   
            \includegraphics[width=\textwidth]{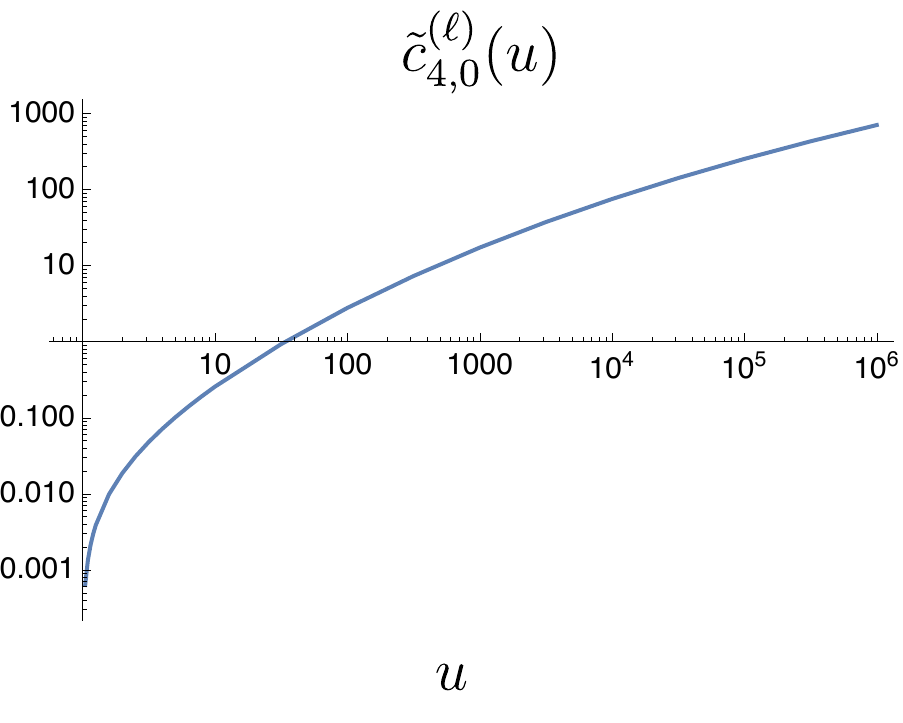}
        \end{subfigure}
        \begin{subfigure}[b]{0.24\textwidth}   
            \includegraphics[width=\textwidth]{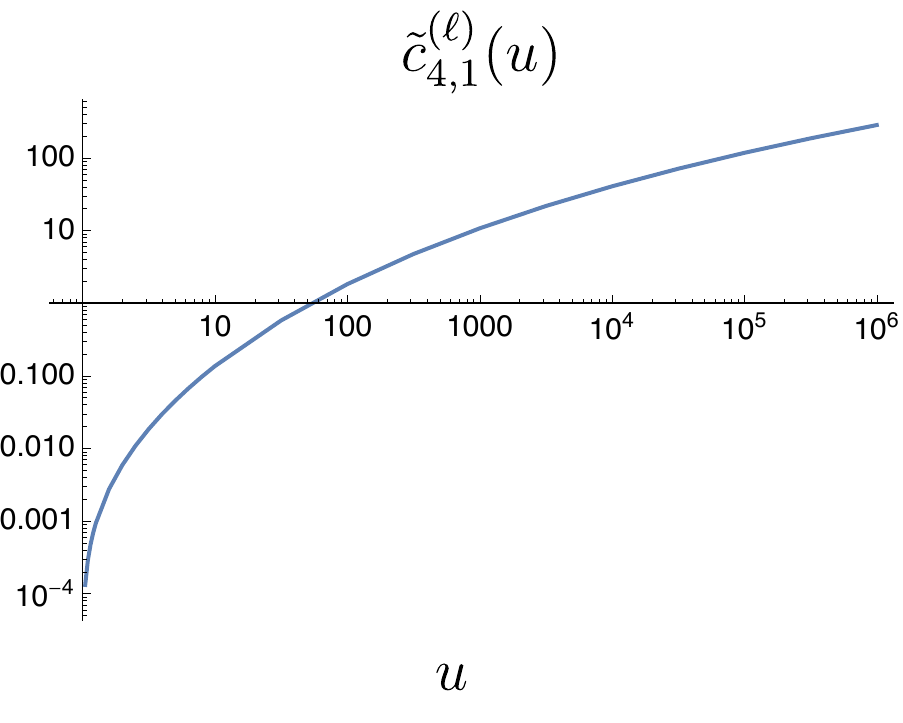}
        \end{subfigure}
        \caption[]
        {The coefficient functions $\tilde{c}^{(\ell)}_{n,k}(u)$ for the $w\to0$ edge of the double-scaling limit at $\ell$ loops. Five loops is shown in blue, four loops in yellow, three loops in green, and two loops in red.}
        \label{ci(u,0)}
    \end{figure*}


\section{Proof that \texorpdfstring{$c_1^{(2)}(u,w)$}{c12(u,w)} is positive and monotonic}
\label{c21monotonicity}

The coefficient function $c_1^{(2)}(u,w)$ has weight 3, which 
guarantees that it can be represented in terms of classical polylogarithms.
From its coproduct representation we found that
\bea
c_1^{(2)}(u,w) &=& - {\rm Li}_3\biggl(\frac{u+w-1}{uw}\biggr)
+ {\rm Li}_3\biggl(\frac{u+w-1}{u}\biggr) 
+ {\rm Li}_3\biggl(\frac{u+w-1}{w}\biggr)
\nonumber\\ &&\hskip0.0cm\null
- \log w \, {\rm Li}_2\biggl(\frac{u+w-1}{u}\biggr)
- \log u \, {\rm Li}_2\biggl(\frac{u+w-1}{w}\biggr)
\nonumber\\ &&\hskip0.0cm\null
- \frac{1}{2} \, \log(uw) 
\, \Bigl( {\rm Li}_2(1-u) + {\rm Li}_2(1-w) - \zeta_2 \Bigr)
\nonumber\\ &&\hskip0.0cm\null
- \frac{1}{2} \, \Bigl( \log^2 u \, \log(1-u) + \log^2 w \, \log(1-w) \Bigr) \,.
\label{c21_polylog}
\eea
Note that it vanishes on the collinear boundary $u+w=1$: $c_1^{(2)}(u,1-u)=0$.
The representation~(\ref{c21_polylog}) is manifestly real
for $u,w>0$ and $u,w<1$. It can acquire an imaginary part in other regions,
so another representation might be preferable in principle.

However, we are going to take its radial derivative now, and write the
result in a manifestly real form:
\be
c_{1,r}^{(2)}(u,w) \equiv (u \del_u + w \del_w) c_1^{(2)}(u,w) = 
\frac{c_0^{(1)}(u,w)}{u+w-1}
- \frac{1}{2} \, \biggl[ \frac{1}{1-u} + \frac{1}{1-w} \biggr]
 \, \log u \, \log w \,,
\label{c21_r}
\ee
where
\be
c_0^{(1)}(u,w) = - C^{(1)}(u,w) 
= - {\rm Li}_2 (1-u) -  {\rm Li}_2 (1-w) - \log u \log w + \zeta_2 
\label{minusdsoneloop}
\ee
is positive and monotonically increasing, from the previous
one-loop analysis.

Although the first term in \eqn{c21_r} is positive in the positive 
double-scaling region~(\ref{ds_positive_region}), the second term can
be negative (say, for $u<1$ and $w<1$).  So we have to show that 
the second term is outweighed by the first term.

Rather than working with dilogarithms, we take another radial derivative.
First we multiply by the quantity $(u+w-1)$, which is uniformly
positive in the positive region.  So if we can show that
$(u+w-1) c_{1,r}^{(2)}$ is positive, it's the same as showing $c_{1,r}^{(2)}$
is positive.  It's easy to see that $c_{1,r}^{(2)}(u,w)$ is regular
on the collinear boundary, because $c_0^{(1)}(u,w)$ vanishes there.
Hence $(u+w-1) c_{1,r}^{(2)}$ vanishes there, which allows a radial flow argument to work.  Multiplication by $(u+w-1)$ before differentiating also
allows the radial derivative to kill the polylogarithms:
\bea
c_{1,rr}^{(2)}(u,w)
&\equiv& (u \del_u + w \del_w) \Bigl[ (u+w-1) c_{1,r}^{(2)}(u,w) \Bigr]
\nonumber\\
&=& - \frac{1}{2} \, \biggl[ \frac{u}{(1-w)^2} + \frac{w}{(1-u)^2} \biggr]
   \, \log u \, \log w
\nonumber\\ &&\hskip0cm\null
- \frac{1}{2} \, \biggl[ \frac{u}{1-w} + \frac{w+2u}{1-u} \biggr] \, \log u
- \frac{1}{2} \, \biggl[ \frac{w}{1-u} + \frac{u+2w}{1-w} \biggr] \, \log w \,.
\nonumber\\
&=& \frac{1}{2} \, \log u \biggl[
- \frac{u \, \log w}{(1-w)^2} - \frac{u}{1-w} - \frac{w+2\,u}{1-u} \biggr]
\ +\ (u\lr w).
\label{c21_rr}
\eea
In the second form, it is enough to show that the term shown is positive
everywhere in the positive region; the same will then be true of the
term obtained by $(u\lr w)$ reflection.

Note that the contribution of the third term in brackets, 
$-(w+2u)(\log u)/(1-u)$,
always has the desired sign, positive.  Suppose first that $u>1$.
Then we combine the first two terms to get
$(-u) \times (\log w + 1-w)/(1-w)^2$.
The last factor is always negative, including $w=1$ where it approaches
a finite limit.  So we are done with the $u>1$ case.

Now let $u<1$.  In this case we have to combine all three terms,
and use the identity,
\be
\frac{u}{1-w} + \frac{w+2\,u}{1-u}\  >\  \frac{u}{w(1-w)} \,,
\ee
which can be established by writing the difference, left minus right, as
\be
\frac{w(w+u)+u(u+w-1)}{w(1-u)}\  >\  0.
\ee
Therefore
\be
\frac{u \, \log w}{(1-w)^2} + \frac{u}{1-w} + \frac{w+2\,u}{1-u} 
\ >\  \frac{u \, \log w}{(1-w)^2} + \frac{u}{w(1-w)}
\ =\  u \times \frac{\log w + \frac{1-w}{w}}{(1-w)^2} \,.
\ee
The last factor is always positive, so the quantity in brackets
in \eqn{c21_rr} is negative for $u<1$. Combined with the fact that 
$\log u < 0$ for $u<1$, we are done proving that $c_{1,rr}^{(2)} > 0$
in the positive region.  This in turn proves that $c_{1,r}^{(2)} > 0$,
and hence that $c_{1}^{(2)}(u,w)$ itself is positive.

For the next simplest quantity, the weight-4 function $c_{0}^{(2)}(u,w)$,
we tried to apply the same method of taking repeated radial derivatives,
but we were unable to remove all the trilogarithms in the second iteration,
because they come with different rational prefactors.  So an analytic
proof would probably require another method.  However,
we could establish numerically that the second such derivative, 
$c_{1,rr}^{(2)}(u,w)$ was positive in the positive region,
consistent with the more general numerical study in
section~\ref{FullDoubleScalingSubsection}.

\vfill\eject


\end{document}